\pdfoutput=1
\documentclass[a4paper,11pt]{article}
\usepackage{jcappub}
\usepackage[utf8]{inputenc} 
\usepackage{amsmath} 
\usepackage{subcaption}
\usepackage{graphicx}
\usepackage{threeparttable}
\usepackage{enumitem}
\usepackage{makecell}

\usepackage{siunitx}
\DeclareSIUnit\year{yr}

\begin{document}

\title{TAROGE-M: Radio Antenna Array on Antarctic High Mountain for Detecting Near-Horizontal Ultra-High Energy Air Showers}

\author[a,b]{Shih-Hao Wang,}
\author[a,b,1]{Jiwoo Nam,\note{\normalsize corresponding author}} 

\author[a,b,c]{Pisin Chen,}
\author[a,b]{Yaocheng Chen,}  
\author[d]{Taejin Choi,}
\author[d,e]{Young-bae Ham,}
\author[a]{Shih-Ying Hsu,} 
\author[a,b]{Jian-Jung Huang,}
\author[f]{Ming-Huey A. Huang,} 
\author[d,e]{Geonhwa Jee,}
\author[g]{Jongil Jung,}
\author[d]{Jieun Kim,}
\author[a,b]{Chung-Yun Kuo,}
\author[d]{Hyuck-Jin Kwon,}
\author[d,e]{Changsup Lee,}
\author[a,b,h]{Chung-Hei Leung,}
\author[b,i]{Tsung-Che Liu,} 
\author[b,j]{Yu-Shao J. Shiao,}
\author[b,k]{Bok-Kyun Shin,}
\author[a,b]{Min-Zu Wang,}
\author[a,b]{Yu-Hsin Wang,}

\author[l]{Astrid Anker,}
\author[l]{Steven W. Barwick,}
\author[m]{Dave Z. Besson,}
\author[n,o]{Sjoerd Bouma,}
\author[o]{Maddalena Cataldo,}
\author[l]{Geoffrey Gaswint,}
\author[p]{Christian Glaser,}
\author[n]{Steffen Hallmann,}
\author[q]{Jordan C. Hanson,}
\author[n,o]{Jakob Henrichs,}
\author[r]{Stuart A. Kleinfelder,}
\author[o]{Robert Lahmann,}
\author[n,o]{Zachary S. Meyers,}
\author[n,o]{Anna Nelles,}
\author[m]{Alexander Novikov,}
\author[l]{Manuel P. Paul,}
\author[n,o]{Lilly Pyras,}
\author[l]{Christopher Persichilli,}
\author[n,o]{Ilse Plaisier,}
\author[l]{Ryan Rice-Smith,}
\author[m]{Mohammad F.H. Seikh,}
\author[s]{Joulien Tatar,}
\author[n,o]{Christoph Welling,}
\author[l]{Leshan Zhao}

\affiliation[a]{Department of Physics, National Taiwan University, Taipei 10617, Taiwan} 
\affiliation[b]{Leung Center for Cosmology and Particle Astrophysics, National Taiwan University, Taipei 10617, Taiwan}
\affiliation[c]{Kavli Institute for Particle Astrophysics and Cosmology, SLAC National Accelerator Laboratory, Stanford University, Stanford, CA 94305, USA}
\affiliation[d]{Division of Atmospheric Sciences, Korea Polar Research Institute, Incheon, Republic of Korea}
\affiliation[e]{Department of Polar Science, Korea University of Science and Technology, Daejeon, Republic of Korea}
\affiliation[f]{Department of Energy Engineering, National United University, Miaoli, Taiwan}
\affiliation[g]{Department of Astronomy, Space Science and Geology, Chungnam National University, Daejeon 34134, Republic of Korea }
\affiliation[h]{Department of Physics and Astronomy, University of Delaware, Newark, DE 19716, USA}
\affiliation[i]{Department of Electrophysics, National Yang Ming Chiao Tung University, Hsinchu 300, Taiwan}
\affiliation[j]{National Nano Device Laboratories, Hsinchu 300, Taiwan}
\affiliation[k]{Ulsan National Institute of Science and Technology, Ulsan 44919, Republic of Korea } 

\affiliation[l]{Department of Physics and Astronomy, University of California, Irvine, CA 92697, USA}
\affiliation[m]{Department of Physics and Astronomy, University of Kansas, Lawrence, KS 66045, USA}
\affiliation[n]{DESY, 15738 Zeuthen, Germany}
\affiliation[o]{ECAP, Friedrich-Alexander-Universität Erlangen-Nürnberg, 91058 Erlangen, Germany}
\affiliation[p]{Uppsala University Department of Physics and Astronomy, Uppsala SE-752 37, Sweden}
\affiliation[q]{Whittier College Department of Physics, Whittier, CA 90602, USA}
\affiliation[r]{Department of Electrical Engineering and Computer Science, University of California, Irvine, CA 92697, USA}
\affiliation[s]{Research Cyberinfrastructure Center, University of California, Irvine, CA 92697, USA}

\emailAdd{wsh4180@gmail.com}
\emailAdd{jwnam@phys.ntu.edu.tw}

\collaboration{TAROGE and ARIANNA collaborations} 
\abstract{ 
    The TAROGE-M radio observatory is a self-triggered antenna array on top of the $\sim$\SI{2700}{\m} high Mt.~Melbourne in Antarctica, designed to detect impulsive geomagnetic emission from extensive air showers induced by ultra-high energy (UHE) particles beyond $10^{17}$\si{\eV}, including cosmic rays, Earth-skimming tau neutrinos, and particularly, the ``ANITA anomalous events'' (AAE) from near and below the horizon.
    %
    The six AAE discovered by the ANITA experiment have signal features similar to tau neutrinos but that hypothesis is in tension either with the interaction length predicted by Standard Model or with the flux limits set by other experiments.
    Their origin remains uncertain, requiring more experimental inputs for clarification.

    The detection concept of TAROGE-M takes advantage of a high altitude with synoptic view toward the horizon as an efficient signal collector, and the radio quietness as well as strong and near vertical geomagnetic field in Antarctica, enhancing the relative radio signal strength. 
    This approach has a low energy threshold, high duty cycle, and is easy to extend for quickly enlarging statistics.
    %
    %
    Here we report experimental results from the first TAROGE-M station deployed in January 2020, corresponding to approximately one month of livetime.
    The station consists of six receiving antennas operating at \SIrange{180}{450}{\MHz},
    and can reconstruct source directions of impulsive events with an angular resolution of $\sim\ang{0.3}$, calibrated {\it in situ} with a drone-borne pulser system.
    %
    To demonstrate TAROGE-M's ability to detect UHE air showers, a search for cosmic ray signals in $25.3$-days of data together with the detection simulation were conducted, resulting in seven identified candidates.
    The detected events have a mean reconstructed energy of $0.95_{-0.31}^{+0.46}$ \si{\exa\eV} and zenith angles ranging from \ang{25} to \ang{82}, with both distributions agreeing with the simulations, indicating an energy threshold at about \SI{0.3}{\exa\eV}.
    The estimated cosmic ray flux at that energy is $1.2_{-0.9}^{+0.7} \times 10^{-16}$ \si{\per\eV \per\square\km \per\year \per\steradian}, also consistent with results of other experiments.
    %
    The TAROGE-M sensitivity to AAEs is approximated by the tau neutrino exposure with simulations, which suggests comparable sensitivity as ANITA's at around \SI{1}{\exa\eV} energy with a few station-years of operation.
    %
    These first results verified the station design and performance in a polar and high-altitude environment, and are promising for further discovery of tau neutrinos and AAEs after an extension in the near future.
    }


\maketitle

\section{Introduction}

    
    %
    The origin and the characteristics of ultra-high energy (UHE, above $10^{17}$\si{\eV}) cosmic rays (CRs) have been long-standing puzzles.
    UHE cosmic neutrinos, expected to be generated as UHECR interact with radiation or matter nearby the sources or with cosmic background photons, can help resolve the mystery of the cosmic accelerators and understand the Universe at the most extreme energies (see reviews in Ref.~\cite{Coleman2022, Ackermann2022}).
    %
    
    An effective way to observe UHE particles is to detect the coherent radio emission emitted by the extensive air showers they induce in the atmosphere.
    The coherent radio signal arises mainly from the geomagnetic emission as electrons and positrons in air shower are deflected in opposite directions by the geomagnetic field; the signal is linearly polarized along the ($\vec{v} \times \vec{B}$) Lorentz force direction\cite{Huege2016}.
    The secondary emission mechanism is the Askaryan effect as the shower develops a $\sim \SI{20}{\percent}$ negative charge excess and is radially polarized from the shower axis.
    The ultra-relativistic air shower leads to forwardly beamed emission and the Cherenkov effect imprints a ring-like lateral profile of about \ang{1} opening angle in air and results in time-compressed pulses of about \SI{10}{\ns} duration \cite{ZHAireS2012}.
    %
    %
    The radio detection of UHE air showers has substantially advanced in the last decade, thanks to the precise characterization of radio emission by different radio observatories such as LOFAR \cite{LOFAR2021} and AERA \cite{AERA2016}, by controlled beam experiments \cite{T-510-2016,T-510-2022}, and also by the development of microscopic shower and radio simulations from first principles \cite{Huege2013, ZHAireS2012}.
    Cross-calibration between different approaches, including conventional particle and fluorescence detectors has made the radio technique capable of reconstructing parameters of primary particles such as direction, energy, and composition with high precision (see, e.g.~Ref.~\cite{Schroeder2017} for a review). 
    %
   %
    %
	%
    %
    The radio detection technique can also be applied to detecting UHE Earth-skimming tau neutrinos via air showers initiated by the decay of tau leptons produced when neutrinos interact with terrestrial rock via charged-current interactions.

    Antenna arrays at high altitude with synoptic views toward the horizon can be particularly efficient detectors for both UHE cosmic rays from the sky and also Earth-skimming tau neutrinos from below the horizon.
    %
    The balloon-borne ANITA experiment, a radio antenna array covering the \SIrange{200}{1000}{\MHz} frequency band at about \SI{35}{\km} altitude above Antarctica, is the pioneer of this approach.
    %
    ANITA detected not only dozens of UHECRs, but also discovered six tau-neutrino like ``anomalous events'' with signatures of upward-going air showers of \si{\exa\eV} energies over the course of four flights \cite{ANITA2016b, ANITA2018,ANITA2021}.
    %
    %
    The first two ANITA anomalous events (AAE) discovered in ANITA-I and III had steep elevation angles \ang{-27} and \ang{-35} below the horizon, while the other four found by ANITA-IV were shallower, within \ang{1} below the horizon.
    The diffuse UHE tau neutrino scenario is highly disfavored as this implies that those two steeper events traversed $\sim10$ times the neutrino interaction length predicted by Standard Model. It is also in tension with the most stringent diffuse flux limits currently set by IceCube \cite{Icecube2018} and the Pierre Auger Observatory \cite{Auger2019Nu} (hereafter Auger), and by ANITA's own lack of detection of Askaryan radiation from ice \cite{Romero-Wolf2019, ANITA2022}.

    The hypothesis of transient neutrino point sources is also in tension with both Auger's limit \cite{ANITA2022} and, at lower energies, those of the IceCube \cite{Icecube2020} experiment.
    Several alternative interpretations of AAE have been proposed, some resorting to physics beyond Standard Model (e.g.~\cite{Anchordoqui2018, Dudas2018, Huang2018, Chauhan2019,Cherry2019,Collins2019, Heurtier2019, Bhupal2020}), while others invoke less exotic explanations, e.g.~by coherent transition radiation \cite{Vries2019} or by subsurface reflections in the ice \cite{Shoemaker2020}, although that explanation contradicts data taken with HiCal-2\cite{Smith2021}.
    Clearly, more events are needed to resolve the mysterious origin and nature of AAE.
    %
    However, the exposure of ANITA is limited mainly by the relatively short duration of a balloon flight, typically of order one month per flight (roughly every three years), making it harder to significantly increase the statistics. ANITA's successor, PUEO, with significantly improved sensitivity \cite{PUEO2021} is currently in preparation.
    %

    To collect more anomalous events and clarify their origin, placing radio detectors on top of Antarctic mountains and looking for near-horizon air showers can provide an alternative way to rapidly collect the  necessary statistics \cite{Nam2020}. 
    %
    %
    Although having a smaller detection volume than a balloon, high-mountain detectors have a lower energy threshold because of the smaller distance to the air showers, a greater duty cycle, are technologically straightforward, and are easily expanded,
    and thus are able to obtain competitive sensitivity.
    High-mountain radio detectors similar to TAROGE \cite{Nam2016}, such as BEACON \cite{BEACON2022} and GRAND \cite{GRAND} were also proposed to detect Earth-skimming tau neutrinos, as detailed in Ref.~\cite{Wissel2020}.
    Similar approaches using optical and particle detectors like Trinity \cite{Trinity2021} and TAMBO \cite{TAMBO2021} have also been proposed.
    %
    Antarctic mountains are ideal places for this approach, not only because of the quiet radio-frequency (RF) background with minimal human activity but also the strong ($>$\SI{60}{\micro\tesla}) and  near vertical geomagnetic field \cite{WMM2020}, which enhances the signal-to-noise ratio and enhances the experimental sensitivity to inclined air showers from all azimuthal directions.  %
    A detection concept and environment similar to ANITA may also be helpful for understanding the origin of AAE's.
    For example, a high-mountain radio detector sensitive to signals from nearly horizontal directions can help to constrain the transition radiation scenario \cite{Vries2019} which requires down-going CR induced showers impacting on surfaces at high altitudes.

    In this article, we report initial results from the first station of the TAROGE observatory deployed atop $\sim$\SI{2700}{\m} high Mt.~Melbourne (hence called TAROGE-M) in Antarctica in 2020, the first component of the proposed, ultimate detection concept.
    %
    %
    The TAROGE-M station is an autonomous and self-triggered antenna array, combining previous experimental efforts of both TAROGE on high mountains in Taiwan \cite{Nam2016}, and the ARIANNA neutrino experiment \cite{ARIANNA2019} in Antarctica.
    The initial TAROGE-M station was installed in March 2019, to conduct an RF noise survey and exercise construction and deployment procedures, as summarized in Ref.~\cite{Nam2020}.
    %
    %
    Herein, we report the first results on UHE air shower detection. The paper is organized as follows. 
    The design of the TAROGE-M station is described in Sec.~\ref{sec:TarogeM}; the 2020 operation is summarized in Sec.~\ref{sec:operation}. In-situ calibration following deployment, including the gain calibration with Galactic noise and the performance of event reconstruction using a drone-borne pulser system are described in Sec.~\ref{sec:Galactic} and \ref{sec:Pulser}, respectively.
    Simulations of expected cosmic ray signals and the resulting sensitivity of TAROGE-M is described in Sec.~\ref{sec:CRSim}, as well as application of simulations to the experimental calibration via air showers.
    As the nature of AAE is still unknown, tau neutrino simulations were performed to estimate sensitivity, and compared to that of ANITA for discovery potential, as summarized in Sec.~\ref{sec:NuTauSim}.
    Background characterization and rejection, and the result of the UHECR search in all 2020 TAROGE-M data are presented in Sec.~\ref{sec:CRsearch}.
    Finally, the detected CR candidates are verified with measurements of polarization, arrival direction, energy, and flux, and are compared with simulation-derived predictions in Sec.~\ref{sec:ObsCR}.
    %

    \begin{figure}
    
        \centering
    	\begin{subfigure}{0.26\textwidth}
    	    \centering
    		\includegraphics[width= \textwidth]{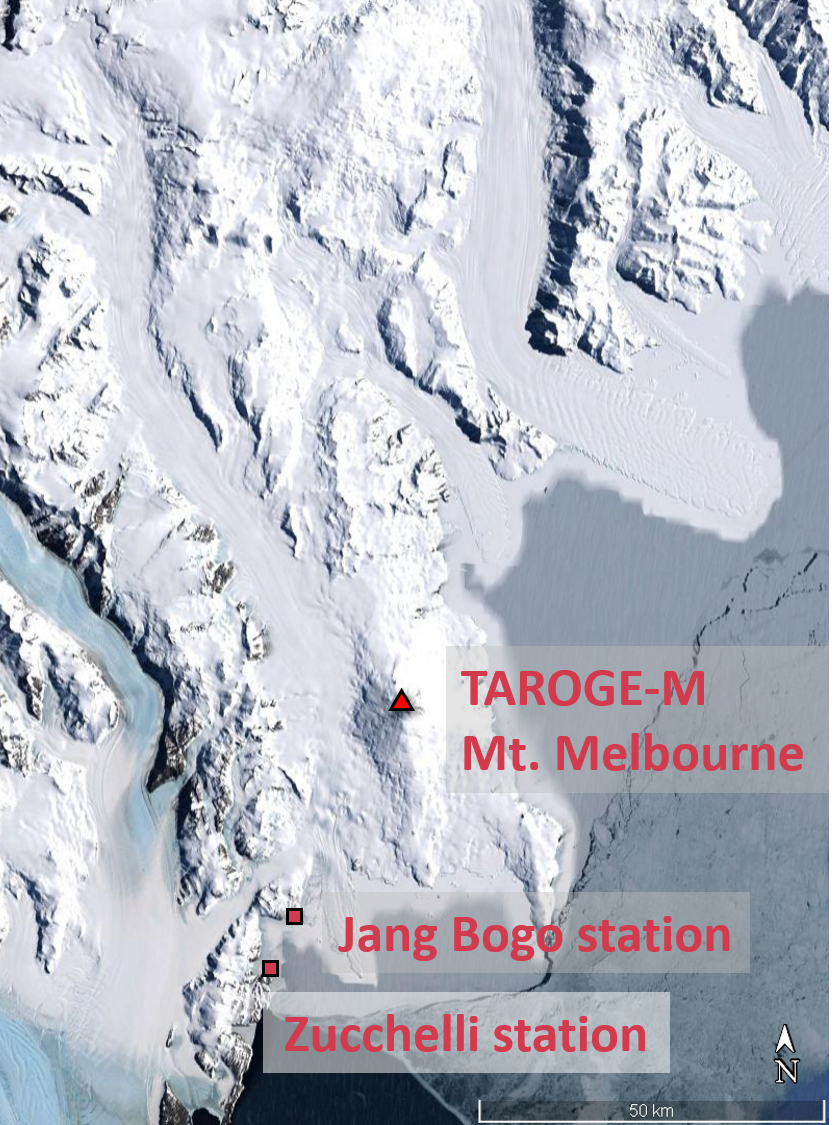}
    	\end{subfigure}
    	\begin{subfigure}{0.7\textwidth}
            \centering
            \includegraphics[width=\textwidth ] 
            {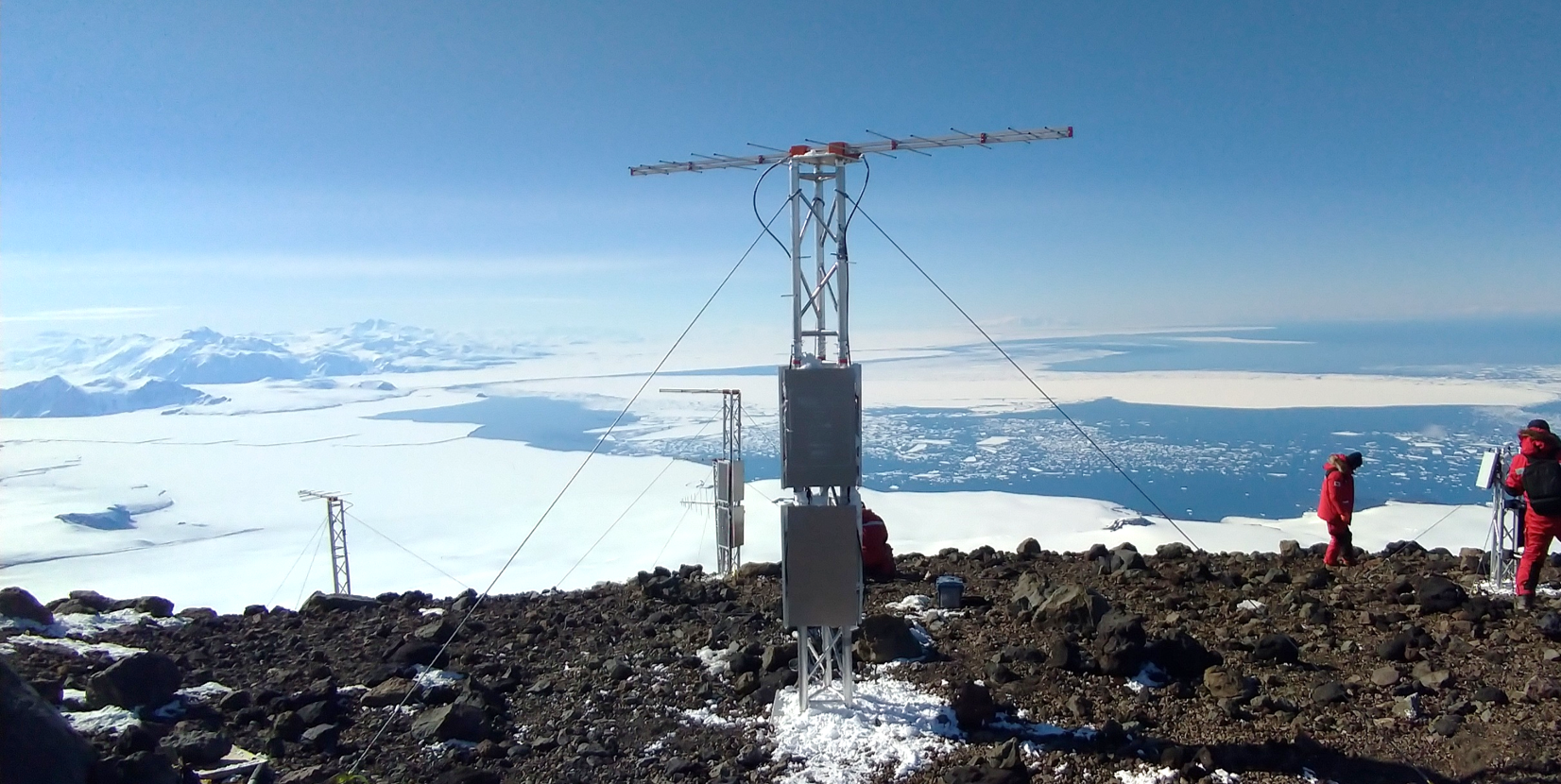}
        \end{subfigure}
    \caption{
        Left: Map in the vicinity of Mt.~Melbourne (from Google Earth Pro \protect\cite{GoogleEarth}).
        Right: the first TAROGE-M station atop Mt.~Melbourne in Antarctica, deployed in January 2020. Six log-periodic antennas are mounted on \SI{3}{m} towers, pointing toward the northern horizon (left), with a separation of approximately \SI{8.5}{m} between the lower three towers, and approximately \SI{19}{m} between the top (including the veto antenna) and bottom ones, 
        The main DAQ system is placed between the top and the middle towers, whereas the satellite communications antenna and wind turbine are installed on a shorter tower at the back side (right).
    }
     \label{fig:TMphoto}
       
    \end{figure}

\section{TAROGE-M Station Design}
   
    \label{sec:TarogeM}
\begin{figure}
        \centering
        \includegraphics[width=0.7\textwidth ]{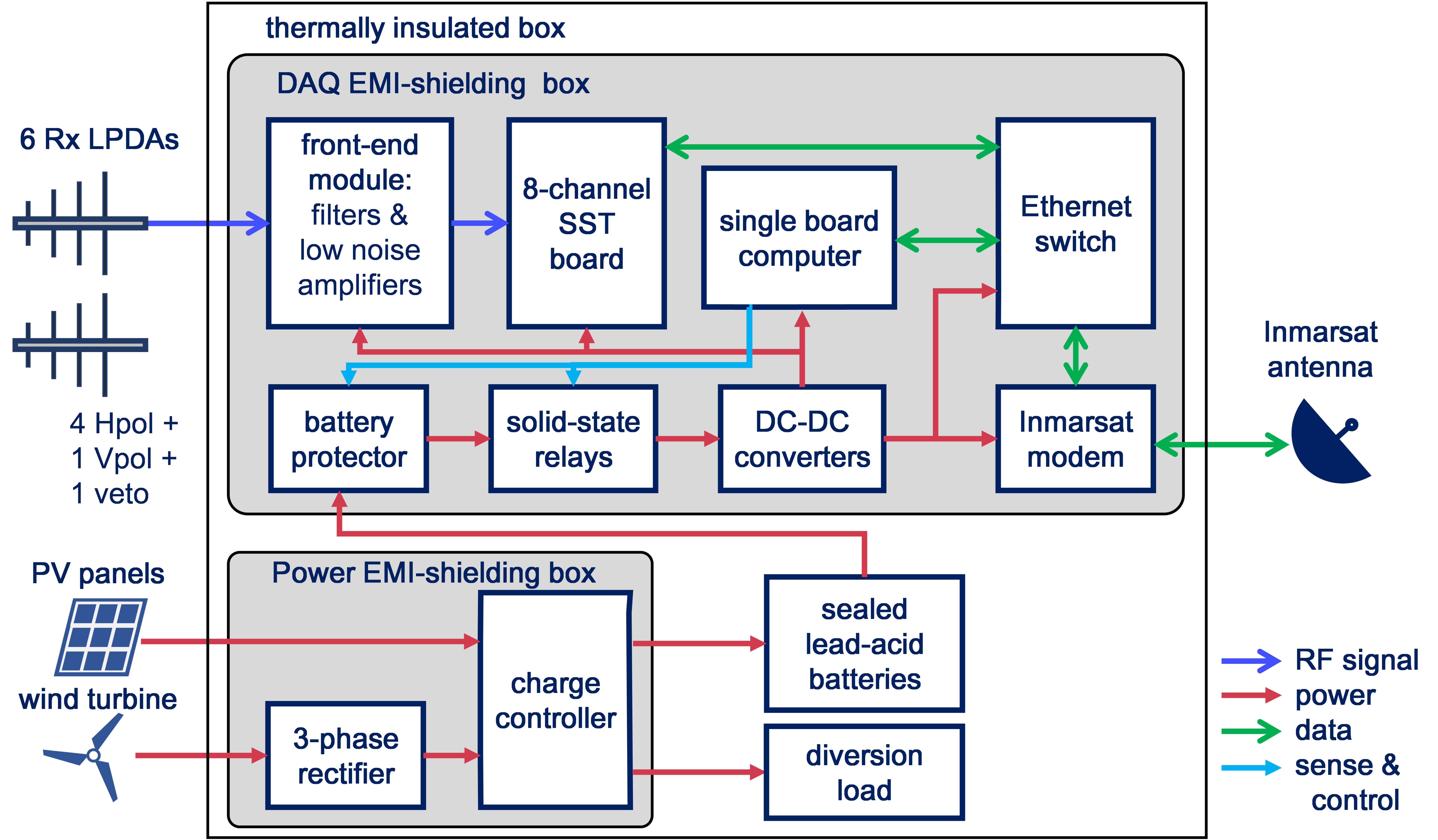}
 
    \caption{   
       Schematic overview of the TAROGE-M system.
    }
      \label{fig:System}
    \end{figure}

    %
    %
    The system design of the first TAROGE-M station on Mt.~Melbourne (Fig.~\ref{fig:TMphoto}) is summarized in the diagram in Fig.~\ref{fig:System}.
    This section begins with a discussion of the environmental conditions at the selected site, followed by details of each system module.
    
    \subsection{Environment at Mt.~Melbourne in Antarctica}
    
        One suitable location is Mt.~Melbourne, an Antarctic volcano of \SI{2720}{m} elevation near the coast at northern Victoria Land (\ang{74;20;55.88} S, \ang{164;41;35.38} E).
        The mountain at Terra Nova Bay is accessible by helicopter and located about \SI{30}{\km} northeast of the Jang Bogo research station (JBS), which is operated by the Korea Polar Research Institute (KOPRI) and provides extensive logistical support, without which deployment would not be possible.
        %
        The map in the vicinity of Mt.~Melbourne is shown in Fig.~\ref{fig:TMphoto}.
        The strength of the geomagnetic field at the experimental site is \SI{63}{\micro\tesla}, with  \ang{-82} inclination (upward) and  \ang{131} declination (southeastward), according to the World Magnetic Model (WMM) 2020 \cite{WMM2020}.
        %
        The surface on the mountaintop is a thin layer of soil above the permafrost, and is only partially covered by snow in summer.

        The TAROGE-M station was built near the top of the northern side of the volcanic crater to avoid potential electromagnetic interference (EMI) from the research stations towards the South; receiver antennas correspondingly face north, with a vast unpopulated area devoid of any radio transmitters within the horizon distance of \SI{185}{\km} (see the map in Fig.~\ref{fig:TMphoto}).
        However, there are existing facilities behind the site at a distance of several hundred meters, including radio repeaters, an automated weather station (AWS), and seismic stations, which may also generate EMI. 
        Therefore, a prototype antenna station was built in March 2019 to both measure the RF background and test the planned construction procedures (more details in Ref.~\cite{Nam2020}). 
        It was found that the RF background is insignificant most of the time and the Galactic noise can be observed (see Sec.~\ref{sec:Galactic}), with occasional active narrow-band, continuous-wave (CW) communications noise observed at \SIrange{140}{160}{\MHz} and \SIrange{340}{360}{\MHz}.
 
        %
        %
        A temperature logger was also installed and recorded data during 2019--2020. The temperature typically varies between \SI{-10}{\celsius} and \SI{-20}{\celsius} during the polar day (summer) from November to February, cooling to \SI{-30}{\celsius} $\rightarrow$ \SI{-40}{\celsius} and occasionally down to \SI{-50}{\celsius} during the polar night (winter) from May to August.
        Hence, a station design with industrial grade devices should meet the environmental conditions.
        The prototype station was upgraded to the first station in January 2020; that design is described in the following sections.

    \subsection{Receiving Antennas}
    \label{sec:LPDA}
    
\begin{figure}
    
        \centering
        \begin{subfigure}{0.37\textwidth}
            \includegraphics[width=\textwidth]{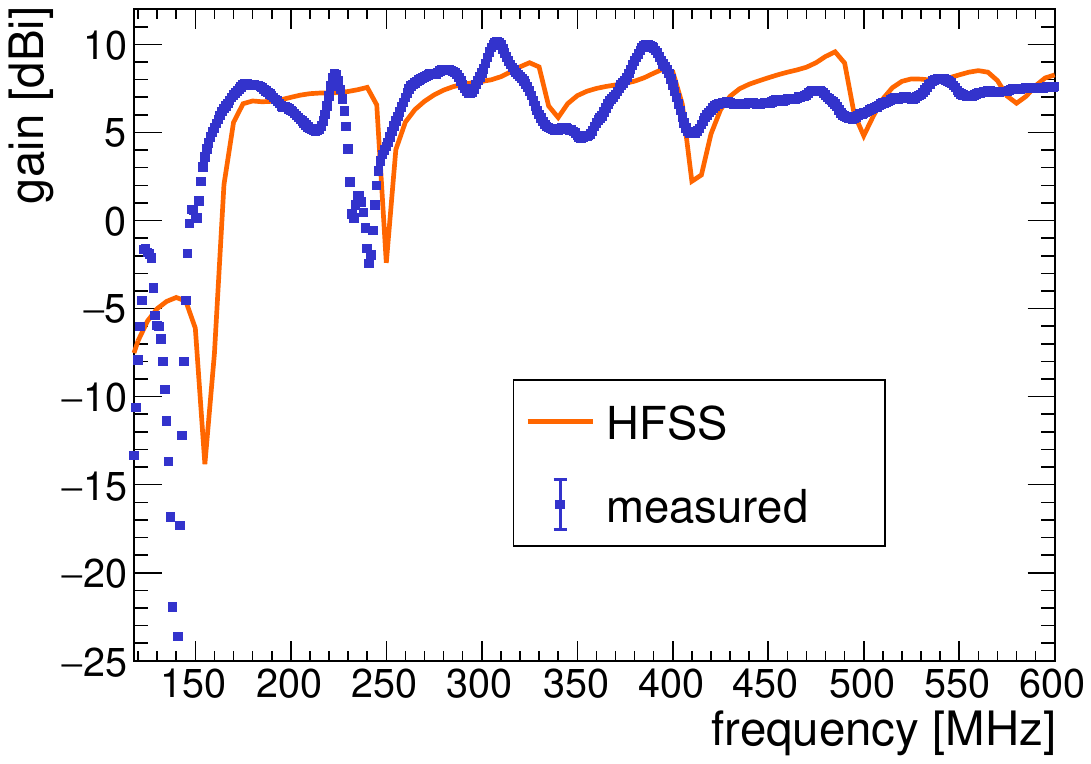}
        \end{subfigure}
        \begin{subfigure}{0.3\textwidth}
            \includegraphics[width=\textwidth]{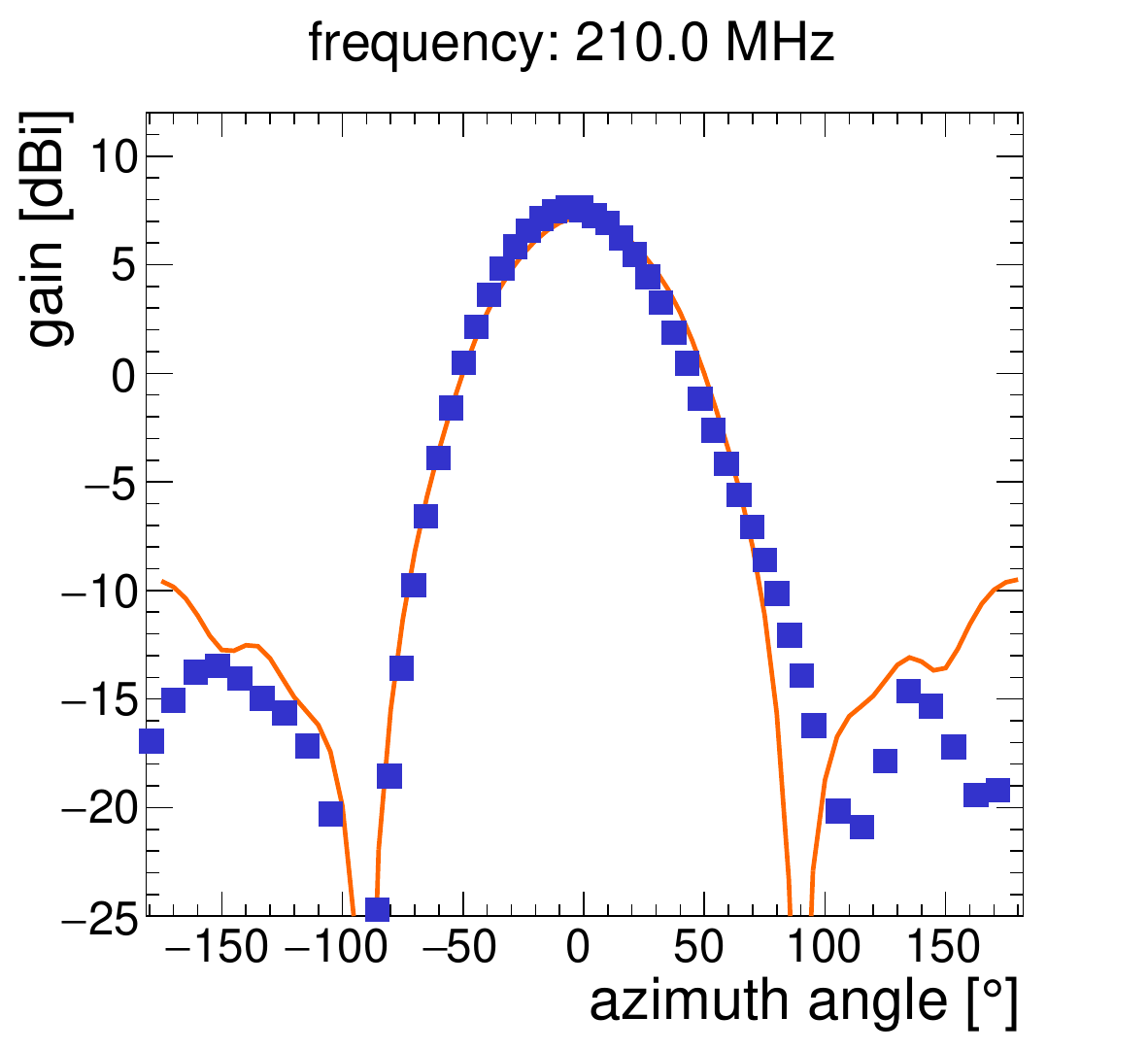}
        \end{subfigure}
        \begin{subfigure}{0.3\textwidth}
            \includegraphics[width=\textwidth]{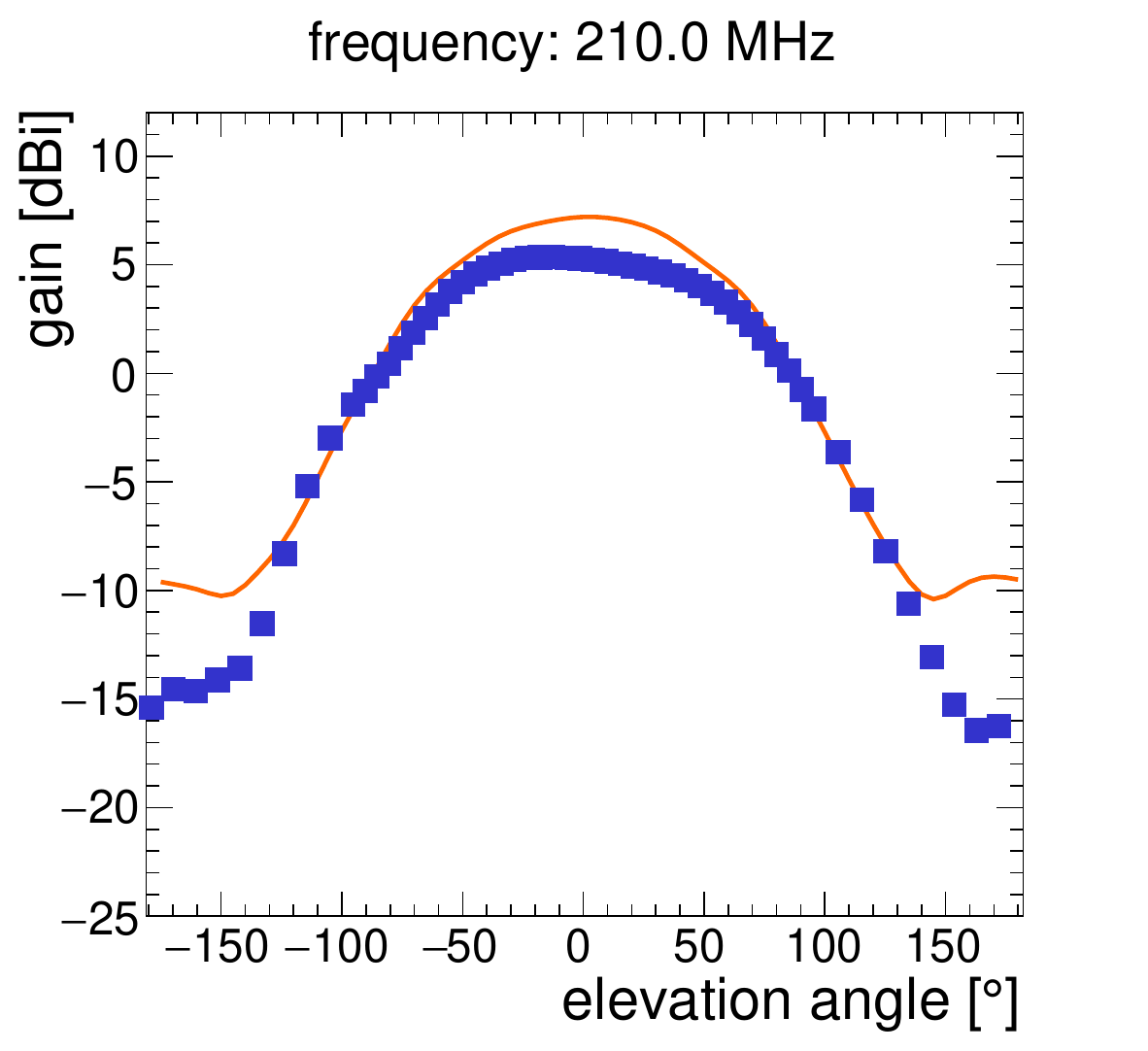}
        \end{subfigure}
        
    \caption{
        Left: realized gain of TAROGE-M LPDA at boresight as a function of frequency, from HFSS simulation (orange curve) and from measurement in anechoic chamber (blue markers).
        Middle and right: the measured (blue markers) and the simulated (orange) antenna radiation patterns at \SI{210}{\MHz} along the E-plane and H-plane, respectively. Angles are relative to the antenna boresight.
    }
     \label{fig:LPDA}
    \end{figure}

    A log-periodic dipole antenna (LPDA) is chosen as the receiving antenna for its directional and broadband properties.
    The custom-designed LPDA has about \SI{7}{dBi} gain at frequencies from \SI{180}{\MHz} to \SI{800}{\MHz}, where the lower cutoff was chosen to both avoid CWs that saturate the receiver and also for the smaller antenna size, which facilitates transport by helicopter.
    The typical measured antenna gain as function of frequency at the boresight and the co-polarized radiation patterns on E-plane and H-plane at \SI{210}{\MHz} are shown in Fig.~\ref{fig:LPDA}, and  compared to simulated results obtained with HFSS software \cite{HFSS}. The \SI{3}{\dB} beam width of the antenna in the E-plane and H-plane directions are \ang{\pm 40} and \ang{\pm 60}, respectively.
    There is about \SI{2}{\dB} gain difference at boresight between measurement and simulation, and this value is used as the systematic uncertainty for later analysis.
    The antenna is less sensitive at around \SI{250}{\MHz}, but this conveniently coincides with the satellite communication band at \SIrange{240}{280}{\MHz}, and is excluded for the final analysis.

    %
    The station has six LPDAs mounted on \SI{3}{\m} towers and pointing horizontally. Four antennas are horizontally polarized (Hpol) to align with the polarization of geomagnetic emission, one vertically polarized (Vpol) for polarimetry, and the other is Hpol but pointing opposite the other antennas, for vetoing potential EMI from behind.  

    \subsection{RF Front-End Module}
    \label{sec:FEE}
    
\begin{figure}
    \centering
    \begin{subfigure}{0.35\textwidth}
        \centering
        \includegraphics[width=\textwidth ]{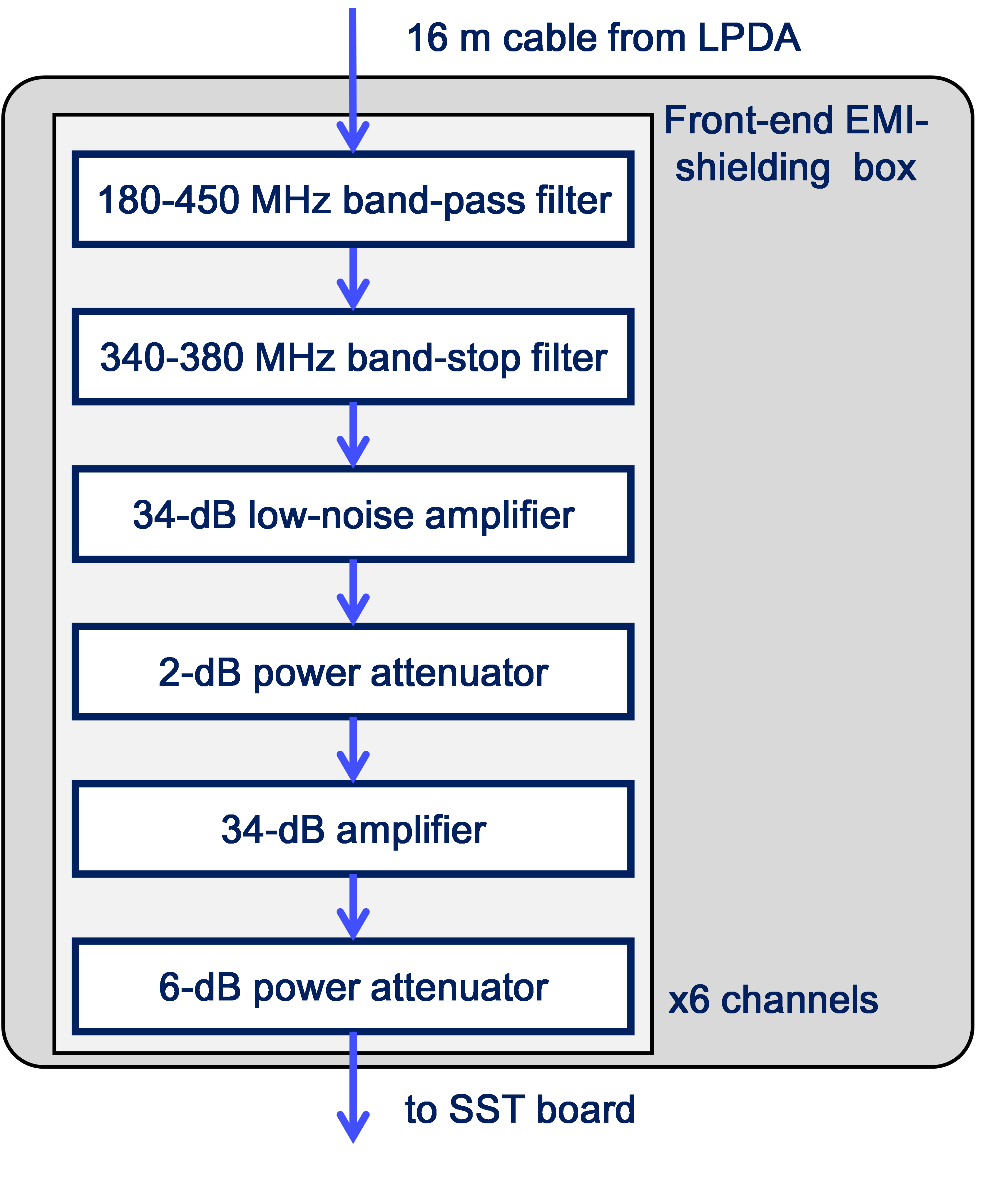}
    \end{subfigure}
    \begin{subfigure}{0.55\textwidth}
        \centering
        \includegraphics[width=\textwidth ]{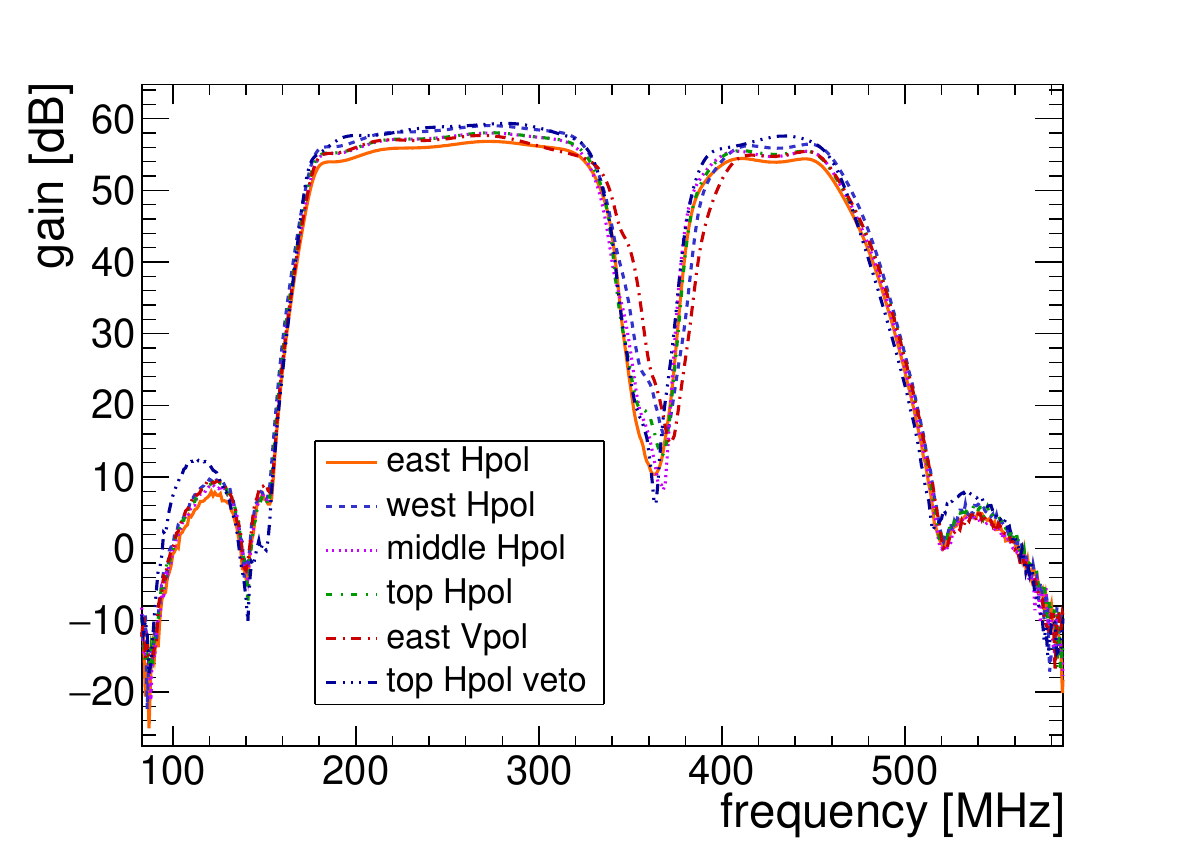}
    \end{subfigure}
        
    \caption{   
        Left: Schematic overview of TAROGE-M front-end module.
       Right: Overall frequency response (in power units) of each receiver channel, including cables and front-end modules, measured in the lab at room temperature. 
    }
      \label{fig:FEE}
    \end{figure}

    The signal received by the antenna passes through a \SI{16}{\m} length of coaxial cable to the RF front-end module located inside the system enclosure, to ensure a stable environment and simplify installation. 
    The module consists of custom-designed \SIrange{180}{450}{\MHz} band-pass and \SIrange{340}{380}{\MHz} band-stop filters,  and two cascaded low-noise amplifiers (LNAs) each of about \SI{34}{\dB} gain, followed by \SI{6}{\dB} attenuators for adjusting the dynamic range.
    The overall gain from antenna output to digitizer input is about \SI{57}{\dB}, as shown in Fig.~\ref{fig:FEE}.
    The receiver bandwidth is chosen based on the previous RF background survey for suppressing observed narrowband continuous-wave (CW) noise at \SI{150}{\MHz} and \SI{360}{\MHz} \cite{Nam2020}. 
    %
    The first-stage LNA has a noise temperature of about \SI{100}{\kelvin}, and cables and filters in front of the amplifier introduce about \SI{3.3}{\dB} of insertion loss. This leads to an equivalent receiver noise temperature of \SI{500}{\kelvin} without the antenna.
    The receiver noise and gain are also calibrated in-situ with Galactic noise, as will be described in Sec.~\ref{sec:Galactic}.

    %
	\subsection{Data Acquisition System}
	\label{sec:SST}
	 An 8-channel SST (Synchronous Sampling and Triggering) board \cite{Kleinfelder2014, Kleinfelder2015} developed for the ARIANNA experiment \cite{ARIANNA2019, ARIANNA2020} is used as data acquisition (DAQ). Its performance has been tested and verified with years of operation of ARIANNA stations at both the Ross Ice Shelf and the South Pole. 
	 The SST board has an analog bandwidth over \SI{1.5}{\GHz} and stores 256 samples with 12-bit dynamic range for each channel, with the sampling rate set to \SI{1}{GHz}. 
	 It continuously samples until stopped by a trigger, which is formed in two stages: first the channel-level trigger requires that the received waveform exceeds both positive and negative (dual-sided) thresholds within a \SI{5}{\ns} window for bipolar pulses. Then the station-level trigger requires a 3-out-of-4 coincidence of Hpol channel triggers within a \SI{32}{\ns} window for the signals to be digitized and read out. 
    %
    %
    The read-out waveforms are further processed by an on-board micro-controller for real-time narrowband noise rejection, designated as the level-one (L1) trigger. The ratio of the peak magnitude of the fast Fourier transformed spectrum to the remaining spectral sum is calculated, and an event is rejected if any of the non-veto channels has a ratio exceeding $0.3$.
    In addition, forced triggers are taken every \SI{100}{\second} for monitoring the RF background and environment.
    All triggered events are saved to a Secure Digital flash memory card.

    The SST board alternates between a data-taking phase for \SI{20}{\minute}, followed by data-transfer to an additional single-board computer (SBC).
    %
    The online event filtering routine on the SBC selects Hpol-dominated impulsive events based on a spectral analysis (with criteria as described in Sec.~\ref{sec:CRsearch}), and the filtered data are transferred to the server in the northern hemisphere via Inmarsat satellite communication. 
    The SBC also monitors the entire system, and can be remotely accessed and configured if necessary.

   \subsection{Power Module}
	The entire system has a power consumption of less than \SI{20}{\W}, and is mainly powered by eight \SI{30}{\W} solar photovoltaic panels which charge sealed lead-acid batteries totaling \SI{150}{Ah} capacity, designed for operation throughout the austral summer from August to April.
	In addition, a wind turbine was installed for investigating the feasibility of extended operation during winter.
	Both solar and wind power are controlled by a charge controller, with which excess power is diverted to a silicone heating pad inside the system enclosure to warm up the system.
   
    A battery protector which disconnects when the voltage is lower than a preset value is installed between the battery bank and DAQ system to prevent overdischarging. 
    The SBC monitors the battery voltage at the protector, and controls the solid-state relays for other components when power cycling is required.
    %
    %
    All active electronic devices are contained within EMI-shielded boxes inside a thermally insulated enclosure on the ground, which maintains a stable internal temperature of \SIrange{10}{30}{\celsius} during operation, roughly \SI{40}{\celsius} higher than the ambient temperature.

    \section{Summary of TAROGE-M in season 2020}
    \label{sec:operation}

    The deployment of the first TAROGE-M station in 2020 was accomplished within about 48 person-hours on site, and also required three helicopter flights, each for transporting 3--4 passengers, and one helicopter flight for instrumentation transport.
    The rapid deployment was made possible by the modularized system design and semi-assembled instrumentation before transportation. 
    Both factors facilitate installation in the field, and minimize the tasks of the installation of towers and antennas as well as cabling.

    The TAROGE-M station operated continuously since deployment on Jan 25, 2020, until Feb 24, 2020, when the system depleted power reserves, and shut down.
    Battery charging stopped after Feb 21, suggesting blockage of sunlight, probably due to icing accumulated on solar panels during a snow storm.
    Unfortunately, the system failed to subsequently cold start. The power problem was later identified as a DC-to-DC converter malfunctioning at low temperatures.
    %
    The power interruption resulted in a reduced total livetime of only $26.5$ days. 

    \begin{table}
        \centering
        \footnotesize
        \begin{tabular}{ |c|c|c|r|r| } 
         \hline
          
run \# & start time (UTC) & \makecell{trigger\\ threshold (mV)}  & \makecell{livetime \\ (day) }  & \makecell{number of \\ events}\\
 	\hline
	21  & 01-25 04:09 & $\pm100$  	& $ 0.230$ & $2144$ \\ 
	22  & 01-25 10:23 & $\pm90 $  	& $ 0.471$ & $0$ \\  
	23  & 01-25 22:30 & $\pm80 $  	& $ 0.971$ & $15$ \\   
	24  & 01-26 23:27 & $\pm70 $  	& $ 2.279$ & $532$ \\  
	25  & 01-29 10:19 & $\pm65 $  	& $ 7.551$ & $164093$ \\  
	26  & 02-06 20:18 & $\pm60 $  	& $10.297$ & $1067303$ \\ 
	27  & 02-19 10:22 & $\pm80 $	& $ 1.259$ & $21762$ \\
	28  & 02-20 20:31 & $\pm60 $	& $ 3.398$ & $2936$ \\
         \hline
        \end{tabular}
         \caption{
            Operation summary of TAROGE-M in 2020. Date format is [MM-DD hh:mm]. Operation was continuous between runs until shutdown at 02-24 12:17.
         }
         \label{tab:run}
    \end{table}

\begin{figure}
    \centering
        \includegraphics[width=\textwidth]{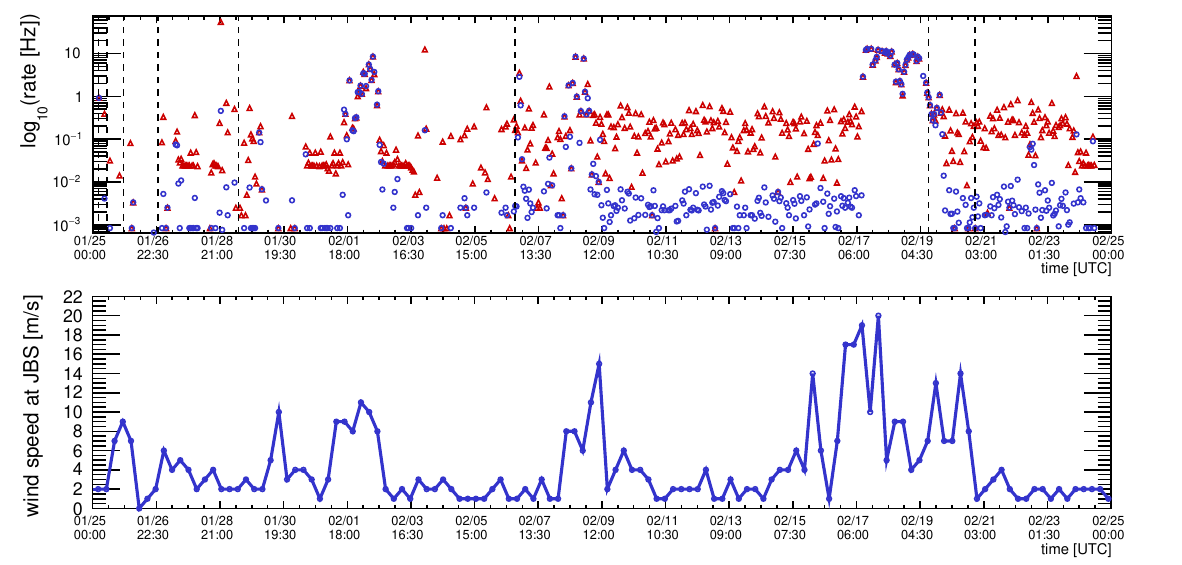}
    \caption{ 
       Top: TAROGE-M station-level event rate (in \si{\Hz}, with suppressed zero) over time on a logarithmic scale before (red) and after (blue) L1 trigger.
       %
       The vertical dashed lines mark the starting time of new run periods with different trigger thresholds (see Table \ref{tab:run}).
       Bottom: wind speed data at the Jang Bogo station (JBS) during TAROGE-M operation, extracted from \protect\cite{KDPC}.
       Periods of elevated event rate ($>$\SI{1}{\Hz}) are correlated with high wind speed over about \SI{7}{\m/\s}. 
    }
      \label{fig:wind_rate}
    \end{figure}

    The data acquisition of TAROGE-M during operation, including the trigger threshold setting, duration of separate running periods, and the number of recorded events, are summarized in Table~\ref{tab:run}. 
    As the RF background was not yet well-understood, the station started from a higher dual-sided trigger threshold in the first few days, at about 7$\times$ the RMS noise voltage ($V_{\rm rms}$, typically \SIrange{15}{17}{\mV}). 
    The threshold was manually relaxed in the following days to \SI{60}{\mV}, around $4 \cdot V_{\rm rms}$, with the exception of a temporary increase to \SI{80}{\mV} (run\# 27) on Feb 19, when the event rate spiked during a high-wind period (Fig.~\ref{fig:wind_rate}).
   The livetime of each run is estimated by the number of forced triggers recorded in a \SI{100}{\second} interval.

    %
    %
    The event rates before and after online CW rejection (L1 trigger) are shown in the top panel of Fig.~\ref{fig:wind_rate}. The L1 trigger effectively reduced the final event rate by about two orders of magnitude, leading to a typical data-taking event rate around \SIrange{1}{10}{\mHz}.
    %
    %
    CW noise predominantly arises from two frequencies at around \SI{150}{\MHz} and \SI{360}{\MHz}, which are likely associated with communication. These were generally concurrent and sporadic in the earlier operation, but became persistent after February 6. 
    Although already known and suppressed by front-end filters, the noise still sometimes was strong enough to trigger the system.

    %
    However, there were episodes of high post-L1 event rates of \SIrange{1}{10}{\Hz}, each lasting for a few hours to a few days, and found to be correlated with high wind speeds (exceeding $\approx$\SI{7}{\m/\s}) at the JBS \cite{KDPC} (Fig.~\ref{fig:wind_rate}),
    likely because the prevailing wind direction at the region around Mt.~Melbourne and JBS became eastward (from the Antarctic Plateau) during those periods.
    This is a similar phenomenon as previously reported by ARIANNA \cite{ARIANNA2017,Wang2019} and other radio neutrino experiments, reviewed in Ref.~\cite{Besson2021}.
    One plausible origin of the noise suggested by Ref.~\cite{Besson2021} is the electrostatic discharge induced by the triboelectric effect between blowing ice crystals and nearby objects, especially those isolated conductive ones with pointy edges. This effect has been reported to become more active when the wind speed exceeds a certain threshold, around \SIrange{7}{10}{\m / \second}, depending on the sensitivity of experiments.
    %
     %
    %
    This interpretation is compatible with our observation, where most of the noise events originated from behind the TAROGE-M station where there are metallic structures associated with other facilities.
    This high-wind induced radio noise (hereafter high-wind events) occurred mostly within periods of roughly $4.5$ days in total (\SI{17}{\percent} of livetime) but account for more than \SI{99.9}{\percent} of recorded events in the TAROGE-M data. Hence, this background must be characterized and effectively rejected in both online filtering and offline analysis. 
    More detailed analysis is described in Sec.~\ref{sec:highwind}.
  
    %
    
    Due to the COVID-19 pandemic and the resulting halt in scientific research at the JBS in the austral summer of 2020--2021, the DAQ system was retrieved from the mountain and sent back to the laboratory for inspection and full data access, and is currently being upgraded.
    %
    %
    The station was also inspected in the 2021--2022 season and no major hardware damage after the winters was found, demonstrating that the design and construction are durable in polar and high-altitude environment.
    %

\section{In-Situ Calibration}
    
    The antenna array mainly relies on the time difference of arrival (TDOA) between signals in receivers to reconstruct the source direction.
    It also requires understanding the receiver response for reconstructing incident electric fields for an energy measurement. Therefore, the position, timing, and the frequency response of each channel must be precisely calibrated.
    %
    Additionally, the trigger efficiency of the DAQ system as a function of signal-to-noise ratio (SNR) has to be calibrated to validate the results of the CR detection simulation (Sec.~\ref{sec:CRSim}).
    
   For the spatial information, the TAROGE-M station and nearby artificial objects were surveyed with photogrammetry and geo-referenced by differential GPS (DGPS) measurements at all antenna towers, following deployment.
   %
   The photographs were then processed using the PIX4Dmapper software \cite{PIX4D} to generate a 3D model of the station, and the position and the orientation of the receiver antennas were determined with positioning precision better than \SI{1}{\cm}.

   The receiver gain is calibrated with Galactic noise as described in the next section, followed by the timing and event reconstruction calibration accomplished with a drone-borne calibration pulser system.

    \subsection{Calibration with Galactic noise}
    \label{sec:Galactic}
    
    The RF background was monitored regularly throughout the operation by taking forced triggers every 100 seconds to obtain unbiased noise samples, yielding a total of $21,919$ recorded forced-trigger events. 
    %
    
	Absent anthropogenic backgrounds, there are three main components to the noise background: the received antenna noise with contributions from sky Galactic noise, thermal radiation from the Earth's surface in the field of view, and the receiver thermal noise generated internally within the receiver electronics.
	%
    The Galactic noise varies over time as the Earth rotates in the Galaxy with a period of one sidereal day.
    The temperature of the Earth's surface, locally mostly covered by ice, is assumed to have only a small variation over the one-month data-taking period.
    %
    %
    The receiver noise is mainly attributed to the insertion loss of the front-end electronics (i.e., before the first-stage amplifier) including cables and filters, for which the temperature also varies slightly ($\sim$\SI{20}{\kelvin}) over time.
	Therefore, by observing the Galactic noise variation in the background events over time and comparing with the expected power profile, the receiver amplitude response and noise temperature can be calibrated {\it in situ}.

	The calibration here follows steps similar to those described in Ref.~\cite{ARIANNA2017,LOFAR2015}.
	%
    %
    Numerically, the observed power versus time $P_{\rm obs}(t)$ is compared with the expected antenna noise power $P_{\rm ant}(t)$ and receiver noise power $P_{\rm rx}(t)$, with an overall gain correction factor $a$:
    \begin{equation}
         P_{\rm obs}(t) = a [P_{\rm ant}(t) + P_{\rm rx}(t)] = a P_{\rm ant}(t) + a \overline{G}_{\rm amp} kT_{\rm rx} W
         \label{eq:NoisePow}
    \end{equation}
    where the last term is independent of time, with $k$ Boltzmann's constant, $T_{\rm rx}$ the equivalent receiver noise temperature determined at the antenna output terminal, $W$ the frequency bandwidth considered,  and $\overline{G}_{\rm amp}$ the average in-band receiver gain.
    The lower passband at \SIrange{180}{240}{\MHz} is chosen for the calibration because both Galactic noise and the sought-after air shower signal are stronger at lower frequencies.

\begin{figure}
        \centering
        \begin{minipage}{0.55\textwidth}
           	\includegraphics[width=\textwidth ]{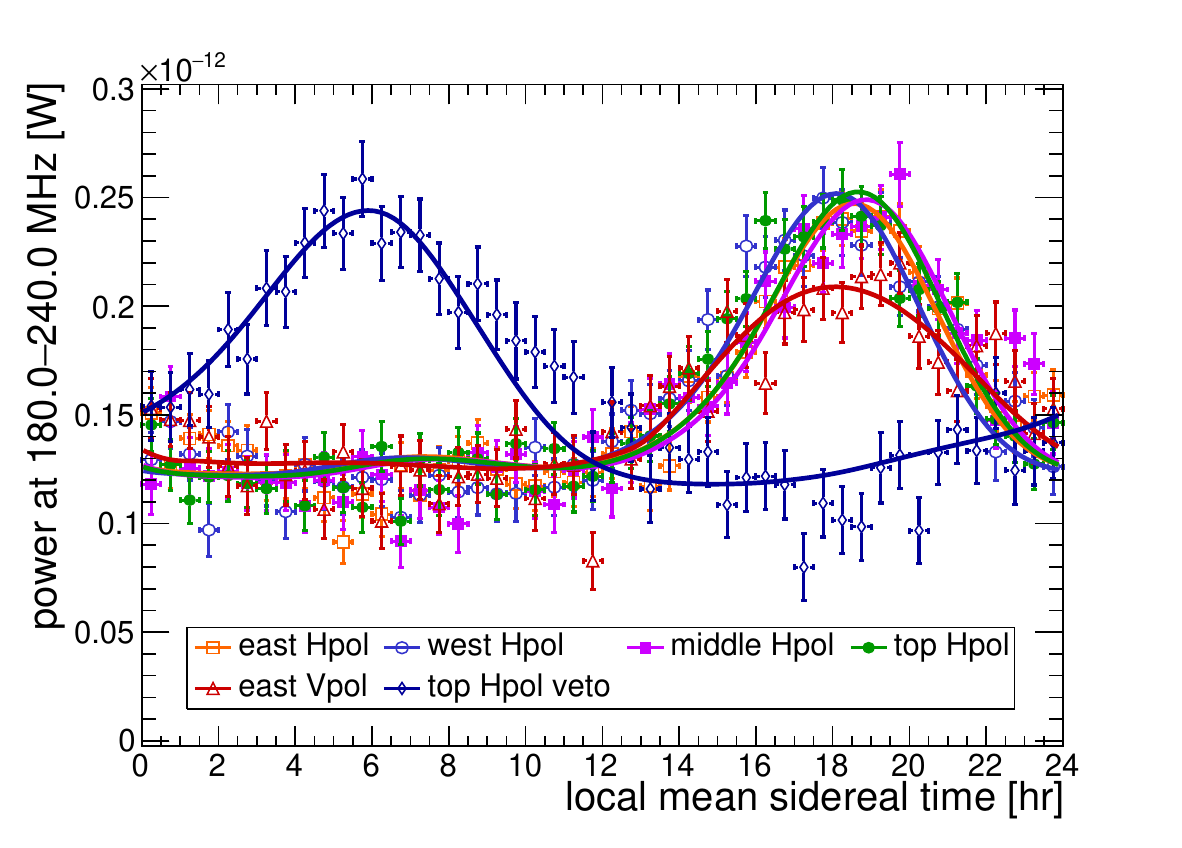}
        \end{minipage}
        \begin{minipage}{0.4\textwidth}
        \centering
        \captionsetup{type=table} 
           \footnotesize
            \begin{tabular}{ |c|c|c|c|c| } 
             \hline
              channel & $a$  & $G$ (dB)  & $\chi^{2}/dof$  \\
             \hline
            	east H & $1.79\pm0.08$ &  57.6  & 1.41 \\
                west H  & $1.22\pm0.06$ &  58.2  & 1.23 \\
                mid  H  & $1.20\pm0.06$ &  57.1 & 1.46 \\
                top  H  & $1.24\pm0.05$ &  57.4 & 1.30 \\
                east V  & $1.11\pm0.08$ &  56.8 & 1.28 \\
                veto  & $0.99\pm0.06$ &  57.6  & 1.06 \\
             \hline
            \end{tabular}
        \end{minipage}
       
        \caption{   
           Left: mean observed (points) noise power in the  \SIrange{180}{240}{\MHz} band of each channel, as a function of local sidereal time, derived from a total of $21,919$ forced-trigger events in $48$ time bins, with receiver gain- and noise- corrected using best-fit values. Error bars indicate the standard error on the mean.
           Curves are the expected profiles simulated with LFMap \protect\cite{LFMap} and HFSS \protect\cite{HFSS} (see text).
            As expected, the veto channel is out of phase compared to the other channels, given its opposite orientation. 
            %
            %
            Right: best-fit values of gain correction factor $a$, corresponding mean gain $G_{\rm amp}$, and reduced $\chi^{2}$ for the fit quality.
        }
        \label{fig:Galactic}
    \end{figure}

	The received voltage waveforms of the forced-trigger events are Fourier-transformed, and their power spectral densities summed over the band.
	%
	To check for a periodic power variation over time consistent with celestial origin, the event timestamp is expressed in local mean sidereal time (LMST) within a sidereal day and divided into 48 time bins (with each bin containing roughly 450 events).
	%
	To reject events contaminated by transient noise (e.g.~high-wind or CW), background events of particularly strong power at each channel are excluded by iteratively computing the median power of each time bin and removing outliers of more than five standard deviation ($5\sigma$) from the median, as the power is expected to be Gaussian-distributed.
	%
	This results in removing less than 4 events for all channels, with the exception of the middle Hpol, for which 34 events are removed.
	%
	The mean received power as function of LMST of each channel is shown in Fig.~\ref{fig:Galactic}.
	The fact that the variation of the veto channel, which points opposite the remaining antennas, is out of phase to other channels, together with the high precision on the mean power and the small number of outliers, further establish the sidereal periodicity and celestial origin of the signal.

      %
    The expected antenna noise power at a given time, $P_{\rm ant}(t)$, equals the integral of the brightness distribution $B(\theta,\phi,f)$ convolved with the measured electronics response (Fig.~\ref{fig:FEE}) and simulated antenna radiation pattern represented by the effective area $A_{e}$ or gain $G_{\rm ant}$, over all directions and frequency band \cite{ARIANNA2017}:
    \begin{align}
       P_{\rm ant}(t) &= \frac{1}{2} \int df G_{\rm amp}(f)
       \int d\Omega B(\theta,\phi,f,t) A_{e}(\theta,\phi,f)  \nonumber \\ 
        &= \frac{1}{2} \int df G_{\rm amp}(f) \int d\Omega [\frac{2 k f^{2}  T_{B}(\theta,\phi,f,t)}{c^{2}}] [\frac{c^{2}G_{\rm ant}(\theta,\phi,f)}{4 \pi f^{2}}] \\
        &\approx \frac{k}{4 \pi} \Delta f \Delta \Omega  \sum_{f} G_{\rm amp}(f) \sum_{\Omega}   T_{B}(\theta, \phi, f,t) G_{\rm ant}(\theta, \phi, f) 
    \label{eq:AntNoise}
    \end{align}
    %
    where the factor of $1/2$ comes from time averaging of unpolarized noise received by a linearly polarized antenna, and the brightness is approximated by the Rayleigh–Jeans law at radio frequencies and is related to the brightness temperature $T_{B}(\theta, \phi, f)$.
    The LFMap package \cite{LFMap} was used to generate the brightness temperature maps of sky Galactic noise in equatorial coordinates at each frequency bin within a \SI{10}{\MHz} interval, whereas a constant ice temperature of \SI{243}{\kelvin} was assumed below the horizon, neglecting any features in the local terrain.
     %
    The angular integral is approximated by summing in equatorial coordinates.
    %
    The frequency integral is approximated by summing the individual contributions in $\Delta f=$\SI{10}{\MHz} bins.
    
   %
    The receiver gain is calibrated by substituting Eq.~\ref{eq:AntNoise} into Eq.~\ref{eq:NoisePow} and fitting the observed power profile versus time, with the result summarized in Fig.~\ref{fig:Galactic}.
    %
    %
    The simulated and observed noise profiles are in good agreement, and the gain is consistent with that measured in the lab (Fig.~\ref{fig:FEE}) to within \SI{25}{\percent} except for the east Hpol channel, which has an offset of about \SI{80}{\percent}.
    The calibrated gains are used in the cosmic ray simulations (Sec.~\ref{sec:CRSim}) and in the signal deconvolution (Sec.~\ref{sec:CRsearch}).
 
\subsection{Calibration with Drone-borne Pulser} \label{sec:Pulser}

    \begin{figure}
       \centering
        \begin{subfigure}{0.48\textwidth}
	    \includegraphics[width=\textwidth]{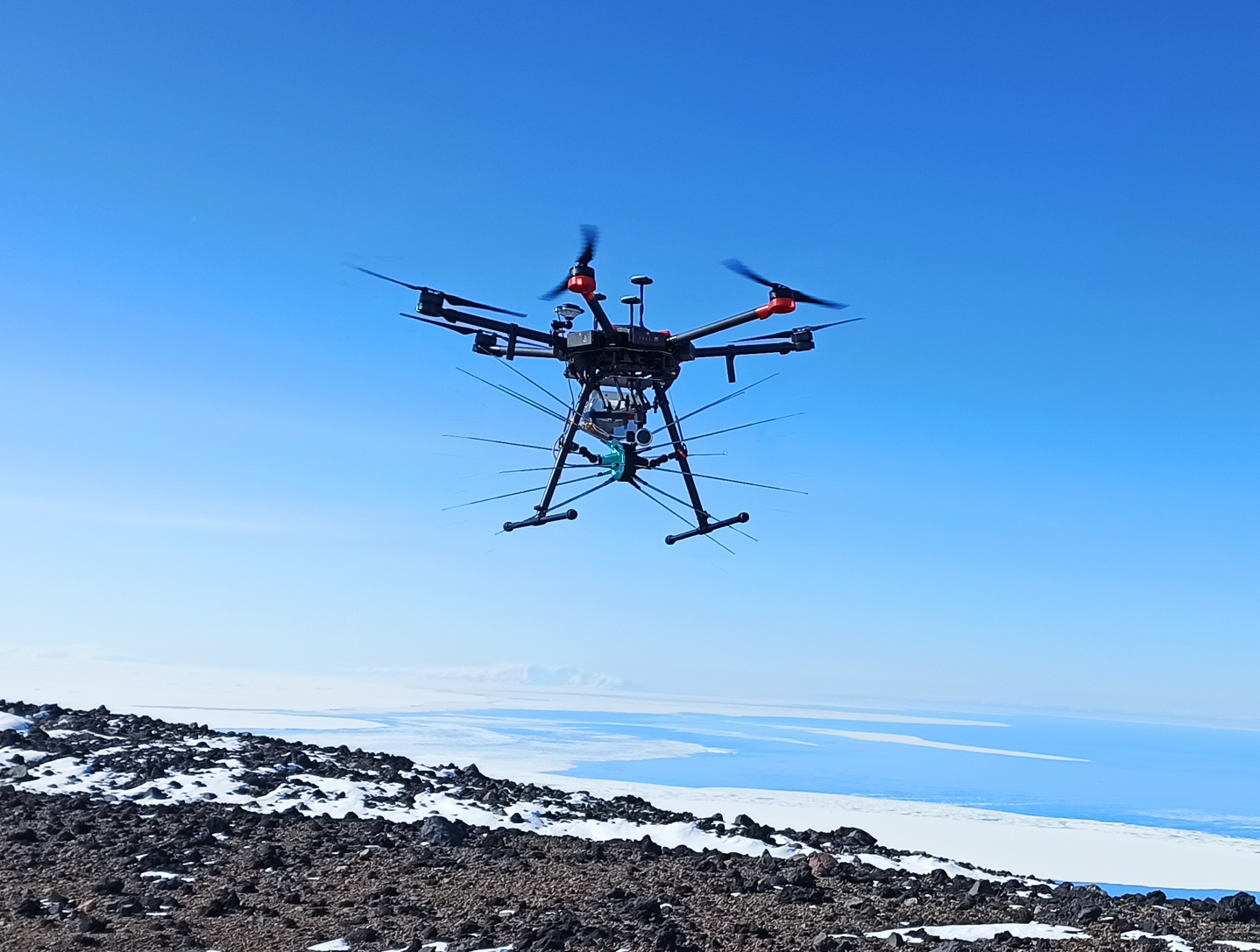}
    	
    	\end{subfigure}
        \begin{subfigure}{0.45\textwidth}
        	\includegraphics[width=\textwidth ]{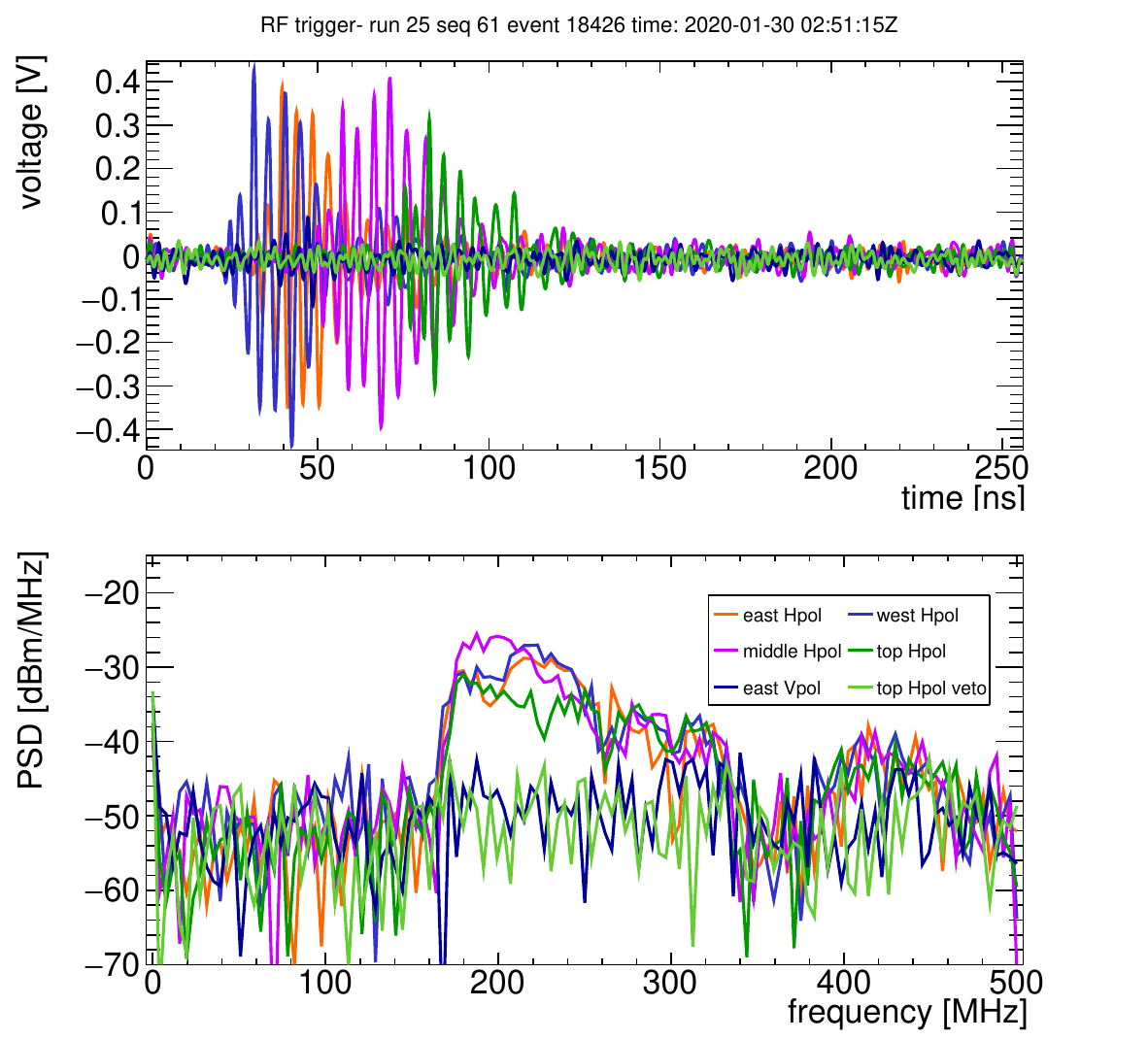}
    	
    	\end{subfigure}
    
        \caption{ 
          Left: photo of the drone-borne calibration pulser system during flight, with the pulser and DGPS modules located inside an EMI-shielded box (middle) and the transmitting biconical antenna is mounted below.
          Right: Example of received waveforms and power spectra of a pulser event; channels are color-coded.
        }
        \label{fig:DronePulser}
    \end{figure}


    %
    A lightweight drone-borne pulser system was developed \cite{Nam2021} (Fig.~\ref{fig:DronePulser}) to calibrate the performance of the TAROGE-M event reconstruction, particularly for near-horizontal directions.
    A drone-borne pulser has the advantage of accessing directions which are difficult for ground-based pulsers, due to glaciers and mountain ranges in the field of view.
    It can also be steered and therefore scan in a more controlled way than balloon-borne or manned aircraft-borne pulsers previously used by other experiments (e.g.~\cite{Prohira:2017sal}).
    %
    The system is fully portable with a total weight less than \SI{1.4}{\kg} and can be used by various experiments for cross calibration (e.g.~other TAROGE stations \cite{Chen2021} and also the proposed IceCube-Gen2 radio array \cite{IceCube2021}).

    The pulser system consists of a DGPS module for positioning and a pulser module, installed on a commercial drone with a maximum payload up to \SI{6}{\kg}. 
    The pulser module includes a solid-state high-voltage pulse generator board also used in the ARA neutrino experiment \cite{ARA2020}, and a digital step attenuator for adjusting the pulse amplitude.
    %
    The DGPS module consists of a base and rover units, where the former is installed at the station whereas the latter on the drone, providing measurement of the pulser position with centimeter accuracy.
    %
    A microcontroller, clock-synchronized with a GPS PPS signal, issues triggers to the pulse generator, programmed to control the pulse rate and amplitude, and can be easily configured in the field for a stepped amplitude scan for calibration of trigger efficiency.
    The generated pulses are transmitted by a telescopic biconical antenna having horizontal polarization with about \SI{2}{dBi} gain across \SIrange{180}{360}{\MHz}.
    %
    The system records the pulse timestamp and strength, and the instantaneous drone position for later analysis.

    Two drone flights were completed on Jan 30, 2020, conducting grid scans in near-horizontal directions at about \SI{500}{\m} distance from the station (Fig.~\ref{fig:Pulser}), each lasting for about 15 minutes; the shortened duration was a consequence of high altitude and low temperature operation.
    %
    The pulse rate was set to \SI{5}{\Hz} with a stepped pulse power scanning over a \SI{18}{\dB} span.
    Nearly $6000$ pulser events were recorded by the station and subsequently identified by matching the event timestamp with the \SI{0.2}{\second} pulsing period.

    \begin{figure}
        \centering
        \begin{subfigure}{0.48\textwidth}
    	    \includegraphics[width=\textwidth]{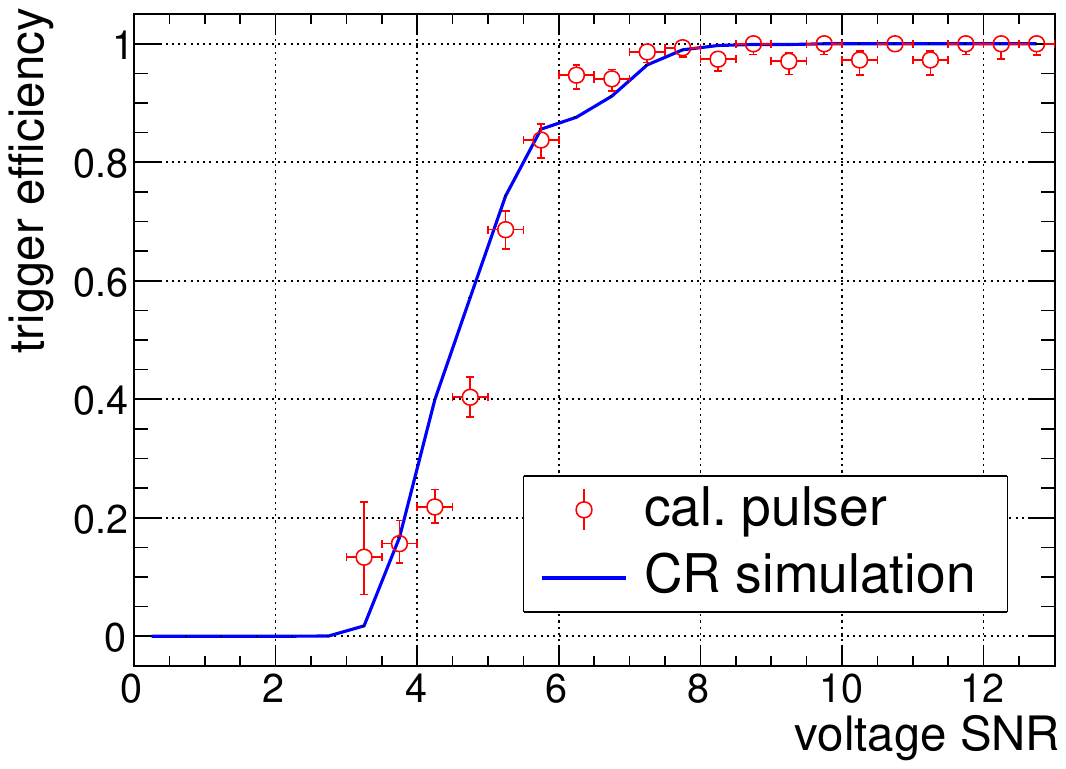}
    	
    	\end{subfigure}
        \begin{subfigure}{0.48\textwidth}
	        \includegraphics[width=\textwidth ]{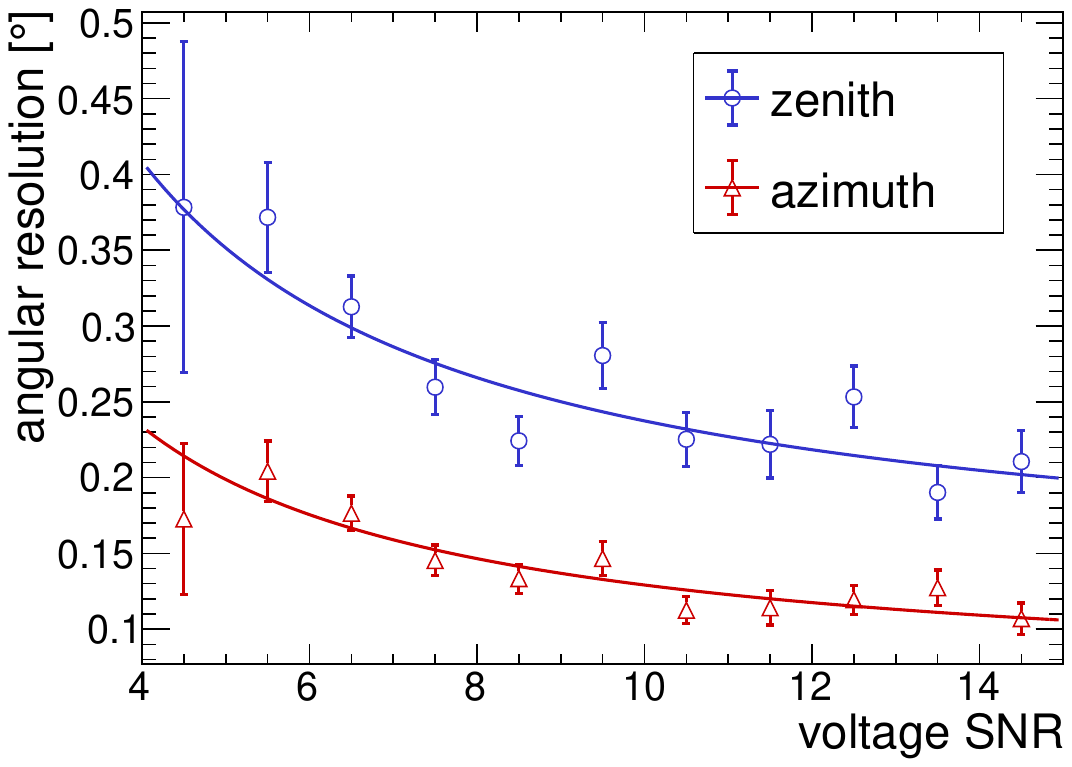}
    	\end{subfigure}
    
        \caption{   
          Left: Measured station-level trigger efficiency as a function of average Hpol voltage signal-to-noise ratio measured with the drone pulser (red markers), compared to the results from the simulation of cosmic ray signals (blue curve, see Sec.~\ref{sec:CRSim}).
          The dual-sided trigger threshold during pulser flights was set to $\rm SNR \sim4$.
         Right: Measured angular resolution in zenith (blue) and azimuth (red) directions as a function of SNR, determined from the drone pulser calibration. Best-fit curves are overlaid. 
         Angular regions with biased reconstructed zenith angles (see text) were subsequently excluded from the cosmic-ray search analysis.
         %
        }
        \label{fig:TrigEff_AngRes}
    \end{figure} 
    
    \subsubsection{Trigger Efficiency}
    
    The trigger efficiency was measured by counting the number of received periodic calibration pulses versus the total number transmitted when the pulser was within the main lobe of the antennas. 
    The pulsing power during the drone flights was configured to scan seven steps over a \SI{3}{\dB} interval, with each step lasting for \SI{2}{\second}.
    Within each step, the average signal-to-noise ratio (SNR) over all events and all Hpol channels was calculated, where the SNR of a waveform is defined as the ratio of its peak voltage to the RMS noise voltage evaluated over the first \SI{50}{\ns} window outside of the signal interval.
    The maximum observed SNR is over $25$; this value is then used to estimate the true SNR of weaker pulses in the same cycle by scaling down by the recorded attenuation value.
    The measured trigger efficiency as function of average Hpol SNR is compared with the distribution of simulated cosmic ray signals passing the trigger in similar directions ($\theta>\ang{75}$, $\phi=\ang{\pm60}$, see Sec.~\ref{sec:CRSim}) and shown in Fig.~\ref{fig:TrigEff_AngRes}.
    %
    The agreement between the two distributions validates the CR detection simulation and our estimate of the CR acceptance.

    \subsubsection{Event Reconstruction using Time Difference of Signal Arrivals}
    \label{sec:EvRec}

	
    The source direction of a TAROGE-M event is reconstructed with an interferometric method based on cross-correlation between received waveforms for extracting TDOAs,  similar to that introduced in Ref.~\cite{RomeroWolf2015}. 
	%
	First, the recorded discrete-time waveform of the $i$-th channel, denoted by $w_{i}[n]$ ($n=1,2,...,N$ for time sequence $t=nT_{s}$, where $N=256$ samples and sampling period $T_s=\SI{1}{\ns}$ ), is band-pass filtered at \SIrange{180}{330}{\MHz} for noise reduction, and upsampled by a factor of 10 (20 for pulser events) for finer time resolution. 
    %
    %
    Then the processed waveforms of the four Hpol channels are cross-correlated with each other in the time domain.
    %
    Because signals are expected to be impulsive, such that most of the non-signal region of the waveform contains only noise, a time window of length $L=L_1+L_2$ is applied between $L_1=\SI{30}{\ns}$ before and $L_2=\SI{50}{\ns}$ after the voltage peak (at $n=n_p$) of $i$-th waveform in performing the cross-correlation, a choice based on the duration of the impulse response of receiver.
    The cross-correlation coefficient $R_{X,ij}[m]$ between the $i$-th and $j$-th channels as a function of the putative time delay $\Delta t_{ij} \equiv t_{j} - t_{i} = mT_{s}'$ ($m=0,\pm1,\pm2,...$), and $T_{s}'$ is the sampling period after interpolation), computed using:
    %
	\begin{equation}
	    C_{ij}[m] = \frac{ \sum_{n=n_p-L_1}^{n_p+L_2} w_{i}[n]  w_{j}[n-m] } { \sqrt{ \sum_{n=n_p-L1}^{n_p+L_2} (w_{i}[n])^2}  \cdot   \sqrt{\sum_{n=n_p-L_1}^{n_p+L_2} (w_{j}[n-m])^2 } },
	    \label{eq:xcor}
	\end{equation}
	%
	where the denominator is a power normalization that ensures that $C_{ij}$ lies between $-1$ and $1$. 
	The measured time delay ($\Delta t_{ij,\rm obs}$) is that value which maximizes the cross-correlation function.

    The source direction in zenith and azimuth angle $(\theta, \phi)$)\footnote{$\phi=0$ at due North and increases counter-clockwise towards the west.} is reconstructed by overlaying the interferometric images derived from the cross-correlation functions for all six Hpol pairs. 
    To construct the interferometric map from the composite of pair-wise interferometric images, the time delay between two channels is calculated by assuming a spherical wavefront for nearby pulser events (with the distance derived from the DGPS record), whereas a plane wave is assumed for all other events.
    The resulting correlation coefficient as function of angles,  averaging over all six Hpol pairs (or baselines, $N_{pair}$), is
    \begin{equation}
        	R_{X}(\theta, \phi) \equiv \frac{1}{N_{pair}} \sum_{i} \sum_{j > i} C_{ij}(\Delta t_{ij}(\theta, \phi) ).
    \end{equation}
	%
    %
	The reconstructed source direction $(\hat{\theta}, \hat{\phi})$ is that with the highest value (i.e., that summed most coherently), and is found by a grid search with iteratively finer steps from \ang{1} to \ang{0.1} for general events, and up to \ang{0.01} for pulser events.

	For pulser or impulsive events of interest, the reconstruction was further improved in a second iteration with the calibrated receiver response (and the transmitter response for pulser events) deconvolved in the currently estimated direction (or the expected one in the case of the pulser) for obtaining a more coherent and sharper cross-correlation function.
	%
	Deconvolution was performed by dividing the complex event spectrum by the receiver frequency response (neglecting the antenna response in the H-plane), with an additional rectangular \SIrange{180}{330}{\MHz} band-pass and \SIrange{240}{260}{\MHz} notch \footnote{due to RFI and a dip in the LPDA frequency response (Sec.~\ref{sec:LPDA}).} filtering to suppress noise amplification in the stopbands.
	Deconvolved Hpol waveforms were once again cross-correlated with each other, but now with a narrower time window of \SI{10}{\ns} ($L_1 = L_2 = \SI{5}{\ns}$ and without power normalization (Eq.~\ref{eq:xcor}) for better alignment of primary signals.
   
    \subsubsection{Result of Reconstruction Calibration}

  	
    For timing and position calibration purposes, a set of $593$ high-quality pulser events were selected.
    The selection criteria required that the events must have a high correlation $R_{X,ij}>0.85$, have the pulser located within the main lobe of the receiver antennas (azimuthal angle between \ang{\pm40}) for which the angle-dependent antenna response does not vary significantly, and are at a zenith angle below \ang{87}, suppressing any possible interference from the reflected signal off the ground.
    The remainder of the events were used to verify the calibration result.

  	As the pulser position is precisely known, the measured time delays of the pulser events extracted from cross-correlation (Eq.~\ref{eq:xcor}) were compared with the expected $\Delta t_{ij}(\theta, \phi)$ values for calibrating the receiver timing and position parameters, including the station orientation, cable delay of receivers, and the phase center of LPDA antennas. 
    These parameters are fitted by minimizing the sum of squared differences between the measured and the putative time delay profiles, for all Hpol pairs.
    The resulting timing resolution of the calibration is determined to be about \SI{50}{\pico\second}.
	%
    The pulser events were then reconstructed using the calibrated receiver timing and positions, using the same steps outlined above.

\begin{figure}
    	\centering
    	\begin{subfigure}{0.48\textwidth}
    		\includegraphics[width=\textwidth]{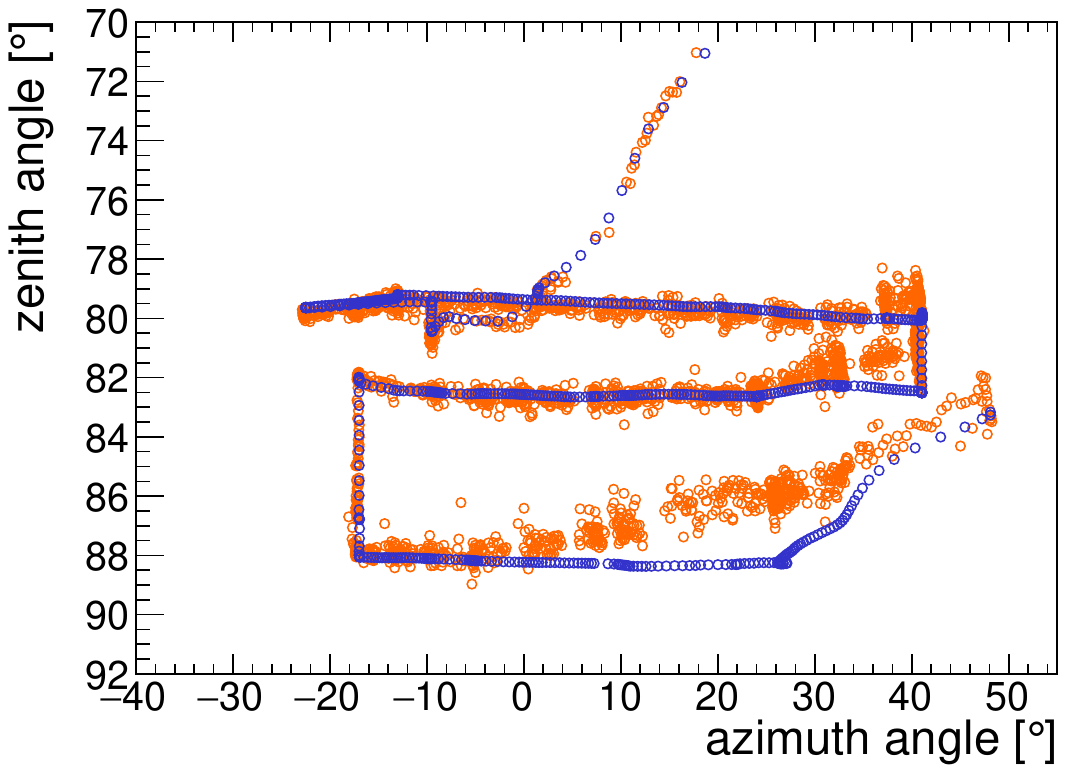}
    	\end{subfigure}
    	\begin{subfigure}{0.48\textwidth}
	        \includegraphics[width=\textwidth]{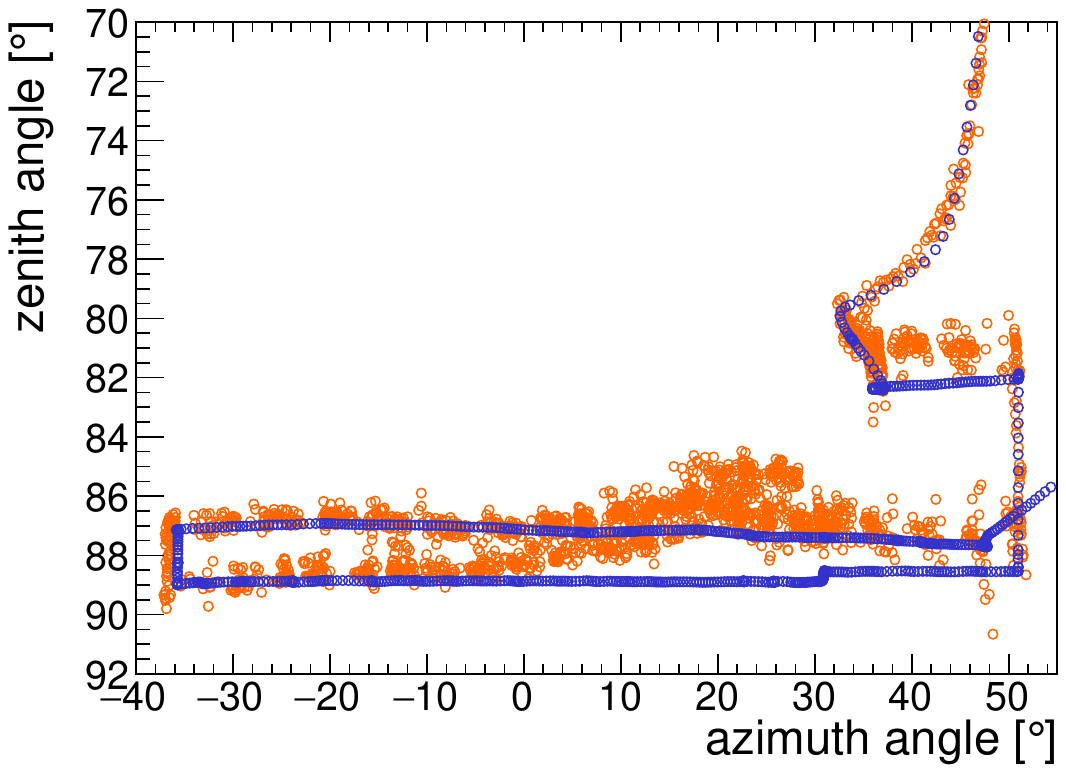} 
    	\end{subfigure}
    \caption{ 
        Recorded track from DGPS record (blue) vs.~reconstructed track (orange) after calibration, for the first (left) and the second (right) drone pulser flights.
        The azimuthal angle is defined as \ang{0} at geographic north and positive westward. LPDA boresight corresponds to $\sim+\ang{10}$.
        %
        }
    \label{fig:Pulser}
    \end{figure}

    The reconstructed versus the expected pulser flight tracks are shown in Fig.~\ref{fig:Pulser} and show overall good agreement. 
    We note two angular regions where there is a noticeable deviation ($>\ang{1}$) between reconstruction and expectation. One is at the edge of the antenna main lobe around \ang{40} in azimuth, and is likely caused by the off-boresight orientation of the antennas (with maximum \ang{13}) and imperfect modelling of the antenna response at these angles where the antenna response is rapidly varying.
    The other region is at high zenith angles above \ang{87} and \SIrange{5}{40}{\degree} azimuth, where the reconstructed angle is biased upward by \SIrange{1}{3}{\degree}.
    We speculate that the discrepancy at these elevation angles arises from the interference of the pulse reflected off the ground (or nearby objects) with the direct signal. Similar effects have been reported in ARIANNA-HCR data, and other TAROGE stations \cite{Wang2017,Chen2021}.
    This interference can potentially lead to mis-identification of an upward-going air shower (neutrino or AAE) as a downward cosmic ray (CR), 
    and may, in principle, be resolved in the future by more detailed mapping of ground terrain, and subtracting the reflection response with more drone pulser scan data. The relevant technique and preliminary results are described in Ref.~\cite{Chen2021}.
    Additionally, the completed flights currently only reached a maximum zenith angle at \ang{89}, which was due to the requirement that the drone operator behind the station have line of sight toward the drone for control. A better location for the operator will be chosen in the future for scanning upward-going directions.

    %
    If both regions with systematic angular offset are excluded, the angular resolution as function of Hpol-average voltage SNR is shown in Fig.~\ref{fig:TrigEff_AngRes}.
    The results are fit with a reciprocal SNR function with constant offset ($b_0 [{\rm SNR}]^{-1} + b_1$) for later estimation of angular uncertainty of detected CR events.
    The TAROGE-M station, in general, has angular resolution of \SI{0.2}{\degree} in azimuth and \SI{0.3}{\degree} in zenith.
    %
    
    \section{Simulation of UHE Cosmic Ray Detection}
    \label{sec:CRSim}
    
     A Monte Carlo simulation was written to model the expected air shower signals from cosmic rays and estimate the detector sensitivity. The simulation is also essential in guiding the event search and characterizing data, as discussed in Sec.~\ref{sec:CRsearch} and \ref{sec:ObsCR}.

    \subsection{Simulation of Cosmic Ray Signals}
     \label{sec:CRsignal}
     
	Signal generation is a modified version of the code used in the ARIANNA cosmic ray analysis \cite{ARIANNA2017}, adapted to TAROGE-M. 
	The simulation begins with the generation of air showers and subsequent evolution with the CORSIKA (version $7.5700$) \cite{CORSIKA} code, with the radio emission provided through CoREAS \cite{Huege2013}.
	In the CORSIKA options, QGSJET-II-04, GHEISHA, and EGS4 were selected for hadronic and electromagnetic (EM) interactions, with hadronic and EM shower thinning factors of $10^{-8}$ and $10^{-6}$, respectively.
    The built-in South Pole atmospheric model (MSIS-90-E) was selected, and the local geomagnetic field around Mt.~Melbourne, with \SI{63.7}{\micro\tesla}, \ang{-82} inclination (upward), and \ang{131} declination (southeastward), was set according to the WMM2020 model \cite{WMM2020}. 
    %

	A total of $992$ proton-initiated air showers were generated with random directions and energies spanning the range \SIrange{0.1}{30}{\exa\eV} ($\log_{10} E = [17.0-19.5]$).
	%
	Proton showers were selected for simulation since they have a larger spread in the depth distribution of shower maximum than showers initiated by heavier primary nuclei of lower energy per nucleon, leading to more diffuse radio signals and therefore a conservative choice for estimating trigger efficiencies.
	The shower directions were limited to those toward the front side of station, i.e., within the azimuthal range $\phi=$ \ang{\pm90} from due North and zenith angle $\theta$ from \ang{0} to \ang{90}. Showers from behind the station were not considered because the receiving antenna response is less sensitive in these directions; such showers are also difficult to model due to the presence of towers and the local mountain slope. 
	%
    Although the azimuthal angles of showers were uniformly distributed, more showers were effectively generated at $E<\SI{1}{\exa\eV}$ and inclination directions $\theta > \ang{60}$ by dividing $(\cos\theta, \log E)$ into evenly spaced bins; within each bin energy and zenith were uniform-distributed.

    For the CoREAS setup, the radio emission from each shower is observed in the horizontal plane at \SI{2700}{\m} altitude, where a star-shaped array of eight arms with a total of $160$ points is used to sample the asymmetric, elliptical radiation profile (electric field vector) around the shower axis, following the method first introduced in Ref.~\cite{Buitink2014}.
    The array was generated with two arms aligned with the $\vec{v}\times \vec{B}$ axis, with suitable point spacing depending on the zenith angle to contain the Cherenkov ring, and is projected from the shower plane (perpendicular to the shower axis) to the horizontal.
    At each sampling position, the electric field waveform is simulated with \SI{0.1}{\ns} sampling.

    The electric field vector at each sampled position obtained from CoREAS is convolved with the simulated antenna response and the calibrated front-end module response (see Sec.~\ref{sec:Galactic} for details) in the frequency domain to generate simulated received voltage waveforms:
    \begin{equation}
        V(f) = G_{\rm amp}(f) \left[ \vec{E}(f) \cdot \vec{H}_{\rm eff}(f) \right] = G_{\rm amp}(f) \left[ E_{\phi}(f)  H_{\text{eff},\phi}(f) + E_{\theta}(f) H_{\text{eff},\theta }(f) \right],
        \label{eq:ReVEL_FD}
    \end{equation}
    where the front-end response $G_{\rm amp}(f)$ includes both amplitude and phase.
    The first term is the inner product between the complex electric field vector $\vec{E}(f)$ and the realized antenna vector effective length (realized VEL) \cite{Balanis,AERA2012}, which takes into account the impedance mismatch with the \SI{50}{\ohm} cable and data acquisition electronics. The realized VEL is related to the realized antenna gain by:
    \begin{equation}
        |H_{ \text{eff},l}(f)| = \frac{c}{f} \sqrt{ \frac{ Z_L }{4 \pi Z_0} G_{r,l}(f) },
    \label{eq:ReVELmag}
    \end{equation}
    where $l$ stands for ${\theta,\phi}$ component, $Z_{0} \approx \SI{377}{\ohm}$ the impedance of free space, and $Z_{L}=\SI{50}{\ohm}$ the load impedance. 
    The realized gain and the phase response are obtained from HFSS.
    %
	Two other sets of simulated events were generated with signal amplitude scaled by \SI{\pm1}{\dB}, respectively, to study the effect of \SI{2}{\dB} receiver gain variation (Sec.~\ref{sec:LPDA}) on the systematic uncertainty in the CR acceptance.

\begin{figure}
        \centering
        \includegraphics[width =.9 \textwidth]{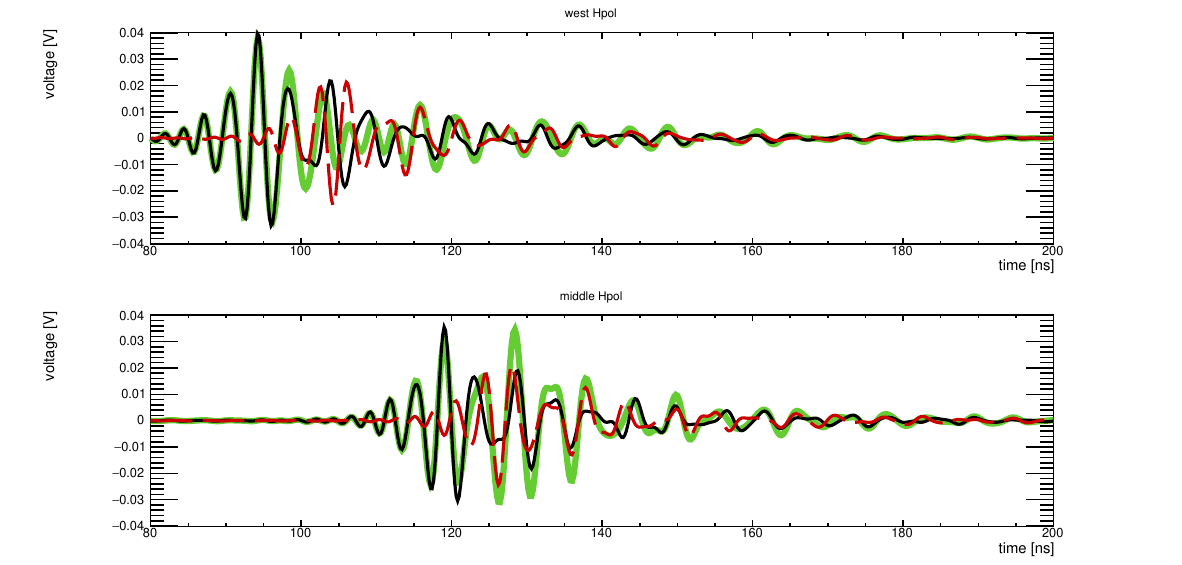}
        \caption{ 
        Simulated cosmic ray signals (with $0.74$ EeV primary energy, \ang{89} zenith, and \ang{-6.2} azimuth) received at Cherenkov angle by two of Hpol channels (top and bottom panels) with different delays in reflected signals because of different modelled ground slope.
        The gray solid curve is the direct signal, red dashed one for reflected one, and the green one for the superposition of the two.
        }
        \label{fig:wav_reflect}
    \end{figure}

    The relative position of the TAROGE-M station to the shower axis is randomly picked from one point of the sampling array (detailed in the next section), and the (assumed plane wave) electric fields arriving at all receiving antennas are assumed to be identical, as the antenna separation ($<$\SI{20}{\m}) is much shorter than the source distance.
    The voltage waveforms were then downsampled to \SI{1}{\ns}, which is the sampling period of the SST board.
    %
    %
    However, as the calibration with the drone pulser suggests interference from ground reflection, an additional event sample was generated, with reflection effects included, for studying the systematic uncertainty.
    The ground surface around the station is modelled by assuming the surface in front of each antenna is an inclined plane, for which the normal vector was estimated from fitting samples of a point cloud from the photogrammetric station model.
   The electric field incident on the surface is decomposed into transversely (S-pol) and parallel (P-pol) polarized components, and the reflection coefficient of each is calculated by Fresnel's equation \cite{Jackson}, assuming the refractive index of permafrost $n \approx \sqrt{5.3} = 2.3$ \cite{Campbell2018}, and specular reflection. 
   The specular S-pol reflection coefficient increases with zenith angle and thus nearly horizontal showers are most affected.
   The specular assumption tends to overestimate the reflected signal strength, as in reality the surface roughness breaks the coherence of reflected waves, and therefore the estimate here can be considered an upper bound. 
   %
   Two examples of simulated cosmic ray signals including reflection interference, one for the constructive case and the other for the destructive, are shown in Fig.~\ref{fig:wav_reflect}.

    \subsection{Detection Simulation and UHECR acceptance}
   \label{sec:CRAA}
    To obtain the expected CR acceptance and event rate as a function of energy and zenith angle, first the detection efficiency of each shower is calculated by passing the generated signals at all sampling positions to the trigger simulation.
    %
    %
    For the CR signal at each position, noise was added by using a randomly selected forced-trigger events from the data (excluding those contaminated by transient noise), and a total of 50 simulated events were generated in this way from the signal to sample event characteristics in the presence of noise. 
    %
    The simulated events were next passed to the trigger simulation including the dual-sided threshold, channel coincidence, and CW rejection (see Sec.~\ref{sec:SST}) criteria. The trigger efficiency at a given position is the fraction of events passing our trigger criteria.
    The star-shaped array is chosen to contain the entire triggerable area.
    %
    In the reference frame of the shower, the location of the antenna station was randomly selected over the elliptical area covered by the sampling array in the horizontal plane. The signals at its closest array point were assigned to the station (i.e., there is no interpolation of the radio footprint).
    %
    Then the effective area $A_{\rm eff}$ for detecting the shower of a given energy and direction is effectively the sum of the product of the trigger efficiency and the sector area of $i$-th array point over the entire array, projected onto the shower plane, $A_{\rm eff}(\log E, \cos\theta, \phi) = \cos \theta \sum_{i} \epsilon_{i} A_{i}$.

    Using the result above, the acceptance at a given CR energy and angular bin is equal to the average effective area over all simulated showers within the bin, multiplied by the solid angle, $\langle A\Omega \rangle (\log E, \cos\theta, \phi) = \langle A_{\rm eff} \rangle \Delta \Omega$. 
    Following the procedure outlined above, the cosmic ray acceptances under various configurations of trigger thresholds, receiver gain, and surface reflection, were calculated.
    %
    %
    The average acceptance is defined as the livetime-weighted average, across data-taking periods separated by different thresholds (Table \ref{tab:run}).
    %
    The systematic uncertainty due to receiver gain, defined as the acceptance difference relative to the mean-gain (without amplitude scaling) configuration, is summed in quadrature with the systematic uncertainty associated with the ground reflection, defined as the acceptance difference with and without reflection.

\begin{figure}
    	\centering
    	\begin{subfigure}{0.49\textwidth}

    		\includegraphics[width=\textwidth]{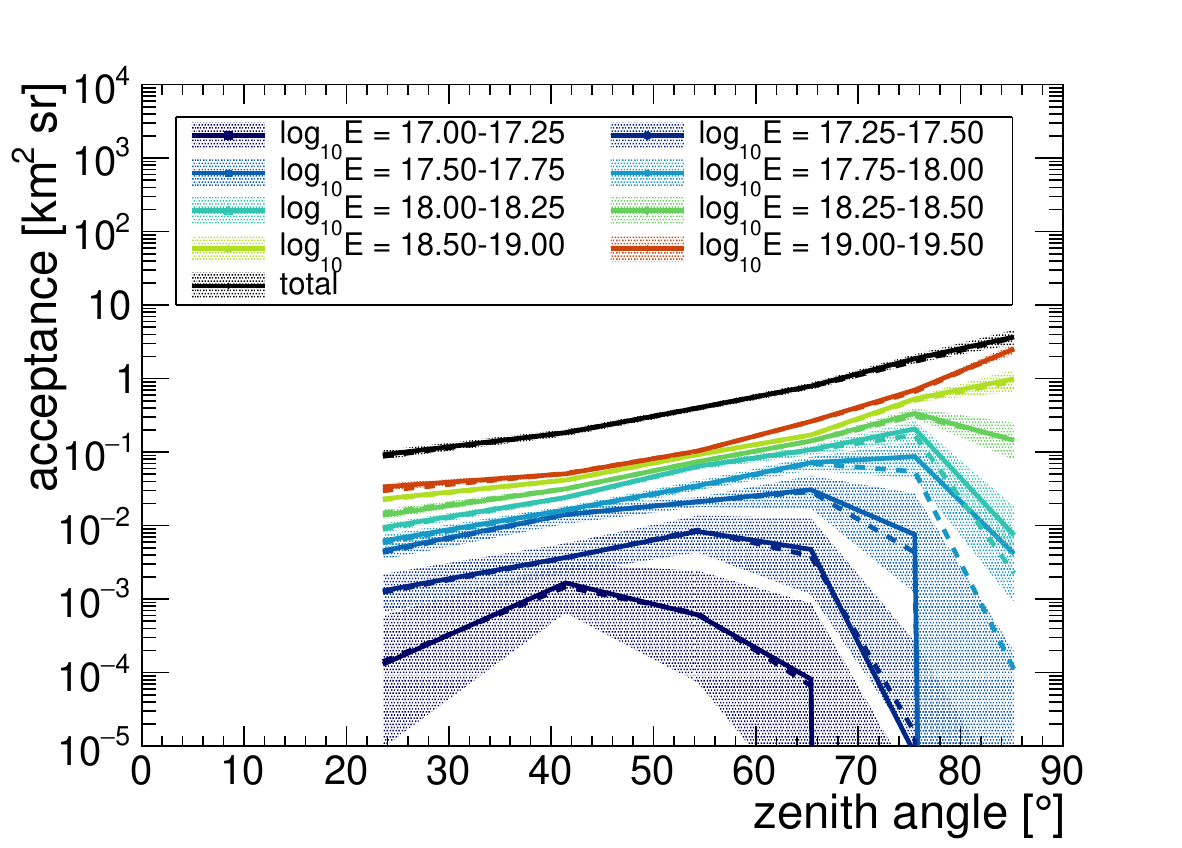} 
    	\end{subfigure}
    	\hfill
    	\begin{subfigure}{0.49\textwidth}
	        \includegraphics[width=\textwidth]{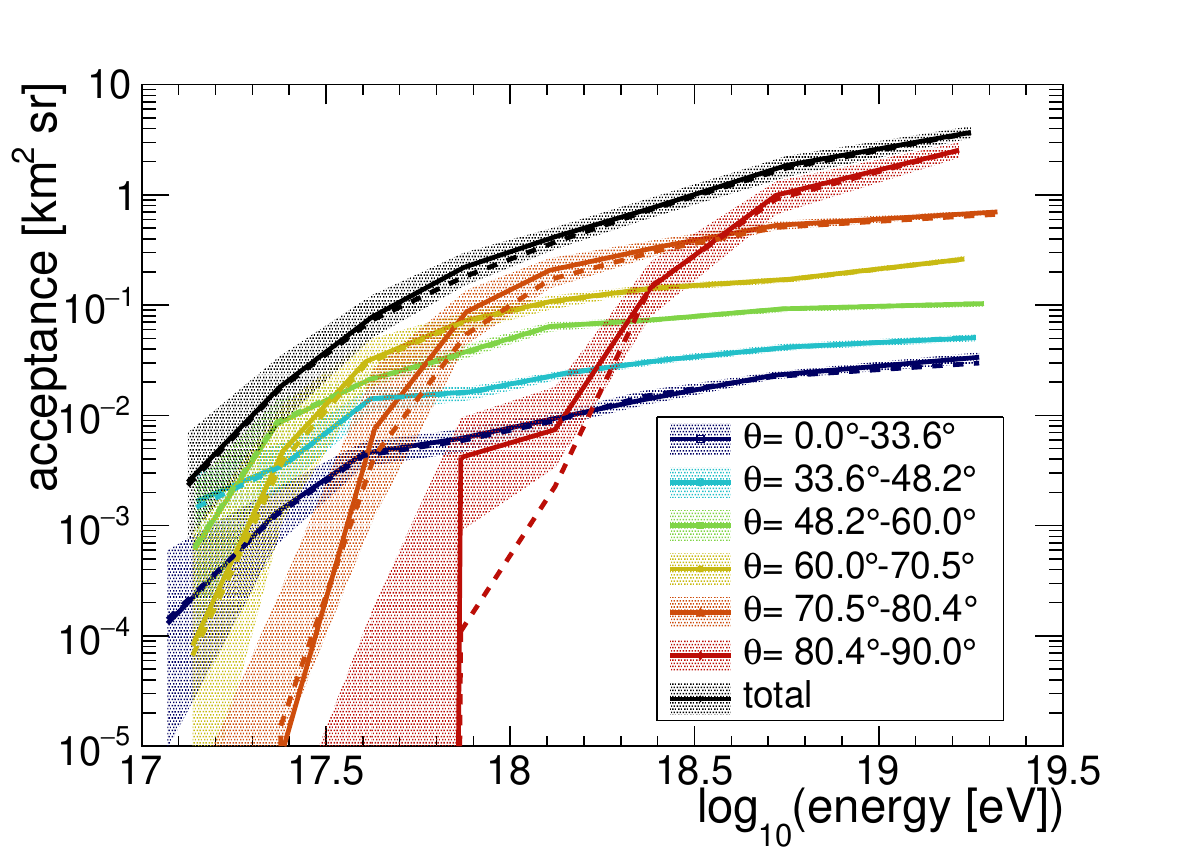}
    	\end{subfigure}
    \caption{ 
        Left: simulated cosmic ray acceptances of the TAROGE-M station as a function of zenith angle for different energy ranges are shown as colored curves; black represents the total. The acceptance is the livetime-weighted average over different trigger thresholds (see Table \ref{tab:run}). 
        Right: the acceptance as a function of primary energy for different ranges of zenith angle of equal $\cos{\theta}$ intervals (colored curves). The total is again shown in black.
        Solid curves are the results including ground reflections, with shaded areas representing the estimated uncertainty in the antenna gain. Dashed curves are calculated without inclusion of reflection effects, for comparison. 
    }
    \label{fig:CRacceptance}
    \end{figure}

    Averaging over all azimuth angles, the resulting cosmic ray acceptance of the TAROGE-M station, as a function of zenith angle and energy, is shown in Fig.~\ref{fig:CRacceptance}. 
    %
    In general, the overall acceptance increases with primary energy and zenith angle.
    %
    The detection energy threshold increases with zenith angle and is approximately \SI{0.3}{\exa\eV}, as the distance to the shower maximum increases from $\sim\SI{3}{\km}$ at $\theta=\ang{45}$ to $\sim \SI{350}{\km}$ at $\theta=\ang{89}$.
    %
    The variation in the receiver gain mainly affects the acceptance at lower energies around the trigger threshold.
    %
    %
    
    The reflection from the ground leads to a slight enhancement in the acceptance for near horizontal showers, and mostly for showers around \SI{1}{\exa\eV} energy, below which the signal is too weak to detect.
    The result can be understood by the three effects involving:
    \begin{enumerate}
        \item The received cosmic-ray waveform after convolution with the receiver response has spectral contents mostly concentrating at lower frequencies ($\sim$ \SI{200}{\MHz}), with an oscillating period of roughly \SI{5}{\ns} (an example shown in Fig.~\ref{fig:wav_reflect}).
        Therefore the constructive interference happens when the time delay of the phase-inverted reflected signal is roughly half-integers of \SI{5}{\ns}, while the destructive one happens at multiples of \SI{5}{\ns}.
        \item \label{TDOA} The time delay of the reflected signal depends on the zenith angle. 
        The delay is $\sim$\SIrange{8}{10}{\ns} at a zenith angle of \ang{80}- \ang{90} and $\sim$\SIrange{10}{14}{\ns} for \ang{70}- \ang{80} for a \SI{3}{\m} high tower on a $\sim$\ang{20} tilted ground slope. 
        \item The reflection coefficient for Hpol signals increases with zenith angle, according to the Fresnel's equation.
    \end{enumerate}
     %
    Because of the above effects, the interference of reflected signals starts from the second cycle of the direct signal or later, leaving the first cycle typically of the highest amplitude less affected (by less than \SI{10}{\percent}).
    Although there are both cases of constructive and destructive interference, it happened that the constructive case is more dominant over the destructive one in inclined directions ($>\ang{70}$ zenith),
    %
    and causes an overall increase in the acceptance, for example, $\sim$\SI{14}{\percent} at \SI{1}{\exa\eV}.

	%
    %
    The expected number of detected events for each trigger threshold $V_{\rm th}$ with livetime  $T_{\rm live}$ can be calculated from:
    \begin{align}
        N_{CR} = T_{\rm live} \int dE \int d\Omega \langle A\Omega \rangle (E,\theta,\phi) \Phi(E),
        \label{eq:rate}
    \end{align}
    assuming an isotropic cosmic ray flux $\Phi(E)$ and the energy spectrum measured by Auger \cite{Auger2019}.  
    The expected number of CR events over the $25.3$-day period is  $4.4_{-0.2}^{+0.3}$, or $0.17$ per day, where the error bars shown reflect the systematic uncertainty.
    The expected zenith-angle and energy distributions of accepted events are shown in Fig.~\ref{fig:CRdist}.
    %
    The majority of detectable events correspond to zenith angles of \SIrange{40}{80}{\degree} and primary cosmic ray energy around \SI{0.4}{\exa\eV}.
    %
    The expected acceptance and event rate are lower than those reported by the ARIANNA HRA and HCR stations (roughly $0.45-1$ per day) of similar detector configuration \cite{ARIANNA2017, Wang2019} for (at least) three reasons:
    %
    \begin{enumerate}
    \item The radio signal at low frequencies, where the signal is strongest, is lost due to the narrower TAROGE-M bandwidth (\SIrange{180}{450}{\MHz} versus \SIrange{100}{500}{\MHz} for ARIANNA).
    \item TAROGE-M is at higher altitude and closer to the showers, leading to a smaller radiation footprint than for a station close to sea level.
    %
     %
    \item The \SI{32}{\ns} coincidence trigger window was not optimized for an antenna separation of \SI{8.5}{\m} ($\sim$\SI{25}{\ns}), causing inefficiency in the trigger for signals in directions along the antenna baselines, as the narrow overlap in trigger pulses between channels may not be captured by the trigger logic processed with finite speed.
  \end{enumerate}

    \section{Simulation of Tau Neutrino Detection for AAE Sensitivity Estimation}
    \label{sec:NuTauSim}
    
	%
    TAROGE-M aims at detecting events similar to the upward-going ANITA anomalous events (AAE). Since the origin of those events is currently unknown and their properties poorly understood, assumptions must be made to estimate the TAROGE-M sensitivity.
    We assume the AAEs indeed originate from air showers at \si{\exa\eV} energies distributed isotropically, as the four detected AAEs were in directions within \ang{1} below the horizon \cite{ANITA2021} while the other two were more than \ang{20} below \cite{ANITA2016b, ANITA2018}.
    As there were no downward-going AAE reported, we assume that AAEs require interactions with the Earth's crust to initiate air showers.
    The most plausible scenario in the Standard Model, given the reported AAE characteristics is that of Earth-skimming tau neutrinos, for which ANITA's exposure is available for comparison \cite{Romero-Wolf2019,ANITA2022}.
    Assuming that they are topologically similar, the tau neutrino sensitivity of TAROGE-M with multiple station-year operation was estimated to approximate that of AAEs, and to assess the TAROGE-M discovery potential. 
    %
    At a lower altitude than ANITA, TAROGE-M is more sensitive to AAEs from near horizontal directions than those from steeper angles.

    Compared to cosmic ray simulations, the tau neutrino simulation uses a simplified approach with a parameterization for signal strength, in order to reduce computing time and circumvent particle-level complications.
    A detailed description of the neutrino simulation can be found in Ref.~\cite{Leung2020}. 
    Our result is also compared to the published tau neutrino sensitivity of ANITA, referenced as a benchmark.

    \subsection{Neutrino Propagation and Tau Decay}
    The simulation starts with neutrino propagation through the Earth toward the station for obtaining the probability of tau  production and subsequent decay in the air, as well as the measured energy distribution as a function of primary neutrino energy and direction.
    %
    %
    This is simulated using SHINIE \cite{SHINIE}, a Monte-Carlo code taking into account the charged-current (CC), neutral-current (NC) neutrino-nucleon interactions, with cross sections taken from Ref.~\cite{Connolly2011}.
    %
    The simulation also propagates secondary tau leptons produced in CC interactions, including stochastic energy losses and decay, and hence the regeneration process via tau decay ($\nu_{\tau} \rightarrow \tau \rightarrow \nu_{\tau}$).

    %
    Tau neutrinos were generated isotropically with initial energies between \SIrange{0.3}{100}{\exa\eV} and random impact parameters, within $R_{D}=$\SI{5}{\km} radius of the station at \SI{2.7}{\km} altitude. 
    The detection radius $R_{D}$ is the range over which the detector is expected to be sensitive, estimated by the size of the radio footprint for the farthest shower at the horizon, roughly equal to the horizon distance of about \SI{200}{\km} times the Cherenkov angle of about \ang{1.4} at sea level.
     %
    Each generated event starts propagation from the entry point on the Earth's surface, and secondary tau leptons and neutrinos are tracked until they either get stopped or are beyond the detection region.
    %
    %
    %
    The Earth is modelled as concentric spherical shells, based on the Preliminary Reference Earth Model (PREM) \cite{PREM}, with local terrain above sea level (assumed to be rock) within \SI{200}{\km} of the station added using the digital elevation model from the Radarsat Antarctic Mapping project \cite{Radarsat}, which has a horizontal and vertical resolution of \SI{200}{\m} and \SI{100}{\m}, respectively.
    From the simulation, the probability of tau neutrinos exiting as secondary tau leptons into the atmosphere, as function of energy and angle are obtained.
    %
    Our results are consistent with those obtained by the other neutrino propagation codes NuTauSim \cite{NuTauSim}, which was used in ANITA's simulation \cite{Romero-Wolf2019} and also the simulation devised by Wissel et al.\ \cite{Wissel2020}.

     \subsection{Radio Signal Parameterization for Tau-Decay Induced Air Shower}
    For those neutrino events with tau leptons decaying in the atmosphere, we used the parameterization given by the ANITA simulation \cite{Romero-Wolf2019} to calculate the strength of radio signals emitted by the induced showers.
    The signal parameterization \cite{Romero-Wolf2019} assumes a tau lepton would only generate air showers via decay into a hadronic mode with \SI{64.8}{\percent} of probability (i.e., neglecting electronic and muonic ones), and the hadronic mode is represented by the most common $\tau \rightarrow \pi^- \pi^0 \nu_{\tau}$ mode (\SI{25.5}{\percent}), with decay products taking \SI{67}{\percent}, \SI{31}{\percent}, \SI{2}{\percent} of the original tau-lepton energy, respectively.
    The resulting air showers take \SI{98}{\percent} of the tau lepton energy, and showers and radio emission induced by  \SI{0.1}{\exa\eV} tau leptons were simulated by ZHAireS \cite{ZHAireS2012} as templates for the parameterization \cite{Romero-Wolf2019}.
    The parameterization provides the peak amplitude of \SIrange{180}{1200}{\MHz} band-pass filtered electric field  $|\vec{E}_{p,0}|$ in the time domain as functions of observing angle relative to the shower axis (off-axis angle, $\psi$), for different tau decay altitudes $h_{\rm dec}$ and emergence angles from the Earth's surface, $\theta_{\rm em}$ (the local elevation angle of tau momentum relative to its exit point at the Earth's surface).
    %
   
   To estimate the peak voltages of received signals from tau-initiated air showers, at all relevant energies, directions, distances, altitudes, and observing angles, the electric field amplitude of a suitable template $|\vec{E}_{p,0}|(h_{\rm dec}, \theta_{\rm em}, R_0, \psi)$ scales linearly with the tau lepton energy and is inversely proportional to the distance to the detector $R$, using the shower parameters determined in the previous section and the following parameterization \cite{Romero-Wolf2019}:
   \begin{equation}
        V_p(E_{\tau}, R, \psi) \approx |\vec{E}_{p,0}|(h_{\rm dec}, \theta_{\rm em}, R_0, \psi) [ \frac{ E_{\tau} }{ \SI{0.1}{\exa\eV} } ] [ \frac{R_{0}}{R} ]  [\frac{c}{f_c} \sqrt{ \frac{ Z_L }{4 \pi Z_0} G_{\rm r, ant}(\theta, \phi)} ] \sqrt{\overline{G}_{\rm amp}},
        \label{eq:parameter}
    \end{equation}
    where $R_0$ is the distance from the detector to the tau decay point for the template, $G_{\rm ant}(\theta, \phi)$ the antenna gain at receiving angles, and $\overline{G}_{\rm amp}$ the overall receiver gain.
    The fourth term in Eq.~\ref{eq:parameter} is the antenna VEL defined in Eq.~\ref{eq:ReVELmag} with a few simplifications as follows.
    %
    The antenna gain $G_{\rm ant}$ is assumed to be constant across its working frequency range and the VEL is evaluated at the central frequency $f_c$.
    %
    The  $\sqrt{G_{\rm ant}}$ scaling of peak received voltage assumes that the antenna impulse response does not vary significantly over the main lobe.
    %
    %
    The HFSS-simulated radiation pattern of the LPDA in the main lobe was modelled by a two-dimensional Gaussian function, with peak gain of \SI{7}{dBi} and standard deviation \ang{27} and \ang{46} for the E-plane and H-plane, respectively.
    Finally, the overall receiver gain $\overline{G}_{\rm amp}$ is assumed to have a constant value of \SI{57}{\dB} over the \SIrange{180}{450}{\MHz} band (see Sec.~\ref{sec:FEE}).
    The variation of signal strength due to the geomagnetic angle between the shower axis and geomagnetic field is neglected in Eq.~\ref{eq:parameter}, because the emerging tau leptons are mostly near horizontal ($<$\ang{3}) while the geomagnetic field is near vertical.

    Since TAROGE-M has a narrower bandwidth at \SIrange{180}{450}{\MHz} than ANITA by a factor of $0.26$, the signal strength calculated with ANITA's parameterization has to be scaled down for evaluating the tau neutrino acceptance of TAROGE-M.
    Two different choices were made for the bandwidth scaling.
    %
    One choice is without scaling, which is expected to be a close approximation as most of the coherent radio emission is concentrated at lower frequencies, except at near the Cherenkov angle.
    The corresponding acceptance of TAROGE-M is assigned as the upper bound.
    %
    The other choice is scaling down the signal amplitude by a factor of $0.4$, a typical value obtained from simulated signals of cosmic ray-induced showers observed at their Cherenkov angles and band-pass filtered to passbands of TAROGE-M and ANITA, respectively.
    This value is higher than the bandwidth ratio of $0.26$ for the reasons enumerated above.
    As radio signals are most coherent at Cherenkov angle, the scaling factor is the lowest possible value, and thus the corresponding acceptance is assigned as the lower bound.
  
    \begin{figure}
	\centering
	\begin{subfigure}{0.48\textwidth}
		\includegraphics[width=\textwidth]{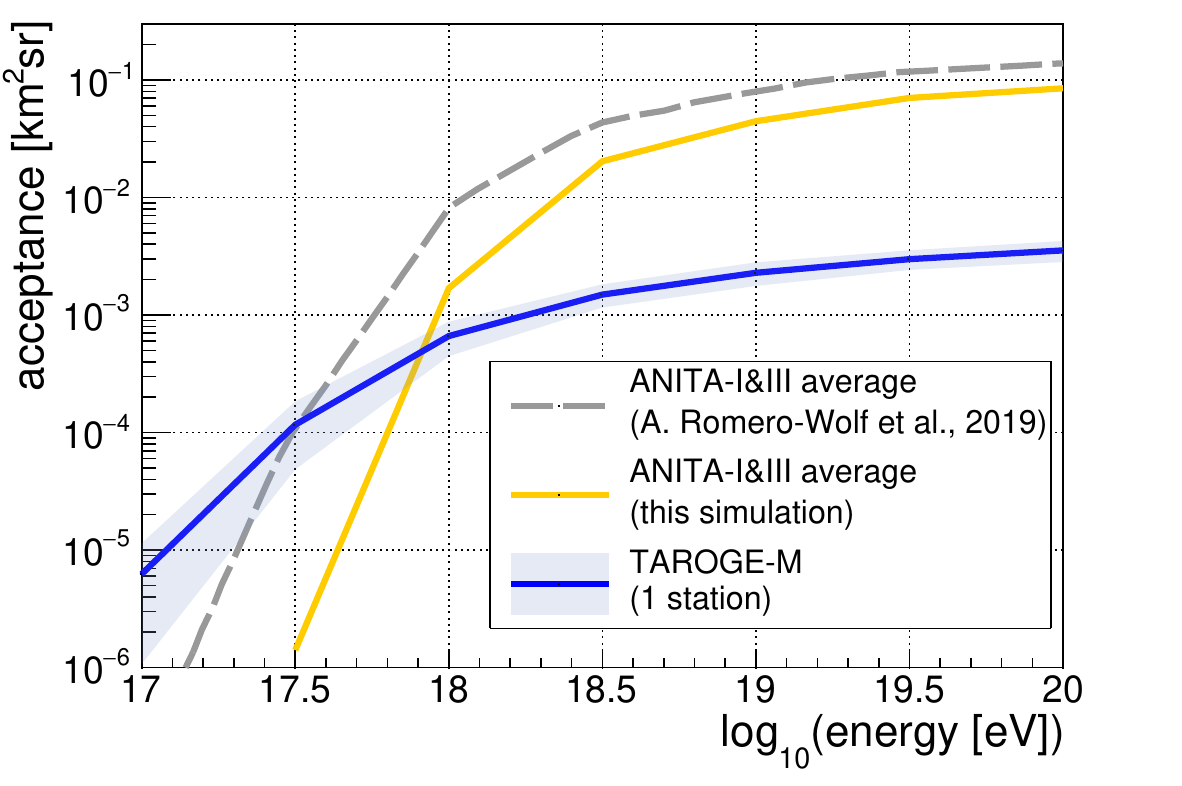} 
	\end{subfigure}
	\begin{subfigure}{0.48\textwidth}
		\includegraphics[width=\textwidth]{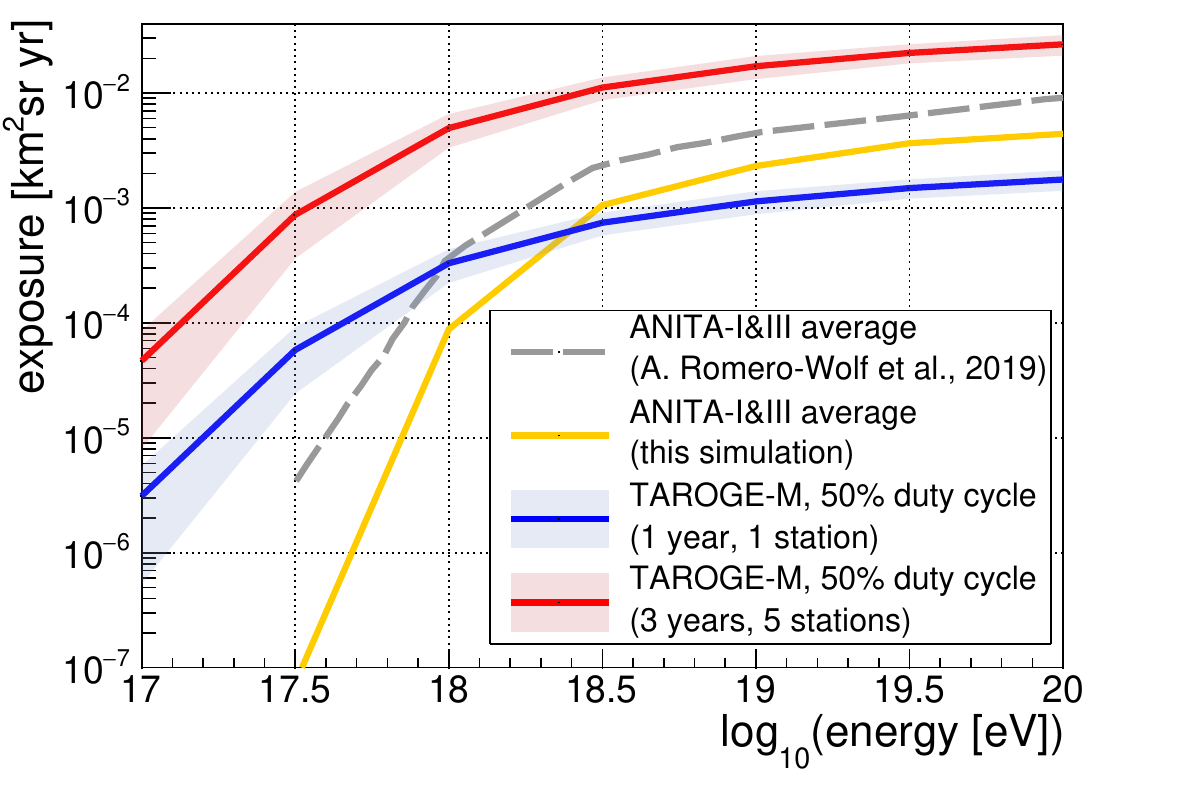}
	\end{subfigure}
    
    \caption{ 
        Tau neutrino acceptance (left panel) and exposure (right panel) as a function of neutrino energy for one TAROGE-M station in one year with \SI{50}{\percent} duty cycle are shown as blue bands, where the curves are the mean values of the upper and the lower bounds (shaded area) assuming different bandwidth scaling (see text for explanation).
        The expected exposure of five stations with three years of operation is shown as the red band.
        The results for ANITA-I and III using a similar simulation setup are shown as the yellow solid curves, and are compared with those from Ref.~\protect\cite{Romero-Wolf2019} (gray dashed).
        Note that the reconstructed air shower energies of ANITA anomalous events are around \SI{1}{\exa\eV} \protect\cite{ANITA2016b, ANITA2018,ANITA2021}, for which a single TAROGE-M station has sensitivity comparable to ANITA. 
        }
    \label{fig:NuTauAcceptance}
    \end{figure}

    \subsection{TAROGE-M Tau Neutrino Acceptance and Exposure}
    A Monte-Carlo simulation was performed to generate tau neutrino events spanning the relevant parameter space, using the experimental parameterization introduced above.
    %
    After event generation, simplified trigger criteria were applied on these events to simulate detection, requiring that an event is detectable if its peak voltage exceeds \SI{48}{\mV} (about three times the RMS noise voltage), a lower threshold anticipated for future operation with multiple stations.
    The detection efficiency $\epsilon$ is calculated as the ratio of the number of detected events to the total number generated, and the mean neutrino acceptance over all angles at a given neutrino energy $E_{\nu}$, $\langle A\Omega \rangle (E_{\nu}) = \epsilon(E_{\nu}) \pi R_{D}^{2} \Omega_{\nu} $,
    %
    with the exposure $\mathcal{E} = T_{\rm live} \langle A\Omega \rangle (E_{\nu})$, and inserting the detector livetime $T_{\rm live}$.
    The resulting acceptance and exposure as a function of neutrino energy for one TAROGE-M station assuming \SI{50}{\percent} duty cycle over one year, i.e., with one full solar-powered Antarctic summer, are shown and compared with those of ANITA-I and III \cite{Romero-Wolf2019} in Fig.~\ref{fig:NuTauAcceptance}.
    The expected exposure for a possible extension to five stations and three years of assumed operation is also shown as the red band in the right panel of Fig.~\ref{fig:NuTauAcceptance}.
    As a cross check, the acceptance and the exposure of ANITA were calculated using our simulation setups with the detector configuration from Ref.~\cite{Romero-Wolf2019}
    %
    %
    The ANITA acceptance obtained with this simulation is systematically lower than the published result, which is larger at lower energies and by a factor of $\sim 4$ at \SI{1}{\exa\eV}, as shown in Fig.~\ref{fig:NuTauAcceptance}.

    Although, given current IceCube \cite{Icecube2018} flux limits, a single TAROGE station is unlikely to detect any tau neutrino events in the near future, TAROGE nevertheless offers promise in illuminating the nature of ANITA's AAE. 
    The sensitivity of a single TAROGE-M station is higher than ANITA's at lower energies ($<$\SI{2}{\exa\eV}), and it can be further improved by a factor of several by deploying additional stations in the near future, as shown in the right panel of Fig.~\ref{fig:NuTauAcceptance}.
    %
    %
    Thus radio antenna arrays on Antarctic high mountains like TAROGE-M are an attractive solution for gaining understanding of ANITA's anomalous events.
    

\section{Search for Ultra-High Energy Cosmic Rays}
\label{sec:CRsearch}
   
	A search for air shower signals induced by UHE cosmic rays was performed on the full data in 2020, with $1,257,122$ triggered events in total within a $26.5$-day livetime.
    All events are reconstructed using the method described in Sec.~\ref{sec:EvRec} assuming a plane wave.
    Thanks to the quiet RF environment in Antarctica, the analysis is relatively simple.
    As most of the events are expected to be noise, especially those induced by high wind, a collection of noise samples is selected and characterized for further rejection in this section. 
    Event selection criteria are developed based on event reconstruction and spectral properties of noise, pulser, and simulated signal samples, with data-driven selection thresholds to avoid systematic errors from simulation.
    We chose not to perform template matching with simulated CR signals at this stage of the analysis, so as to avoid discarding potentially interesting impulsive events.
    %
    Events passing the selection criteria are inspected and evaluated, and
    their properties compared with those of cosmic rays.
    
	\subsection{Characteristics of High-Wind Events}
    \label{sec:highwind}
    
    As mentioned in Sec.~\ref{sec:operation}, radio noise associated with high-winds comprise the majority ($>\SI{99.9}{\percent}$) of the data.
    %
    To characterize and reject high-wind triggers, events during two periods of persistently high event rate above \SI{1}{\Hz} (Fig.~\ref{fig:wind_rate}), one between 2020-02-01 16:00Z and 2020-02-02 16:00Z and the other during 2020-02-18 00:00--04:00Z, were selected as the representative collection (hereafter high-wind samples), with $154,907$ and $110,840$ triggered events in total, respectively.
    %
    It was found that the noise events during different high-rate periods share similar properties, and that similar events occasionally also appeared outside the periods, suggesting a common production mechanism.
    These selected periods of $1.17$ days are excluded from the final UHECR search, resulting in an insignificant loss in the expected number of cosmic ray events by about $0.2$.  

\begin{figure}
        \centering
        \begin{subfigure}{0.45\textwidth}
    		\includegraphics[width=\textwidth]{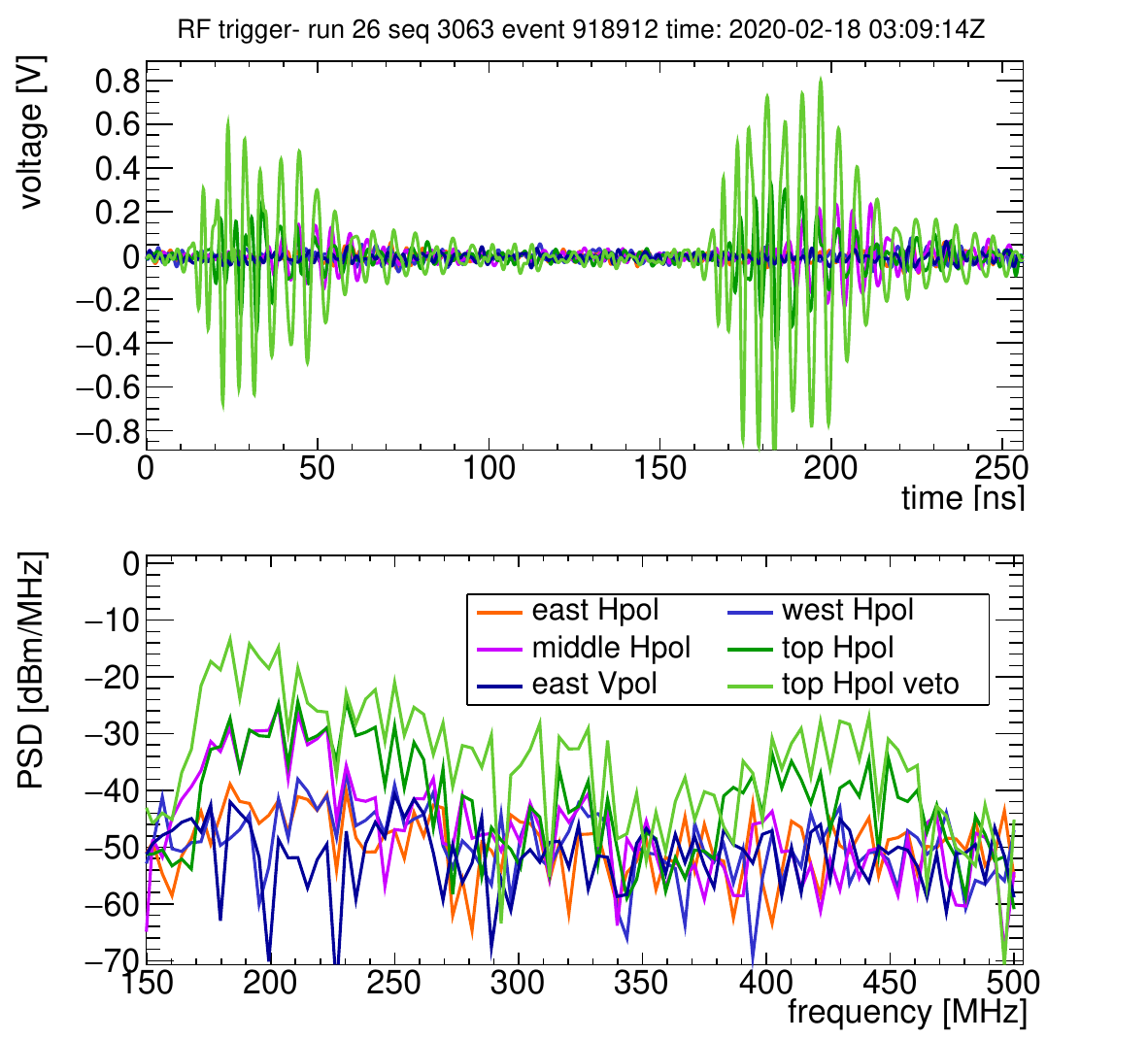}
    
        \end{subfigure}
         \begin{subfigure}{0.45\textwidth}
    		\includegraphics[width=\textwidth]{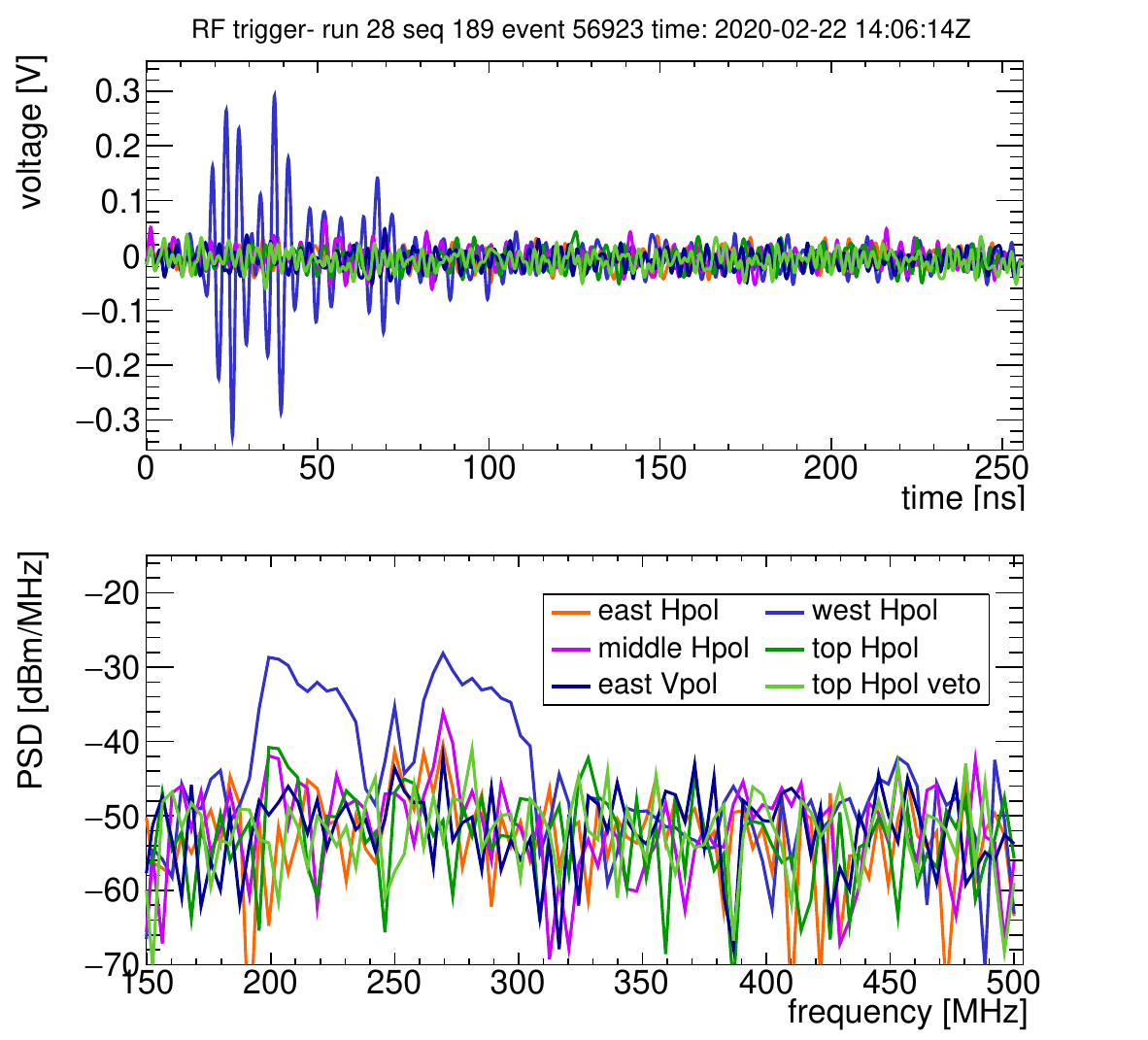}
        \end{subfigure}
        \caption{
        Two typical examples of high-wind noise events, which have one channel with a particularly strong amplitude, implying a local origin. Top and bottom panels show the waveforms and power spectra, respectively.
        Events shown in the left panel, with the strongest signal in the veto channel are most prevalent, suggesting a source location behind the station. 
        }
        \label{fig:highwind}
    \end{figure}

    The high-wind events are impulsive, typically with one of the active channels exhibiting peculiar signal waveform shapes, and also having high amplitude, implying a nearby source. This is similar to events detected by the HCR station \cite{Wang2017} during high-wind periods. Two examples are shown in Fig.~\ref{fig:highwind}. 
    These characteristics are different from that of cosmic ray-induced shower signals with more distant source locations, for which received signal amplitudes across channels are expected to be comparable, as the distance to the shower maximum is typically from several to hundreds of kilometers (except for near vertical showers for which TAROGE-M has much less sensitivity).
    %
    %
    In addition, a large portion of the high-wind events have the strongest impulse in the veto channel, likely from nearby pointy metallic objects (e.g.~antennas and poles) a few hundred meters behind the station.
    %
    %
      %
    These two distinctive features are used for the noise rejection, summarized in next section. 

    \begin{figure}
        \centering
    
    	\includegraphics[width=.9\textwidth]{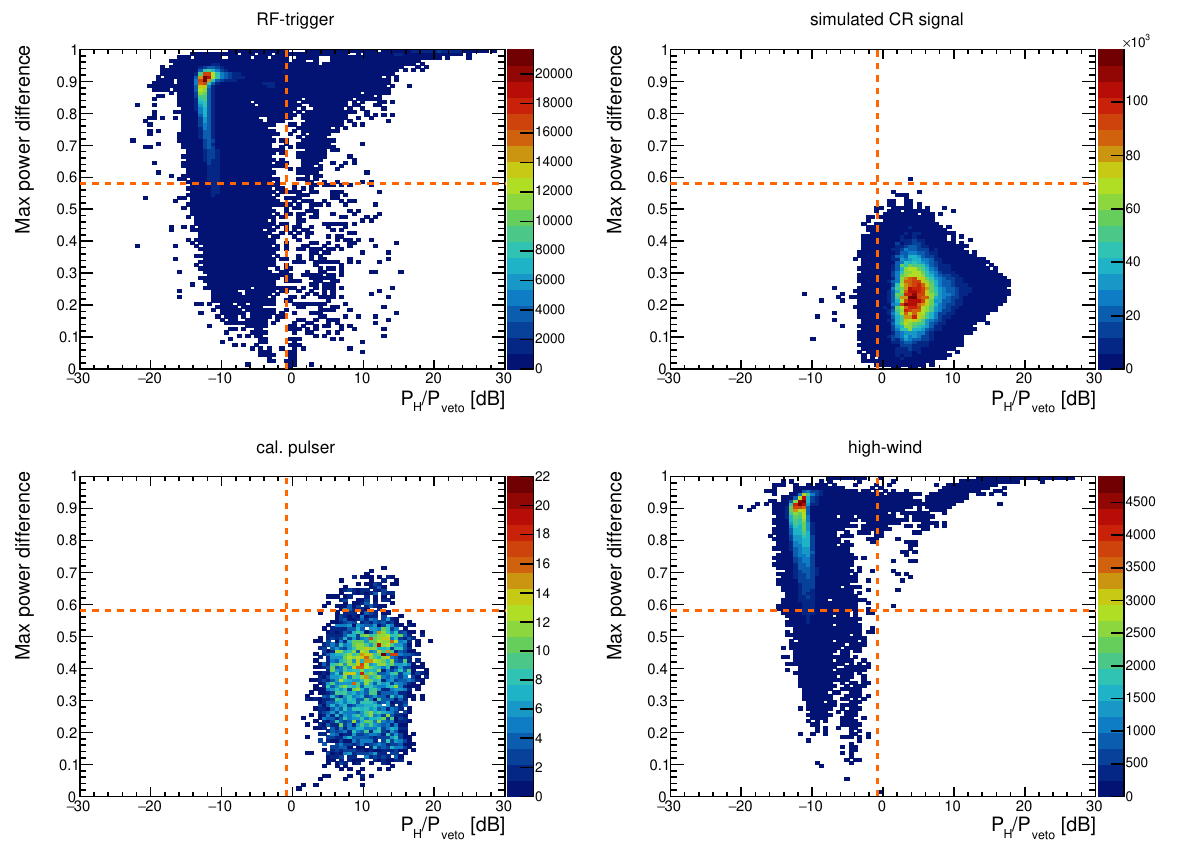}
    
        \caption{Maximum fractional power difference among Hpol channels ($r$ defined in text) versus Hpol-to-veto ratio (in \si{\dB}). Distributions of RF-trigger data (top left), simulated CR events (top right), drone pulser events (bottom left), and high-wind noise samples (bottom right). Orange dashed lines show the selection cuts; the signal region is in the fourth quadrant. The corresponding rejection power and selection efficiency are listed in Table \ref{tab:selection}.
        }
        \label{fig:specCut}
    \end{figure}
	
	\subsection{Event Selection Criteria and Result}
	\label{sec:selection}
    %
    %
    %
    Besides the high-wind noise samples, forced-trigger events were used to characterize the randomly fluctuating Galactic and thermal noise (hereafter thermal noise) which may occasionally pass the trigger. 
    The thermal noise samples also provide a reference for the spectral selection threshold for low SNR events.

    For signal samples, both calibration pulser data and simulated CR signals passing the trigger criteria outlined in Sec.~\ref{sec:CRSim} are used.
    %
    Pulser events from forward directions were selected as they provide samples with true detector response, but with limited angular ranges and higher SNRs.
    The simulated cosmic ray samples with proper weighting are used to assess the analysis efficiency.
	Each simulated event for a given shower observed at a specific point in the star-shaped array (see Sec.~\ref{sec:CRSim}) is weighted by three factors: a correction factor accounting for uneven energy and angular sampling of air showers in the simulations, an energy scaling of $E^{-3.3}$ for the CR energy spectrum (following the spectral index from Auger \cite{Auger2019}), and a scaling factor for the detection probability, equal to the area of the angular sector of the elliptical annulus comprising the sampled point. 
	%


	The passband power of the four Hpol channels are useful indicators of Hpol-dominated broadband geomagnetic emission.
	As air shower signals are typically characterized by a falling frequency spectrum, the total received power in the \SIrange{180}{240}{\MHz} and \SIrange{280}{320}{\MHz} frequency bands is defined as the passband power for the spectral selections, while \SIrange{240}{280}{\MHz} and higher frequencies are excluded due to interference from satellite communication (SatCom) and the expected lower signal SNR, respectively.
	%
	
\begin{figure}
        \centering
    		\includegraphics[width=\textwidth]{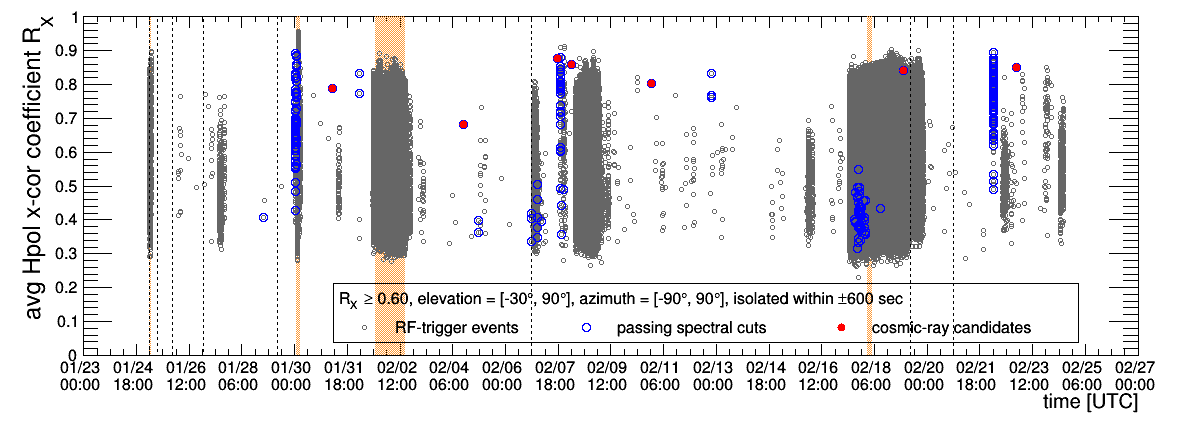}
    
        \caption{Average Hpol cross-correlation coefficient from the event reconstruction of all TAROGE-M data in 2020; gray dots indicate all RF-trigger events, open blue circles highlight the $252$ events passing the spectral cuts (selection criteria \ref{itm:Hveto} to \ref{itm:HV}), and red circles indicate the seven identified cosmic ray candidates passing all the selection criteria.
        Most events are densely clustered in high-wind periods with $R_X$ around $0.3-0.85$ (Fig.~\ref{fig:wind_rate}).
        The shaded areas in light orange indicate periods excluded from the cosmic ray search, including those when the field team visited on January 25th and 30th, and two high-wind periods (Feb 1st--2nd and Feb 18th) that comprise tagged noise samples (Sec.~\ref{sec:highwind}).
        The vertical dashed lines demarcate times when the trigger threshold value was adjusted (see Table \ref{tab:run}).
        }
        \label{fig:xcor_time}
    \end{figure}

\begin{figure}
        \centering
        
        \begin{subfigure}{0.48\textwidth}
    		\includegraphics[width=\textwidth]{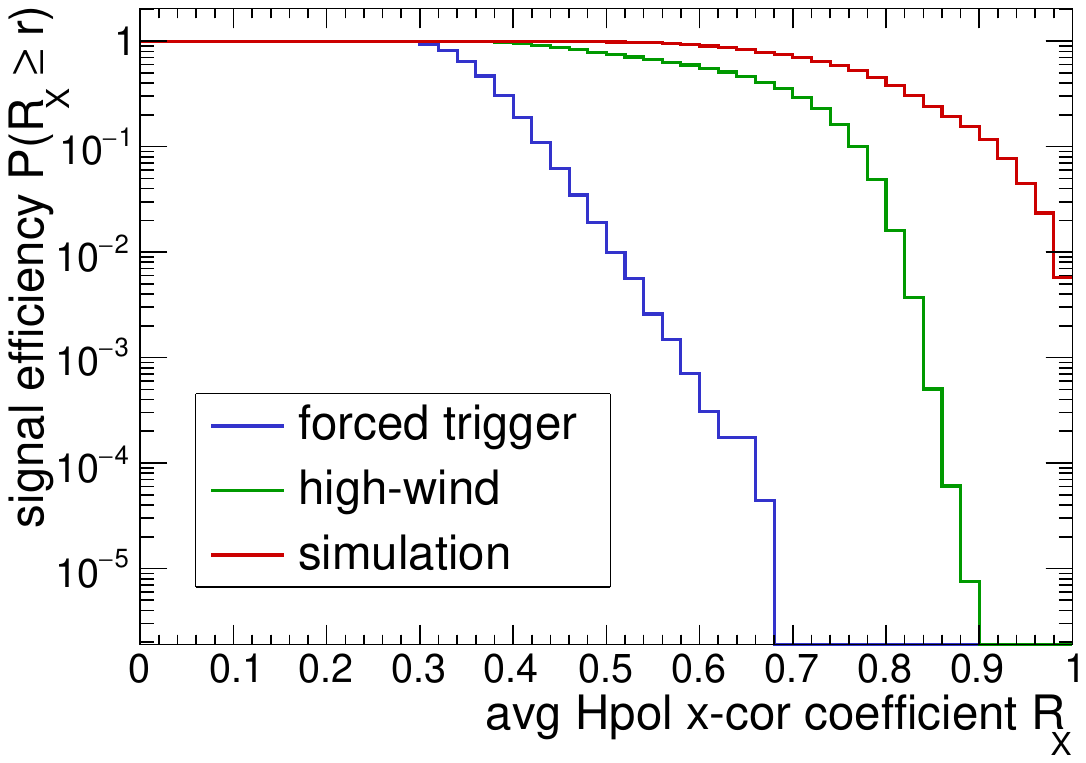}
        \end{subfigure}
         \begin{subfigure}{0.48\textwidth}
    		\includegraphics[width=\textwidth]{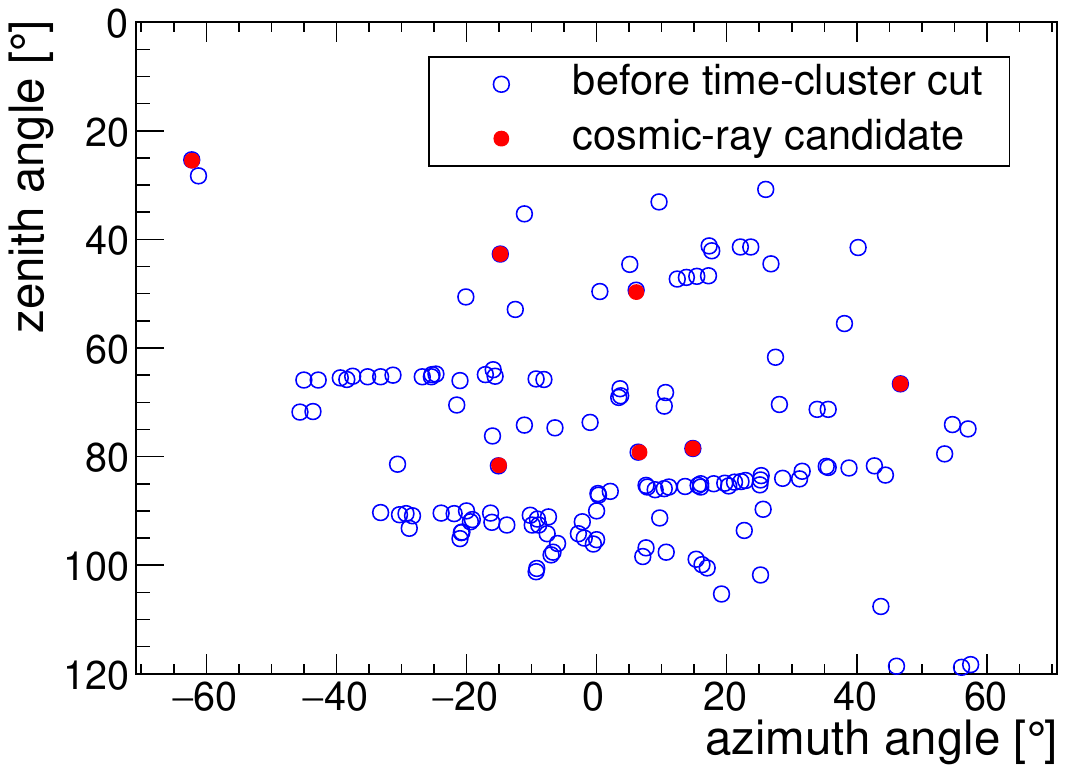}
        \end{subfigure}
        
        \caption{ 
        Left: Selection efficiency as function of cross-correlation coefficient for forced-trigger thermal noise events (blue), high-wind noise (green), and simulated CR signals (red).
        %
        Right: Angular distribution of selected events before the temporal cluster rejection (blue); tracks are attributed to aircraft. The red markers denote the seven CR candidate events passing the selection criteria.
        }
        \label{fig:EvRec}
    \end{figure}

    The event selection criteria were applied in the following order:
	\begin{enumerate}
	    \item \textbf{Hpol channels must not be saturated}: an event should have peak voltages in all four Hpol channels below the saturation voltage of the SST board at \SI{0.8}{\V} to avoid signal distortion.
    	%
	    \item \label{itm:Hveto} \textbf{High power ratio of Hpol to veto channels}: to suppress impulsive events originating from behind the station, mostly due to high-wind noise, an event is rejected if the passband power ratio of Hpol average to the veto channel $P_{\rm H}/P_{\rm veto}<1$, based on the observed distribution of high-wind samples (Fig.~\ref{fig:specCut}).
	    \item \textbf{Comparable signal strength between Hpol channels}: consistent with the expectations for a distant source location, we require approximately equivalent illumination of the Hpol channels, in contrast to the observed skew in signal amplitudes observed for the high-wind noise (Sec.~\ref{sec:highwind}).
	    The similarity is quantified by the maximum fractional power difference between any of two Hpol channels, $r \equiv (P_{\rm max}-P_{\rm min})/(P_{\rm max}+P_{\rm min})$, which is bounded between $0$ and $1$.
	    %
	    However, the differential interference of ground reflection at each antenna, as suggested by the calibration pulser, complicates the situation, as shown in Fig.~\ref{fig:specCut}, where there are some events with high dissimilarity.
	    Therefore a looser $r > 0.58$ threshold was set for rejecting events, i.e.~$P_{\rm max}/P_{\rm min}>3.8$.
	       
        \item \textbf{Exclusion of interference from satellite communication}: the SatCom interference mainly at \SIrange{240}{280}{\MHz} is present in all the data events. In some cases, the interference can trigger the data acquisition system.  To reject potential SatCom noise, the event selection requires all Hpol channels to have an average ratio of power spectral density at the passband to the satellite communication band higher than $0.16$, corresponding to 3 standard deviations below the mean value of thermal noise (forced triggers) samples.
        %
	    \item \label{itm:HV} \textbf{Exclusion of Vpol-dominated events}: as the geomagnetic emission is primarily Hpol, we use an additional criterion based on the average passband power ratio of Vpol to Hpol channels. 
	    To preserve weaker CR signals, a loose threshold at Vpol-to-Hpol ratio of $P_{\rm V}/P_{\rm H} > 2.2$ is set, corresponding to 4 standard deviations below the mean value of un-polarized thermal noise distribution. 
	    The polarization measurement is left as a verification for selected CR events later.
	   \item \textbf{High cross-correlation coefficient between Hpol channels}: Given the similarity in the Hpol channel response, we expect the waveforms for true UHECR signal to also be similar. The minimum required average cross-correlation coefficient over six Hpol pairs is $R_{X}  > 0.6$ (defined in Eq.~\ref{eq:xcor}).
	   %
        The overall distribution of correlation coefficient versus time is shown in Fig.~\ref{fig:xcor_time}.

	   This allows separation of random fluctuations of the thermal noise environment, which typically have less correlations between channels, and random hit time differences from that of plane wave signal propagation. The cross-correlation distribution can also reject high-wind noise produced nearby, or out of the field of view that can not otherwise be well-reconstructed.
	   This cut retains \SI{89.9}{\percent} of CR signal, while passing only \SI{0.03}{\percent} of thermal noise triggers and about \SI{55}{\percent} of high-wind triggers (left panel of Fig.~\ref{fig:EvRec}).
	   %
	    %
	    \item \textbf{Temporal isolation between selected events}: Whereas the expected CR events are relatively rare (we expect about $0.2$ event per day), arthropogenic and wind-induced noise tend to cluster in time. 
	    
	    A temporal clustering cut requires that, of the remaining sample of 174 events, there be no more than one candidate event within a time window of \SI{\pm600}{\second}. The  effect of this cut is illustrated in the angular map in Fig.~\ref{fig:EvRec}.
	    A total of about $1.6$ hours of livetime is masked due to the temporal clustering.
	    \item  \textbf{Reconstructed direction within field of view}: to further exclude high-wind noise and mis-reconstructed events (e.g.~thermal noise), we define a fiducial azimuth comprised by the angular range as viewed from the front side of the station, to which the antenna is sensitive: $\phi=[\ang{-90}, \ang{90}]$. A maximum zenith angle of \ang{120} for the mountain slope is also required; events with inclination angles below the value are likely mis-reconstructed.
	    Additionally, the angular region at zenith angle above \ang{87} and azimuth above \ang{0}, which was found to be biased in the drone pulser calibration, is excluded from the analysis (Sec.~\ref{sec:Pulser}).
	    %
	    %
	\end{enumerate}
    The number of events passing successive selection cuts and the analysis efficiency for CR signals are summarized in Table \ref{tab:selection}.
    The spectral selections \ref{itm:Hveto}--\ref{itm:HV} were implemented in the online filtering of the DAQ program for data transfer via satellite (in Sec.~\ref{sec:TarogeM}), with the threshold set empirically based on the forced-triggered and high-wind events.

    The Hpol-to-veto ratio and the channel similarity cuts rejected all the high-wind noise samples, as shown in Fig.~\ref{fig:specCut}, and also effectively rejected most of the events recorded during high-wind periods. This is illustrated in Fig.~\ref{fig:xcor_time}, for example, by the evident cluster around Feb 18th--19th.
    Though some of the high-wind noise survived the spectral selections (all of which had the highest amplitude signal in the middle Hpol channel), they failed in the reconstruction selection, likely because of their local origin around the antenna.
    
\begin{figure}
    \centering

        \begin{subfigure}{0.32\textwidth}
    		\includegraphics[width=\textwidth]{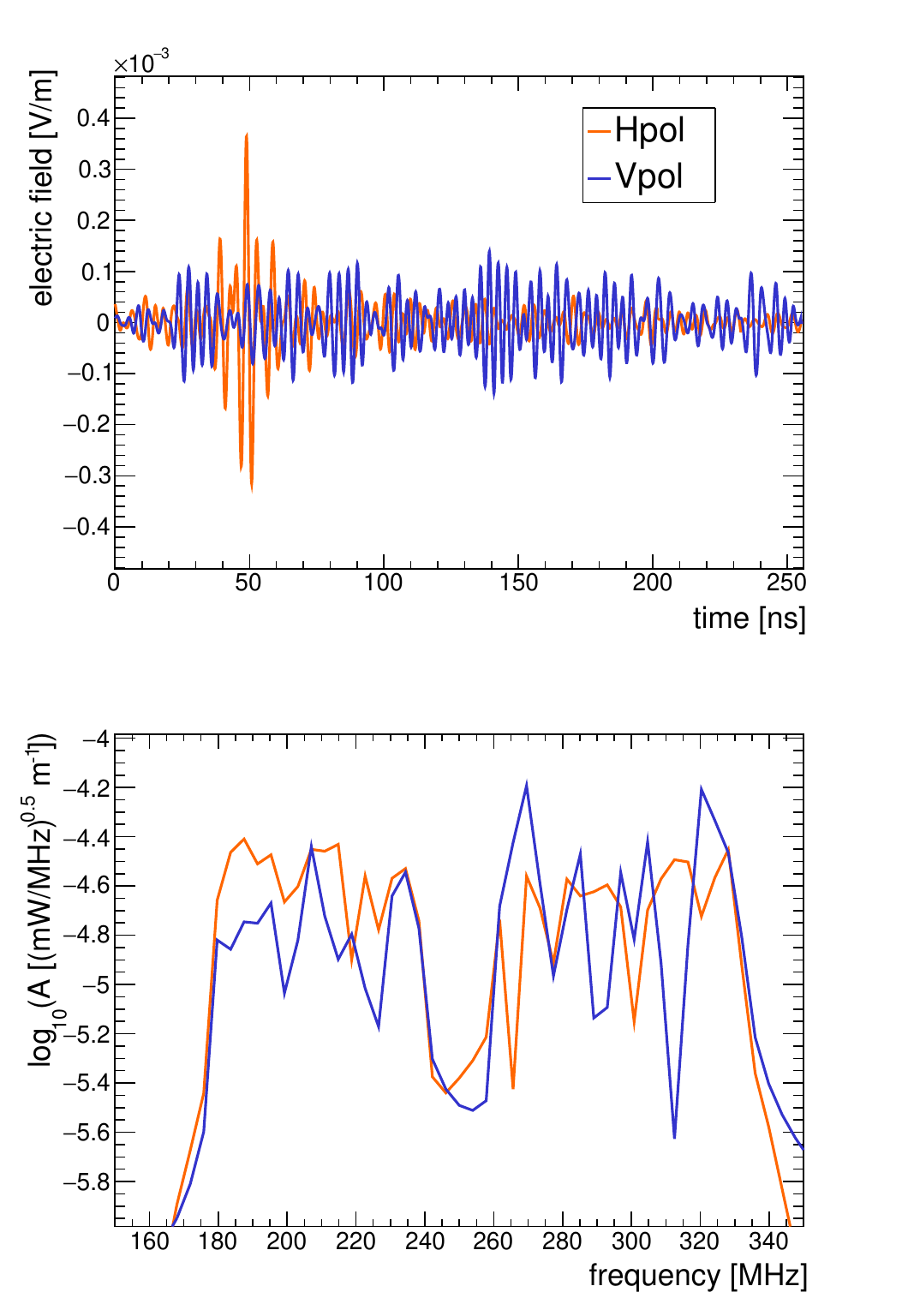}
    	\end{subfigure}
    	\begin{subfigure}{0.32\textwidth}
    		\includegraphics[width=\textwidth]{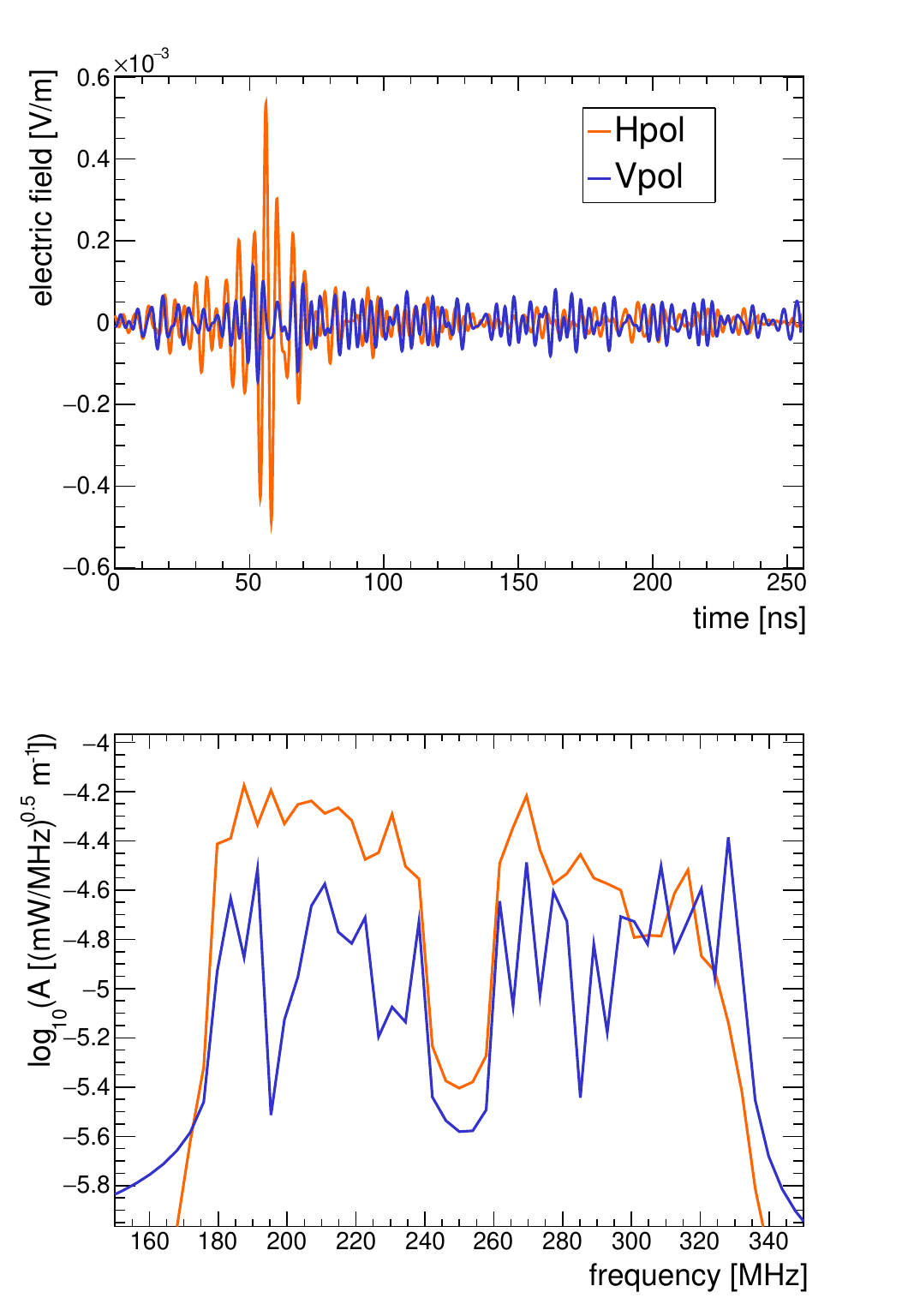}
    	\end{subfigure}
    	\begin{subfigure}{0.32\textwidth}
    		\includegraphics[width=\textwidth]{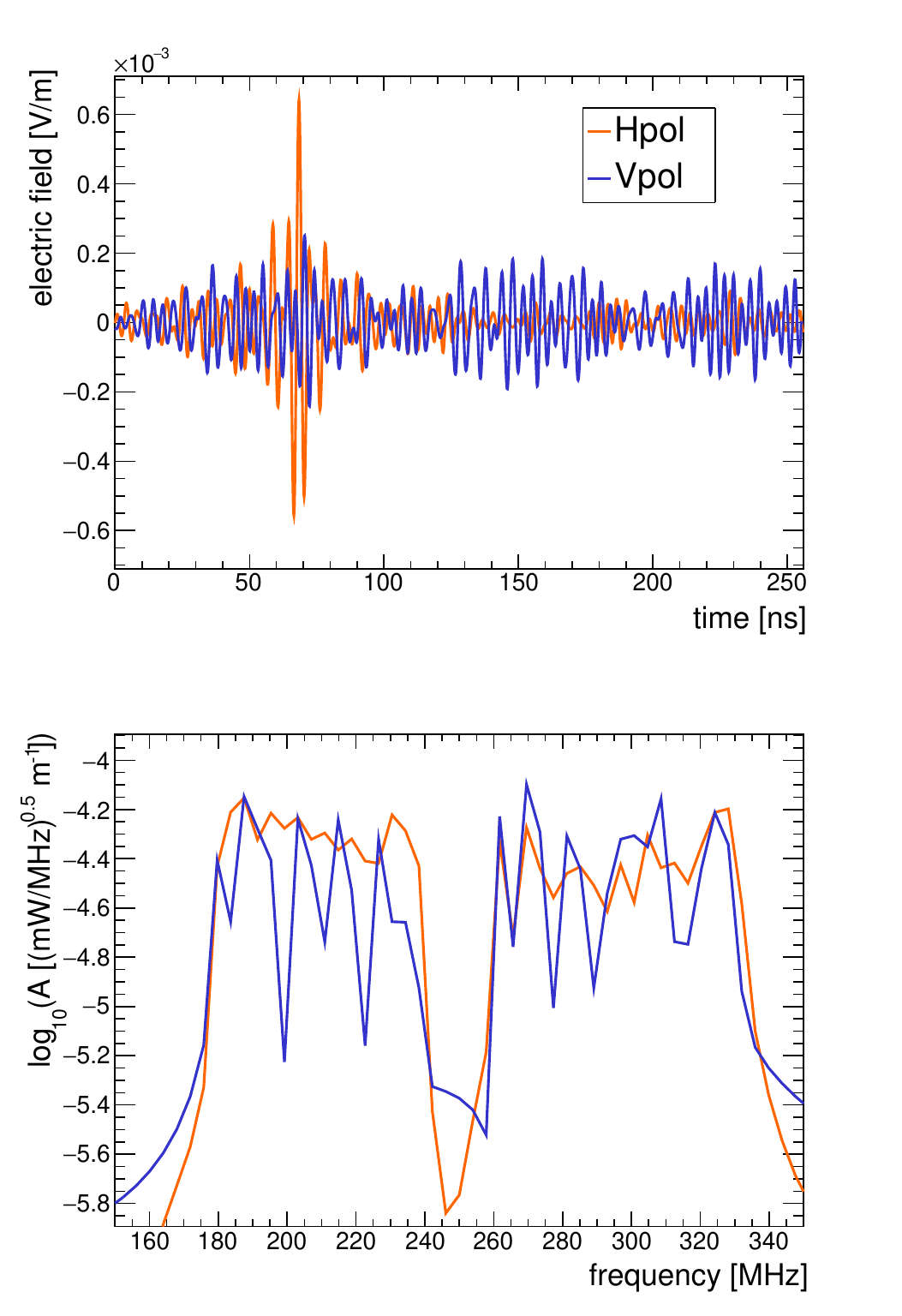}
    	\end{subfigure}

        \caption{    
        The filtered, deconvolved, coherently summed Hpol (orange) and Vpol (blue) waveforms (top panels) and power spectra (bottom) of three detected CR events in the TAROGE-M data in 2020.
        The reconstructed zenith and azimuth angles are (from left panel to right), $(\theta, \phi)=$(\ang{81.6}, \ang{-15.0}), (\ang{79.2}, \ang{6.6}), and (\ang{49.7}, \ang{6.1}), respectively. See Table \ref{tab:CR} for further information.
        }
        \label{fig:CRcandidate}
    \end{figure}

	After all selections are applied, seven cosmic ray candidates are retained, as summarized in Table.~\ref{tab:CR}. 
    The overall analysis efficiency for CR signals is \SI{89.5}{\percent}.
	%
	The Hpol waveforms of each candidate event are now shifted in time, aligned, summed, and averaged for reducing the noise. These coherently-summed waveforms (CSW) and their frequency spectra are used for further study.
	Two of the candidates are shown in Fig.~\ref{fig:CRcandidate}.
	These events were all detected during periods with the lowest trigger thresholds ($\rm SNR \sim 4$), and all have a strong Hpol component (Table \ref{tab:CR}).
	%
	%
    The zenith-angle distribution, as shown in Fig.~\ref{fig:CRdist}, is roughly flat with a slight skew in inclination, and consistent with expectation.
	The characteristics of these candidate events are scrutinized in the next section (Sec.~\ref{sec:ObsCR}) to verify their cosmic ray origin.

    \begin{table}
    \centering
    \begin{threeparttable}
    \begin{tabular}{ |l||r|r|r|r| } 
      \hline
        & RF-trigger  & high-wind & cal pulser & MC signal efficiency \\
    	 \hline
	total 				& $987832$ 	& $265747$ 	& $5080$ & $100\%$  \\
	saturation 			& $844150$ 	& $251564$ 	& $5080$ & $99.93\%$  \\
	Hpol to veto 		& $20230$ 	& $1768$ 	& $5080$ & $99.91\%$  \\
	channel similarity 	& $319$ 	& $0$ 		& $4924$ & $99.91\%$  \\
	sat. comm. 			& $275$ 	& $0$ 		& $4924$ & $99.91\%$  \\
	Hpol to Vpol 		& $252$ 	& $0$ 		& $4924$ & $99.91\%$  \\
	cross-correlation 	& $173$ 	& $0$ 		& $4906$ & $89.78\%$  \\
	time clustering 	& $7$ 		& $0$ 		& N/A & $89.54\%$\tnote{\dag} \\
	angular cut 		& $7$ 		& $0$ 		& $4865$ & $89.54\%$  \\
	final		 		& $7$ 		& $0$ 		& $4865$ & $89.54\%$  \\
     \hline
    \end{tabular}
    \begin{tablenotes}
     \footnotesize
     \item [\dag] estimated by Poission probability of \SI{0.24}{\percent} that at least one event occurs in $20$ minutes given the true expected CR event rate of $0.17$ per day (Sec.~\ref{sec:CRAA}).
   \end{tablenotes}
   \end{threeparttable}
    
    \caption{Summary of event selection (see text), showing the number of events and the expected signal efficiency from CR simulation, after successive selection cuts are applied.  
    Each column stands for different event categories: RF-trigger, high-wind noise samples (selected in Sec.~\ref{sec:highwind}), calibration pulser, and simulated CR signals.
    }
    \label{tab:selection}
    
    \end{table}

    \begin{table}
    \footnotesize
    \centering
    \begin{tabular}{ |c|r|c|c|r|r|r|r|c| } 
     \hline
		run \# & event \# & \makecell{timestamp \\ (UTC)} & $R_X$ & zenith [\si{\degree}]  & azimuth [\si{\degree}] & \makecell{V/H amp.\\ ratio} & $\log (E [\si{\eV}])$ \\
     \hline
		25	& 54906		&	\makecell{2020-01-31 \\ 05:45:10} & 	$0.79$ & $25.5\pm 0.4$  &	$-62.2 \pm 0.2$  	& $0.58 \pm 0.30$ & $18.79 \pm 1.64 \pm  0.10$ \\ 
    	25	& 319509	&	\makecell{2020-02-04 \\ 14:23:46} & 	$0.68$ & $81.6\pm 0.4$  &	$-15.0 \pm 0.2$  	& $0.27 \pm 0.09$ & $17.97 \pm 0.43 \pm  0.10$ \\ 
    	26	& 50916		&	\makecell{2020-02-07 \\ 17:11:37} & 	$0.88$ & $66.6\pm 0.3$  &	$ 46.7 \pm 0.1$ 	& $0.00 \pm 0.17$ & $18.06 \pm 0.40 \pm  0.10$ \\
    	26	& 68712		&	\makecell{2020-02-08 \\ 04:36:31} & 	$0.86$ & $42.7\pm 0.3$  &	$-14.8 \pm 0.2$  	& $0.27 \pm 0.01$ & $17.74 \pm 0.79 \pm  0.09$ \\
    	26	& 244803	&	\makecell{2020-02-10 \\ 20:33:35} & 	$0.80$ & $78.5\pm 0.3$  &	$ 14.8 \pm 0.2$ 	& $0.28 \pm 0.04$ & $17.88 \pm 0.23 \pm  0.10$ \\
    	26	& 1399188	&	\makecell{2020-02-19 \\ 05:02:30} & 	$0.84$ & $79.2\pm 0.3$  &	$  6.6 \pm 0.2$ 	& $0.15 \pm 0.03$ & $18.24 \pm 0.40 \pm  0.10$ \\
    	28	& 69900		&	\makecell{2020-02-22 \\ 22:51:36} & 	$0.85$ & $49.7\pm 0.3$  &	$  6.1 \pm 0.2$ 	& $0.00 \pm 0.15$ & $17.85 \pm 0.68 \pm  0.10$ \\
     \hline
    \end{tabular}
    \caption{ 
        Summary of cosmic ray candidate events found in TAROGE-M 2020 data.
        Columns from left to right are: run number, event number, timestamp in UTC (in ``YYYY-MM-DD hh:mm:ss'' format), average Hpol cross-correlation coefficient of event reconstruction with six Hpol pairs ($R_X$, Eq.~\ref{eq:xcor}), reconstructed azimuth and zenith angles, Vpol-to-Hpol amplitude ratio (defined in Sec.~\ref{sec:pol}), and estimated primary energy assuming proton (described in Sec.~\ref{sec:ObsCR}).
        We use a convention where the azimuthal angle is defined as \ang{0} at due north and increases counter-clockwise towards the west.
        The angular uncertainty is estimated  from the calibration results based on the drone pulser (Fig.~\ref{fig:TrigEff_AngRes}).
    }
    \label{tab:CR}
    \end{table}

    \subsection{Impulsive Events from Below the Horizon}
    
    Besides the seven CR candidates, the events rejected by temporal clustering were inspected.
    %
    In the temporal and angular distributions of selected events shown in Fig.~\ref{fig:xcor_time} and \ref{fig:EvRec}, we find three temporal clusters each comprising several tens of high $R_X$ events (blue circles) and duration less than an hour corresponding to tracks across the sky, which we attributed to aircraft.

\begin{figure}
    	\centering
    	\begin{subfigure}{0.325\textwidth}
    	    \includegraphics[width=\textwidth]{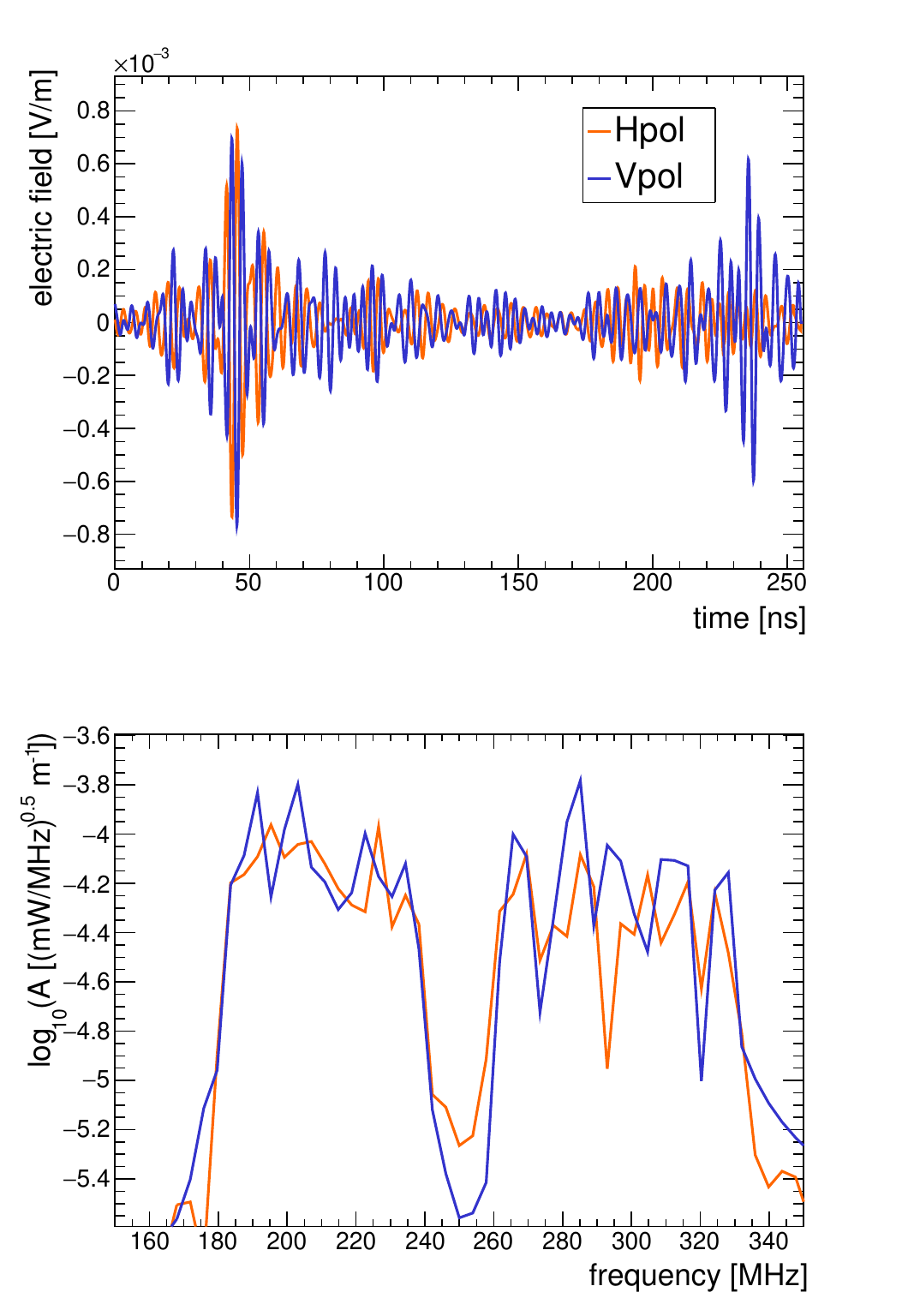} 
    	\end{subfigure}
    	\begin{subfigure}{0.325\textwidth}
    		\includegraphics[width=\textwidth]{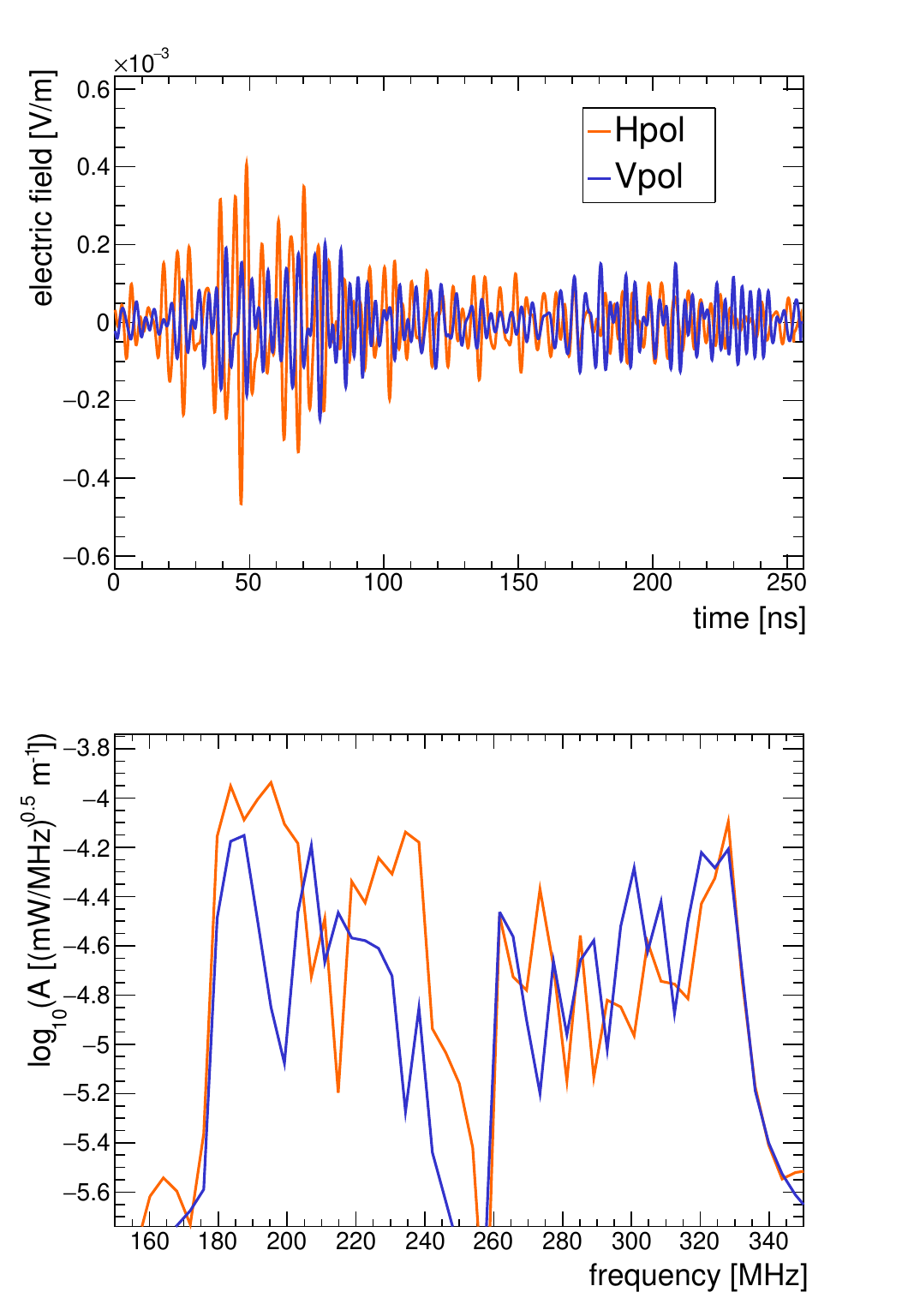} 
    	\end{subfigure}
    	\begin{subfigure}{0.325\textwidth}
    	    \includegraphics[width=\textwidth]{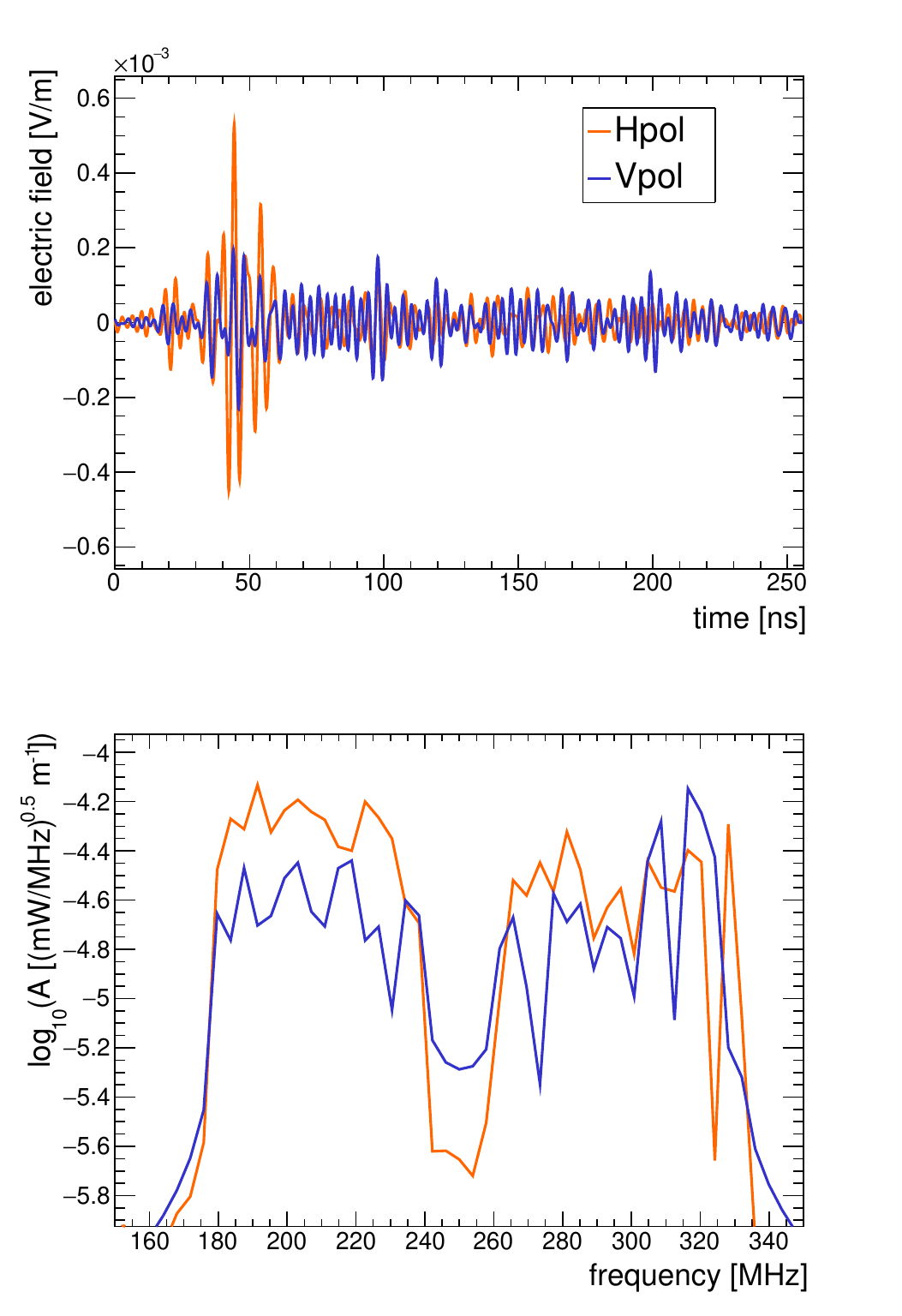} 
    	\end{subfigure}
    
        \caption{ 
            Filtered, deconvolved, coherently summed waveforms (top panels, see Sec.~\ref{sec:EvRec} for details) and power spectra (bottom panels) of the ``triplet'' impulsive events (left to right, sequential in time) which temporally and spatially cluster from a common direction below the horizon. 
            }
        \label{fig:triplet}
    \end{figure}

\begin{figure}
    	\centering
        \begin{subfigure}{0.48\textwidth}
        	\includegraphics[width=\textwidth]{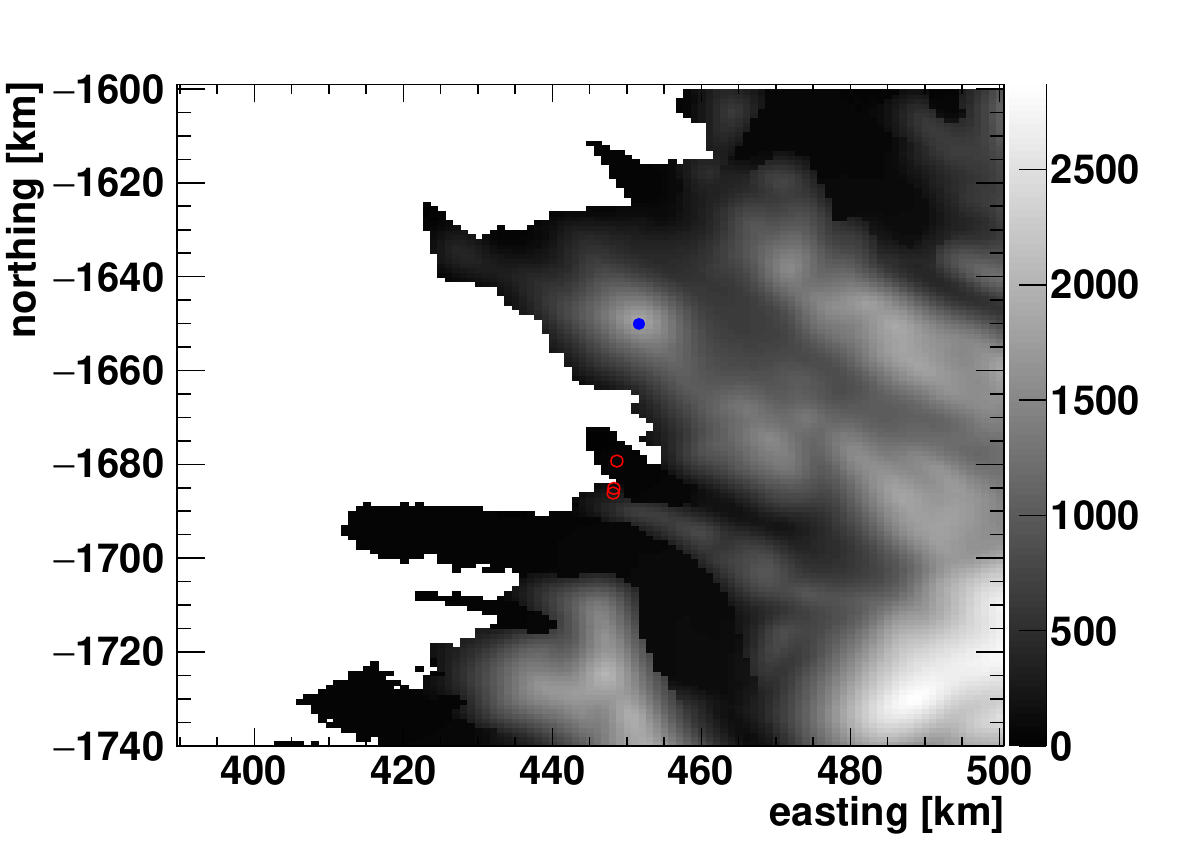}    
        \end{subfigure}
        \begin{subfigure}{0.48\textwidth}
        	\includegraphics[width=\textwidth]{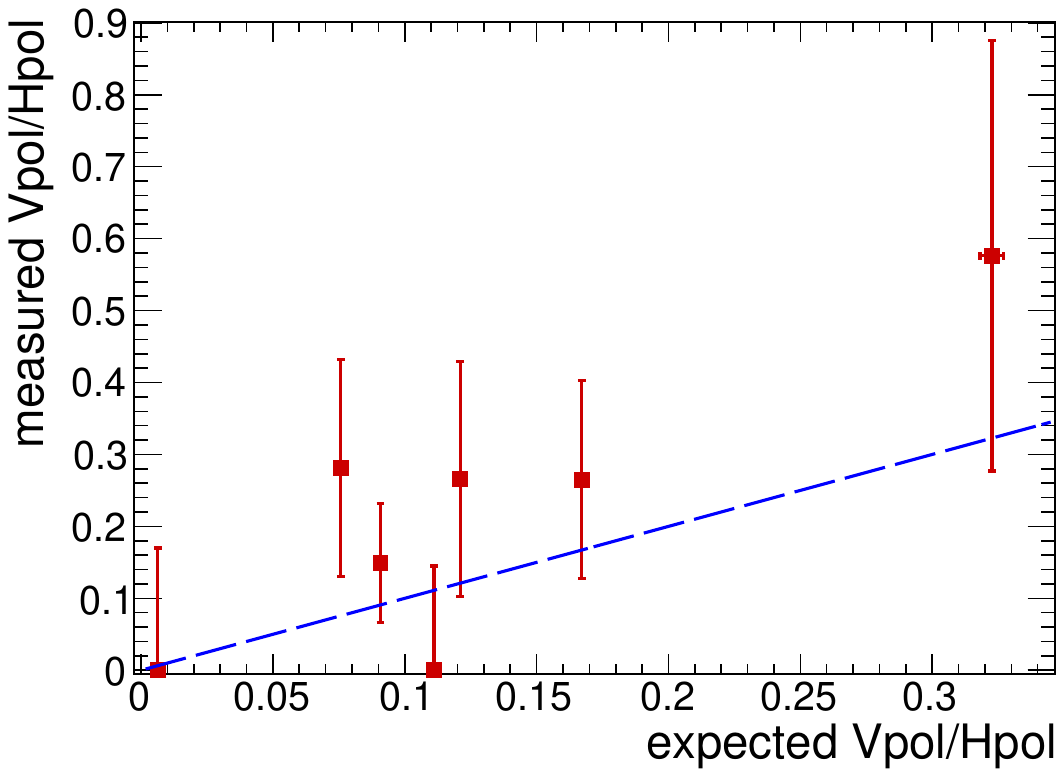}    
        \end{subfigure}
    
        \caption{ 
            Left: Reconstructed directions of the impulsive triplet events projected on the ground (red circles) from the TAROGE-M station (blue circle), using Antarctic surface elevation model Bedmap2 \cite{Bedmap2} ($\SI{1}{\km}\times\SI{1}{\km}$ grid) in Antarctic Polar Stereographic projection, where the gray scale indicates the altitude in meters, and the northing and the easting are toward geographic south and west, respectively.
            %
            Right: the measured polarization (as Vpol-to-Hpol amplitude ratio, Eq.~\ref{eq:VHAmpRatio}) of seven detected CR events (markers), compared with expected values for geomagnetic emission ($\vec{v} \times \vec{B}$).
            }
        \label{fig:bedmap2_VH}
    \end{figure}

    %
    It is notable that three impulsive events with high correlation values $\sim0.8$ and SNR \numrange{6}{9} survive all selections except the temporal cluster requirement. These events are shown in Fig.~\ref{fig:triplet} and summarized in Table \ref{tab:triplet}. 
    They were detected within \SI{23}{\second} at Feb 12, 2020 when the station was not in a high-rate period nor was the Jang Bogo station (JBS) in high wind conditions (Fig.~\ref{fig:wind_rate}), making them less likely to be of high-wind origin.
    Their reconstructed directions are from below the horizon ($\theta \sim \ang{95}$)  toward a glacier near Wood Bay with angular separation $\sim\ang{1}$ (Fig.~\ref{fig:bedmap2_VH}), where no known artificial object exists.
    %
    %
    The first event has comparable amplitude in both antenna polarizations with a Vpol afterpulse at around \SI{220}{\ns} (left panel of Fig.~\ref{fig:triplet}), making it unlikely to originate from an air shower, while the other two are Hpol-dominated. 
    Currently we have no explanation about which process could generate such polarized pulses. More events gathered in future operation might help clarifying the source.

    \begin{table}
        \footnotesize
        \centering
        \begin{threeparttable}
        \begin{tabular}{ |c|r|c|c|r|r|r|r|l| } 
         \hline
          run & event & timestamp (UTC) & \makecell{avg.~x-cor \\ coefficient } & zenith (\si{\degree})  & azimuth (\si{\degree})  & \makecell{V/H amp.\\ ratio} \\
            \hline
        	26 	& 310500	& 	2020-02-12 20:18:47	&	$0.75$ & $93.8 \pm 0.4$ & $-20.8 \pm 0.2$ & $1.16 \pm 0.60$\tnote{\dag} \\
        	26 	& 310501	& 	2020-02-12 20:18:48	&	$0.81$ & $94.0 \pm 0.4$ & $-20.8 \pm 0.2$ & $0.48 \pm 0.25$ \\
        	26 	& 310502	& 	2020-02-12 20:19:10	&	$0.76$ & $95.1 \pm 0.4$ & $-21.1 \pm 0.2$ & $0.34 \pm 0.18$ \\
         \hline
        \end{tabular}
        \begin{tablenotes}
            \footnotesize
            \item[\dag] the 120th--160th \si{\ns} of the waveforms are used as noise window to avoid afterpulses (Fig.~\ref{fig:triplet}). 
        \end{tablenotes}
        \end{threeparttable}
        \caption{ 
            Summary of the impulsive triplet events originating from below the horizon.
        }
        \label{tab:triplet}
    \end{table}

    \section{Characterizing the Detected Cosmic Ray Events}
    \label{sec:ObsCR}
    
    In this section, the seven identified cosmic ray candidates are inspected in detail to verify their consistency with expectations for UHE air showers.
    
   \subsection{Polarization Measurement}
   \label{sec:pol}
   
    Although only a loose polarization selection was imposed to reject Vpol-dominated events in Sec.~\ref{sec:CRsearch}, all seven candidate events are Hpol-dominated.
    The measured Vpol-to-Hpol amplitude ratio of each event was compared to the expected ratio of geomagnetic emission.
    For the measured polarization, the signal amplitude of each polarization is estimated with its CSW. 
    The signal power of each polarization is estimated within the signal window, defined as \SI{-30}{\ns} to \SI{+50}{\ns} around the peak amplitude $V_{\rm rms,s}^2 = (\sum_{n_p-L_1}^{n_p+L_2} w[n]^{2})/N_s$. 
    The noise power $\sim V_{\rm rms,n}^2$, as estimated by the last \SI{40}{\ns} of the waveform, is now subtracted from the signal portion of the waveform.
    %
    The power is calculated without deconvolving the receiver response, as the precision is currently limited by the single Vpol channel with amplitude close to the noise level, unlike the multiple Hpol channels with reduced noise after coherent summing.
    The result using the adopted deconvolution method tends to be biased by the noise. 
    Numerically, the Vpol-to-Hpol amplitude ratio is calculated by
    \begin{align}
         E_{\theta}/E_{\phi} \approx  \sqrt{ V_{\rm rms,s,V}^{2} - V_{\rm rms,n,V}^{2}} / \sqrt{ V_{\rm rms,s,H}^{2} - V_{\rm rms,n,H}^{2} },
         \label{eq:VHAmpRatio}
    \end{align}
    %
    %
    The ratio is unsigned as the antenna polarity has not yet been calibrated.
    The statistical uncertainty due to noise fluctuations within the finite signal window is estimated from the standard error of the noise power.
    If the signal power is less than the noise power, zero is assigned and the uncertainty of the noise RMS voltage is assigned to that datum.
    The systematic uncertainty due to \SI{\pm2}{\dB} antenna gain is also included in the error budget.

    %
    The measured Vpol-to-Hpol ratio is compared with the dominant geomagnetic emission polarization, expected along the $\vec{v} \times \vec{B}$ direction.
    Neglecting the secondary contribution from the Askaryan charge-excess emission is expected to cause only O(\ang{1}) uncertainty in polarization angle, with a contribution that decreases at high zenith angles for nearly vertical geomagnetic field, as recently shown by the CR measurement and simulation study by ARIANNA \cite{ARIANNA2022}. 
     %
    The uncertainty in the expected ratio is propagated from the angular uncertainty of each event. 
    The result is shown in the right of Fig.~\ref{fig:bedmap2_VH}. 
    We obtain good agreement between expected and measured polarization.

   \subsection{Cosmic Ray Energy Measurement}
    To estimate the primary energy of the detected cosmic ray candidates, unlike large-scale arrays which can broadly sample the radiation profile of radio emission on the ground and fit both the depth of shower maximum and the energy,
    a compact and standalone antenna array like TAROGE-M (with antenna spacing less than \SI{20}{\m})  must exploit the encoded information in the signal sampled from a single spot in the radio profile.
    Hence we followed the same method described in Ref.~\cite{ANITA2016a,ARIANNA2017,Welling2019}, by fitting the measured Hpol electric field amplitude spectrum $A(f)$ with an exponential function, or equivalently a linear fit in a logarithmic scale:
    %
    \begin{equation}
        \log_{10} A(f) = \log_{10}A_{200} + \gamma (f-200\si{\MHz}),
        \label{eq:SpecFit}
    \end{equation}
    with spectral slope $\gamma$ and amplitude intercept at \SI{200}{\MHz}, $\log_{10}A_{200}$ (the base $10$ will be omitted hereafter).
    The frequency range of fitting for TAROGE-M was chosen between \SIrange{180}{320}{\MHz}, with \SIrange{240}{260}{\MHz} excluded because of SatCom interference and reduced antenna response.
    %
    The method is based on the fact that the radio emission is most coherent when observed on the Cherenkov ring of an air shower (for which the corresponding off-axis angle $\psi$ is denoted as $\psi_c$), as the radiation from different parts of the shower arrive almost simultaneously, leading to higher amplitude intercept and flatter slope. 
    The radio coherence decreases if the observer is off the Cherenkov angle, with increasing angular offset $|\psi - \psi_{c}|$, resulting in a reduction of the frequency cut-off at which coherence is maintained, leading to both decreasing amplitude intercept and also steeper spectral slope. 
     %
    Thus, it is possible to separately estimate the energy and the off-Cherenkov angle from the spectral intercept and slope.

    \subsubsection{Fitting the Measured Spectra}
\begin{figure}
    \centering
    
        \begin{subfigure}{0.45\textwidth}
    		\includegraphics[width=\textwidth]{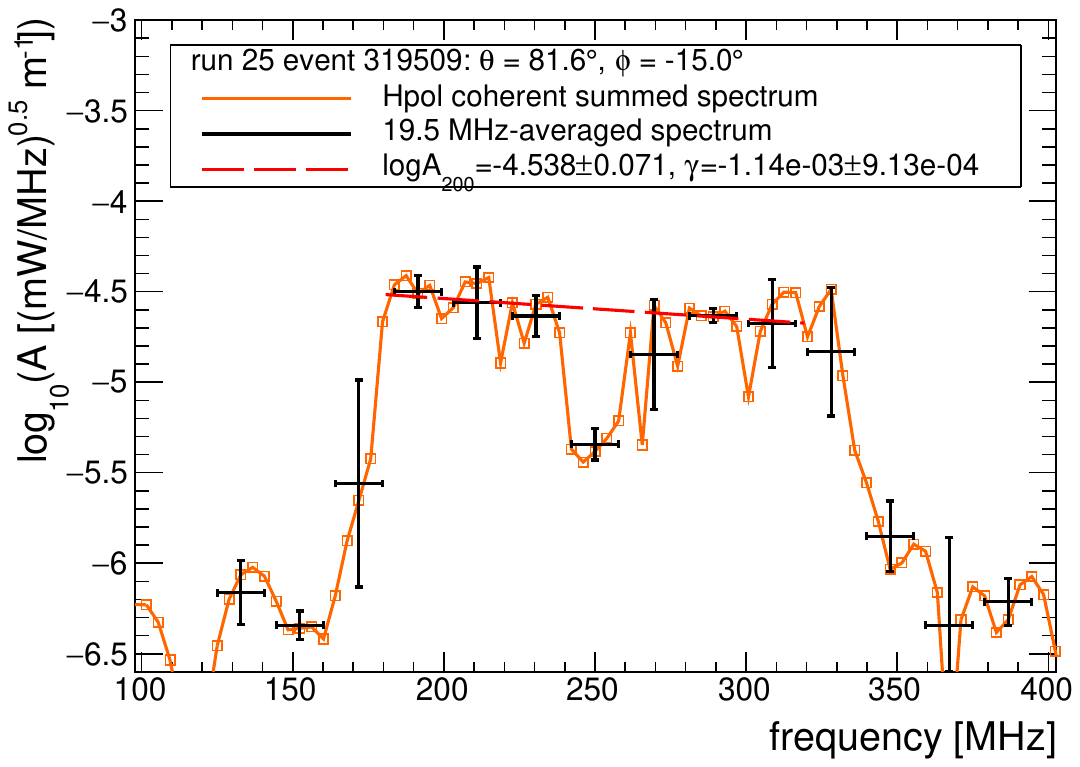}
    	\end{subfigure}
    	\hfill
        \begin{subfigure}{0.45\textwidth}
    		\includegraphics[width=\textwidth]{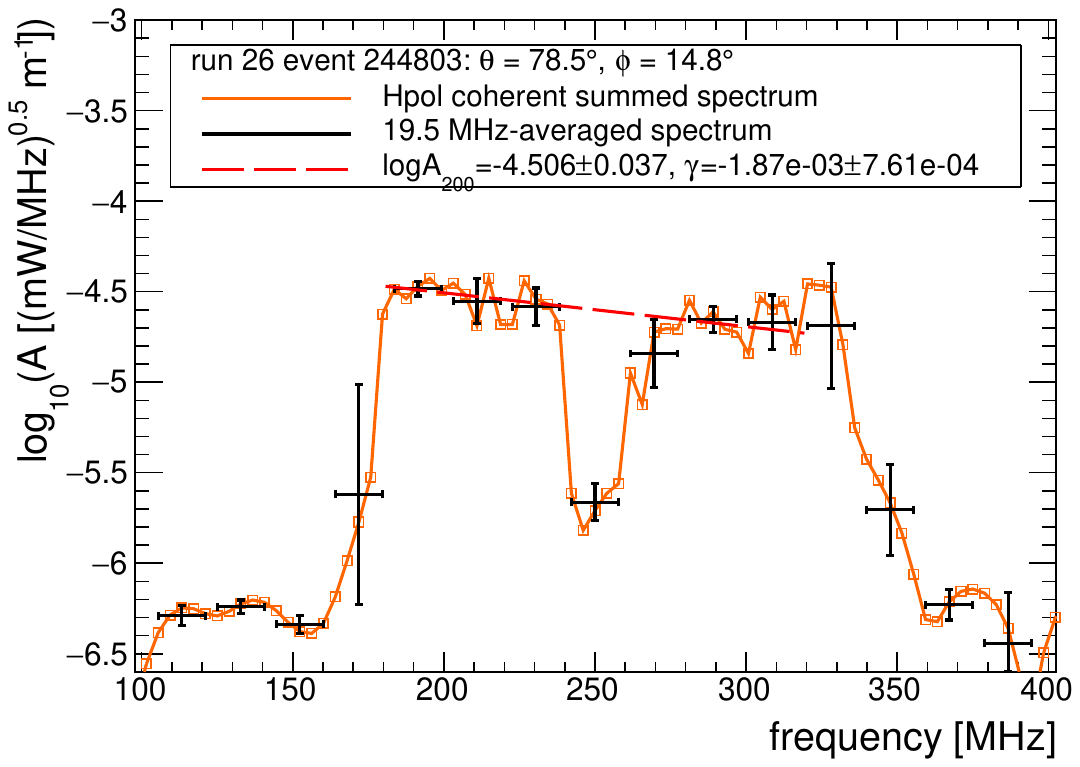}
    	\end{subfigure}

    	\begin{subfigure}{0.45\textwidth}
    		\includegraphics[width=\textwidth]{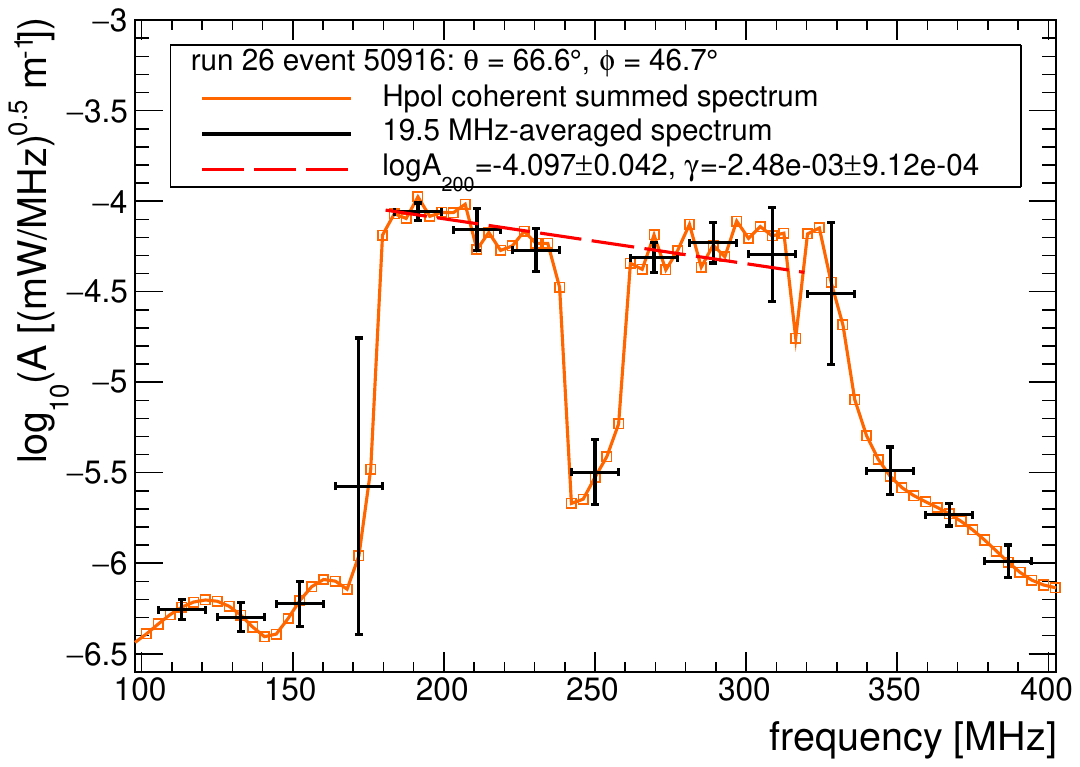}
    	\end{subfigure}
    	\hfill
    	\begin{subfigure}{0.45\textwidth}
    		\includegraphics[width=\textwidth]{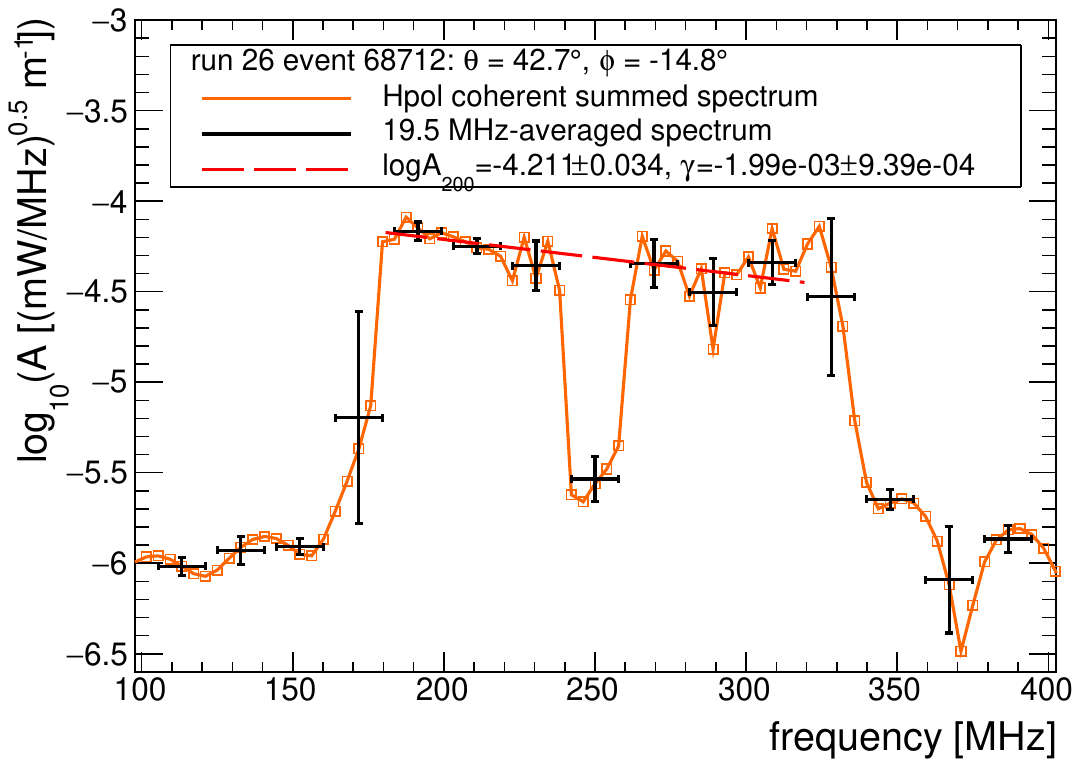}
    	\end{subfigure}
        \caption{Best fit (red dashed line) of \SI{20}{\MHz}-averaged (black cross) coherently summed Hpol spectra (orange curves) of four of the cosmic ray events used for the energy measurement, shown in order of decreasing reconstructed zenith angles: \ang{82} (top left), \ang{79} (top right), \ang{67} (bottom left), and \ang{43} (bottom right).
        }
        \label{fig:SpecFit}
    \end{figure}

      %
    The measured spectra used in the analysis are transformed from the coherently summed Hpol waveforms with deconvolution and filtering, and after subtracting
    average noise amplitude spectrum derived from the forced-trigger events. 
    %
    %
    However, as the noise spectrum is roughly flat and the amplitude is reduced due to the intrinsic incoherence of noise (with $\log A(f) \approx -5.4$), this correction does not affect the measured slope and intercept of detected events significantly.  
    %
    The signal amplitude spectrum is averaged in \SI{20}{\MHz} bins and the RMS value is added in quadrature to the statistical uncertainty for estimating the total signal uncertainty. 
    All spectra of the candidate events are consistent with that of an air shower; numerical results for four candidates are presented in Fig.~\ref{fig:SpecFit}.

    \subsubsection{Energy Fitting with Simulated Signals}
    To obtain the expected distribution of spectral parameters for the energy measurement, additional proton-induced showers at reconstructed directions $(\tilde{\theta},  \tilde{\phi})$ of the CR candidates were simulated, with three showers per energy bin of $\log E = 0.25$ between $\log E = 17.0-19.0$. 
    Each candidate has to be studied separately because the shower direction, mainly the zenith angle, affects the detection geometry including the distance to the shower and its altitude, which dictate the Cherenkov angle, resulting in different distributions of spectral parameters.
    %
    For example, inclined showers reach their maximum at higher altitude of lower air density and thus smaller Cherenkov angles and farther distances, from $\psi_{c}\approx\ang{1.2}$  at $\theta=\ang{45}$ to $\psi_{c}\approx\ang{0.4}$ at $\theta=\ang{89}$.
    Showers previously generated for the acceptance estimation in Sec.~\ref{sec:CRSim} with zenith angles within $\tilde{\theta} \pm \ang{1}$ (\ang{\pm3} for $\tilde{\theta} < \ang{60}$) and azimuth angles $\tilde{\phi} \pm \ang{45}$ were also included in the analysis, as it is found that the change in shower azimuth angle for zenith angles considered here only causes slight variations in the geomagnetic radiation for the near vertical geomagnetic field at the experimental site.
    %
    %
    The expected spectral parameters were extracted by fitting the spectra of the Hpol component of the simulated electric field.

\begin{figure}
    \centering
       
    	\begin{subfigure}{0.47\textwidth}
    		\includegraphics[width=\textwidth]{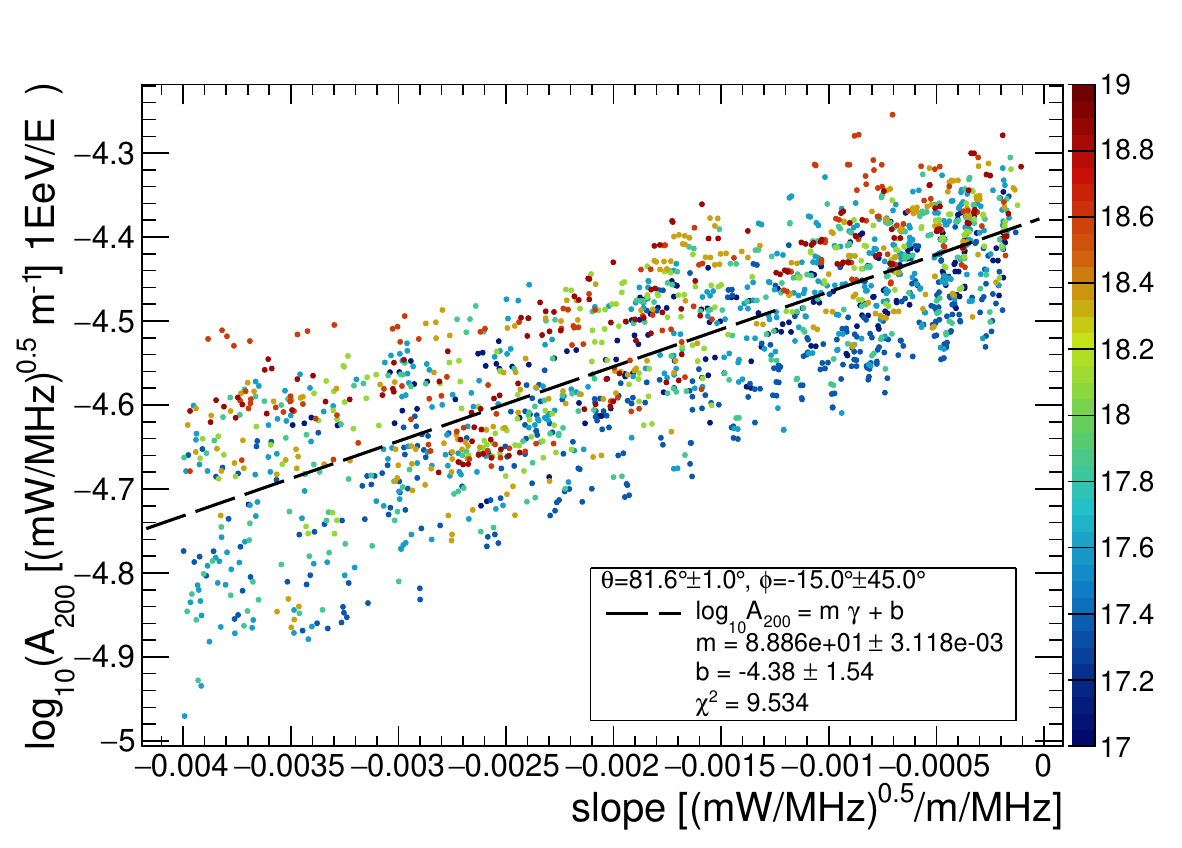}
    	\end{subfigure}
        \hfill
        \begin{subfigure}{0.47\textwidth}
    		\includegraphics[width=\textwidth]{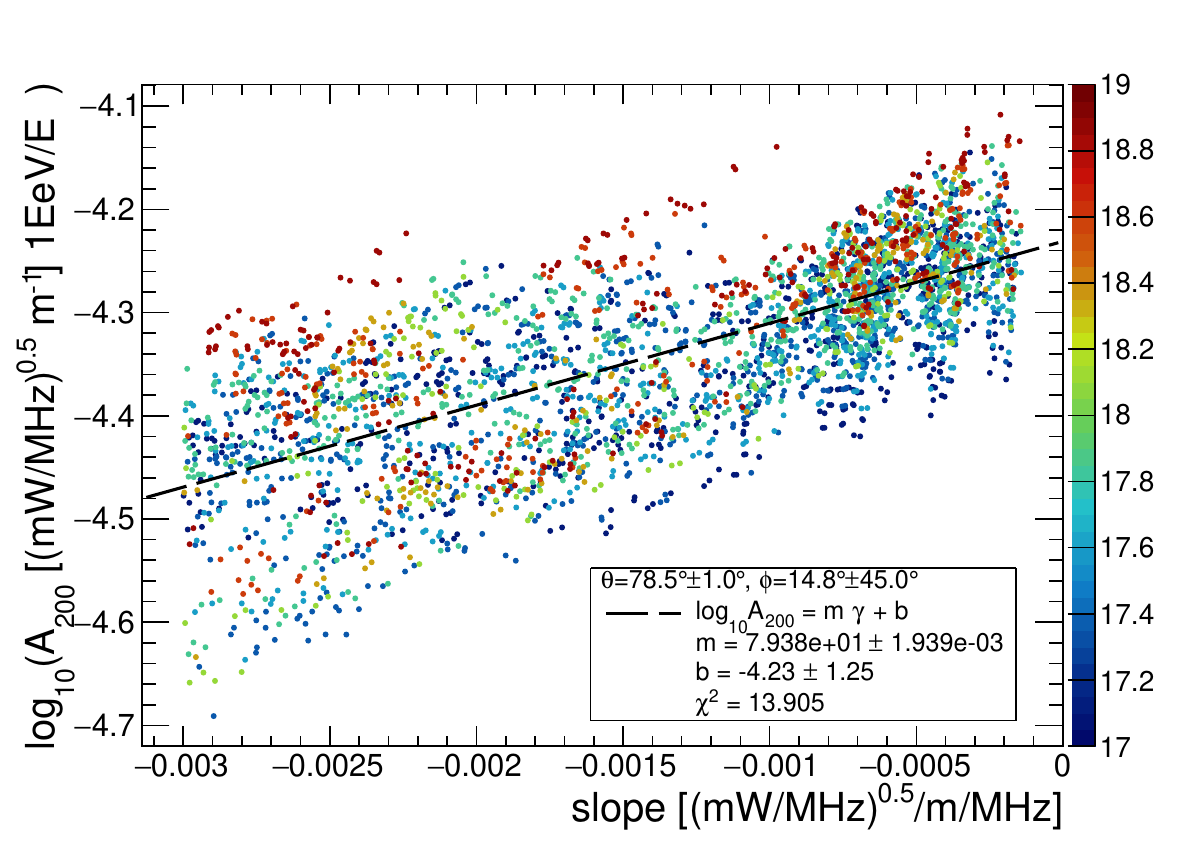}
    	\end{subfigure}
    	
    	\begin{subfigure}{0.47\textwidth}
    		\includegraphics[width=\textwidth]{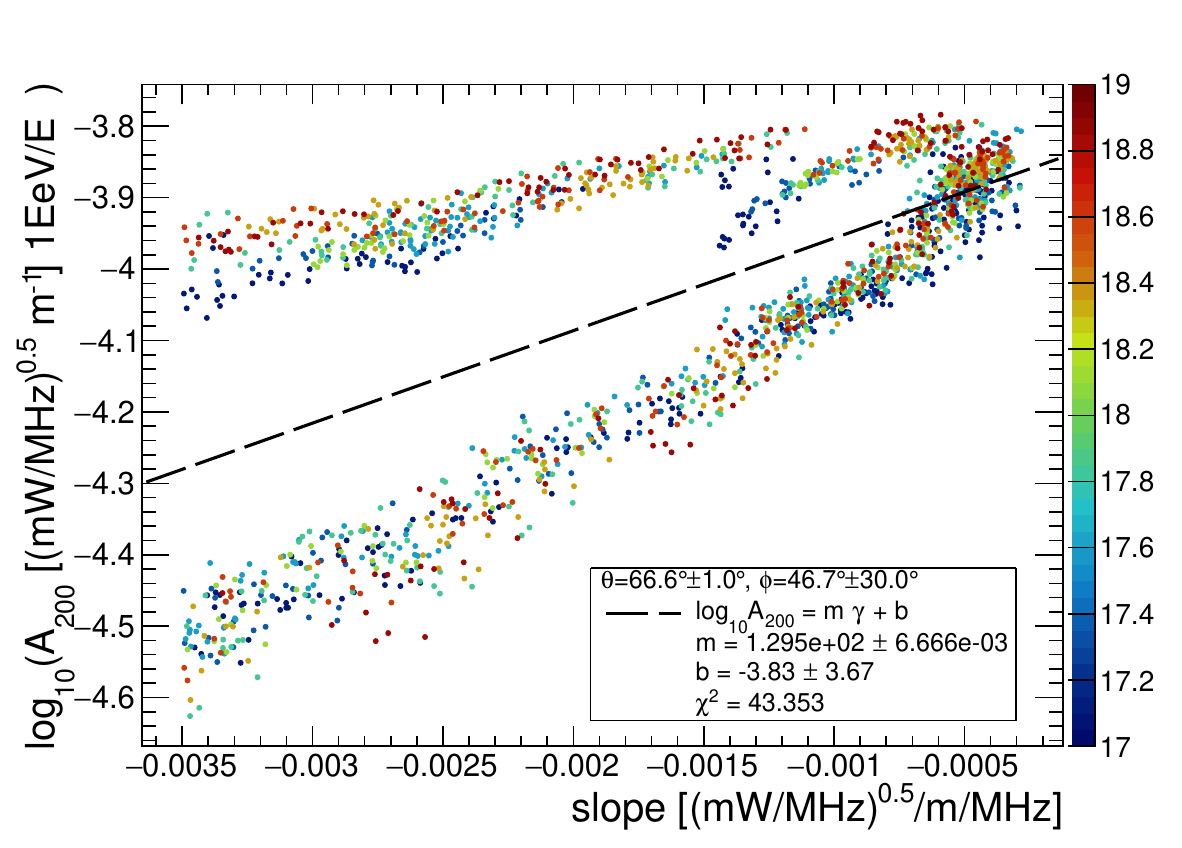}
    	\end{subfigure}
    	\hfill
    	\begin{subfigure}{0.47\textwidth}
    		\includegraphics[width=\textwidth]{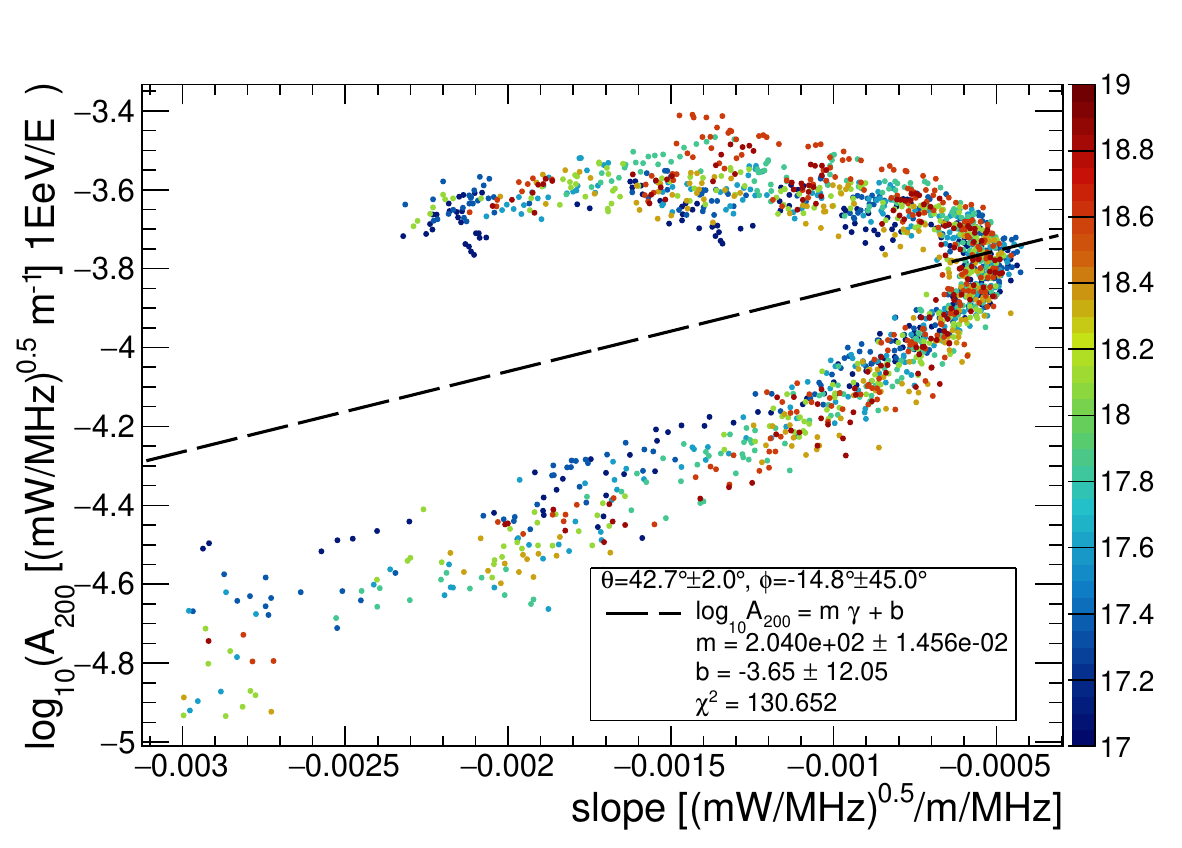}
    	\end{subfigure}

        \caption{ The distribution of spectral slopes and energy-normalized intercepts (to \SI{1}{\exa\eV}) for simulated  showers (markers with colors indicating the energy $\log E$) at the reconstructed directions of the four cosmic ray events in Fig.~\ref{fig:SpecFit}. 
        The global linear fit (dashed line) using Eq.~\ref{eq:logA_slope} is also indicated. Note that the linear fit only approximates the behavior reliably for zenith angles above \ang{70}, below which the ambiguity in viewing angle around the Cherenkov cone becomes more evident, with the upper one corresponding to $\psi < \psi_{c}$ and the lower one to $\psi > \psi_{c}$.
        }
        \label{fig:CREFit_Enorm}
    \end{figure}

    %
    %
    The energy fitting relies on the coherence of the emission at the Cherenkov angle, and hence is verified first by checking if the signal amplitude is proportional to the primary energy of cosmic rays,
    \begin{equation}
        \log  A_{200,c} = p_0 + p_1 \log  E,
        \label{eq:coherence}
    \end{equation}
    where the subscript $c$ denotes quantities at the Cherenkov angle and parameters $p_0$ and $p_1$ are fitted from the distribution of simulated signals.
    %
    The linearity was verified ($p_1\approx1$) within a \SI{10}{\percent} error for all considered shower directions. 
    %
    It was found in Ref.~\cite{ANITA2016a,ARIANNA2017} that the distribution of intercept and slope ($\log A_{200}, \gamma$) roughly follows a linear relation when observing near the Cherenkov angle,
    \begin{equation}
        \log A_{200} = \log A_{200,c} + m (\gamma - \gamma_{c}),
        \label{eq:logA_slope}
    \end{equation}
    where $\log A_{200,c}, m$ are determined from fitting and $ \gamma_{c}$ are estimated by the maximum sampled values of each simulated shower.
    %
    %
    The spectral intercept $\log A_{200}$ is found to roughly scale with the primary energy and the ($\log A_{200}, \gamma$) distribution does not vary significantly, after energy normalization. 
    Therefore a global linear fit with Eq.~\ref{eq:logA_slope} of the energy-normalized distribution was performed for all the showers at a given direction, using the average value of $\gamma_{c}$.
    As shown in Fig.~\ref{fig:CREFit_Enorm}, the linear relation is a good approximation for showers at inclined angles $\theta > \ang{70}$ for TAROGE-M.

    %
    However, in Fig.~\ref{fig:CREFit_Enorm}, for showers at lower zenith angles $\theta < \ang{70}$, the spectral distribution deviates from a straight line and exhibits two branches with turning point at the Cherenkov angle, where the upper branch corresponds to observations inside the Cherenkov ring ($\psi < \psi_{c}$) and the lower one to observations outside the ring ($\psi > \psi_{c}$).  
    This branching results in an ambiguity in determining the energy.
    %
    One possible explanation for the asymmetry across the Cherenkov ring is that the distance to shower maximum becomes shorter at smaller zenith angles ($R\sim \SI{20}{\km}$ at $\theta=\ang{70}$), and the angular size of the longitudinal shower profile (about \SI{100}{\m}) becomes non-negligible.
    The observers inside and outside the ring see different parts of the shower at different angles and delays, and hence in general the two receive different signals.
    %
    %
    This phenomenon was also reported and investigated previously in Ref.~\cite{Welling2019}, where it was suggested that the geometric ambiguity may be resolved by fitting the spectrum with a quadratic, rather than linear function, thereby improving the energy estimate. 
    However, in this analysis, the fitting did not yield an improvement in energy resolution, perhaps due to the different electric field reconstruction applied here, as well as the frequency band for this analysis (\SIrange{180}{240}{\MHz} and \SIrange{280}{320}{\MHz}) being both higher and also narrower than that used in Ref.~\cite{Welling2019} (\SIrange{80}{300}{\MHz}). 
    %
    Therefore, we conservatively took the largest deviation from the fitted line in the relevant parameter range as part of the energy systematic uncertainty in the current analysis.


    Combining Eq.~\ref{eq:coherence} and \ref{eq:logA_slope}, the primary energy can be estimated by
    \begin{equation}
        \log E = \frac{1}{p_1}[\log A_{200} - p_0 - m(\gamma - \gamma_{c})],
        \label{eq:CREFit}
    \end{equation}
    %
    The systematic uncertainty in the energy is propagated from the errors on the ($p_0$,$p_1$) and ($\log A_{200,c}, m$) parameters returned from the fit, the spread of $\gamma_c$, and the maximum deviation from the line fit.
    The statistical uncertainty is determined from fitting the uncertainty in ($\log A_{200}, \gamma$) of the measured spectra.
    Both uncertainties are added in quadrature.
    %
    An extra scale uncertainty of $\sigma_{\log E}=0.1$ due to the \SI{2}{\dB} antenna gain uncertainty is included and added coherently to the others.

    \begin{figure}
    \centering
       
    	\begin{subfigure}{0.47\textwidth}
    		\includegraphics[width=\textwidth]{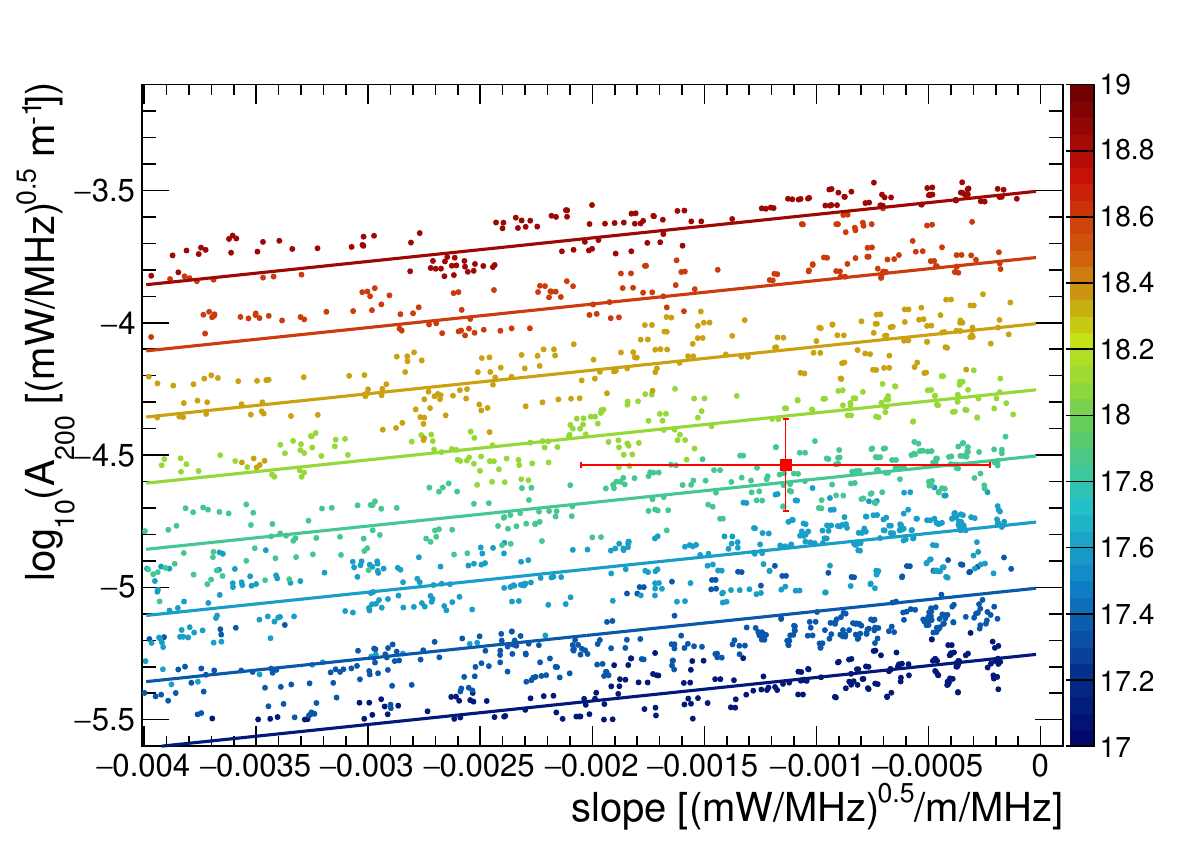} 
    	\end{subfigure}
        \hfill
        \begin{subfigure}{0.47\textwidth}
    		\includegraphics[width=\textwidth]{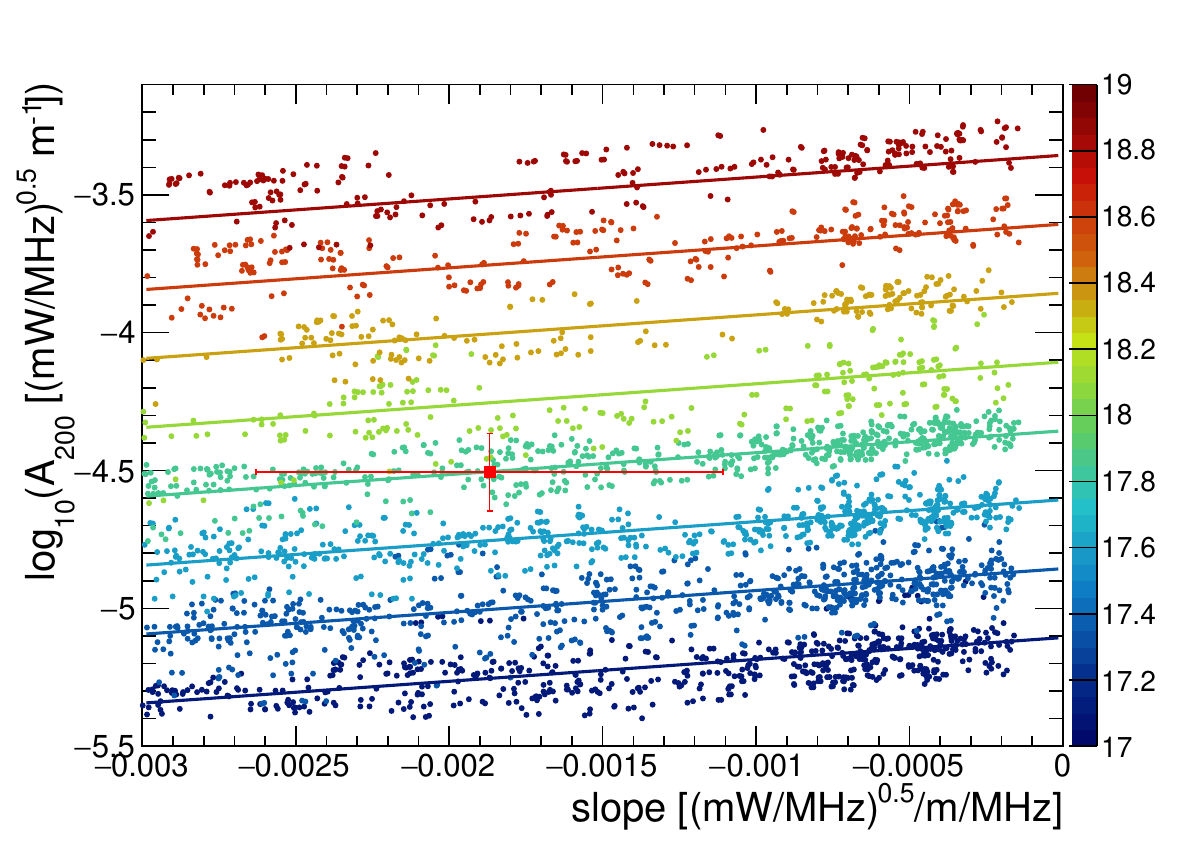} 
    	\end{subfigure}
    	
    	\begin{subfigure}{0.47\textwidth}
    		\includegraphics[width=\textwidth]{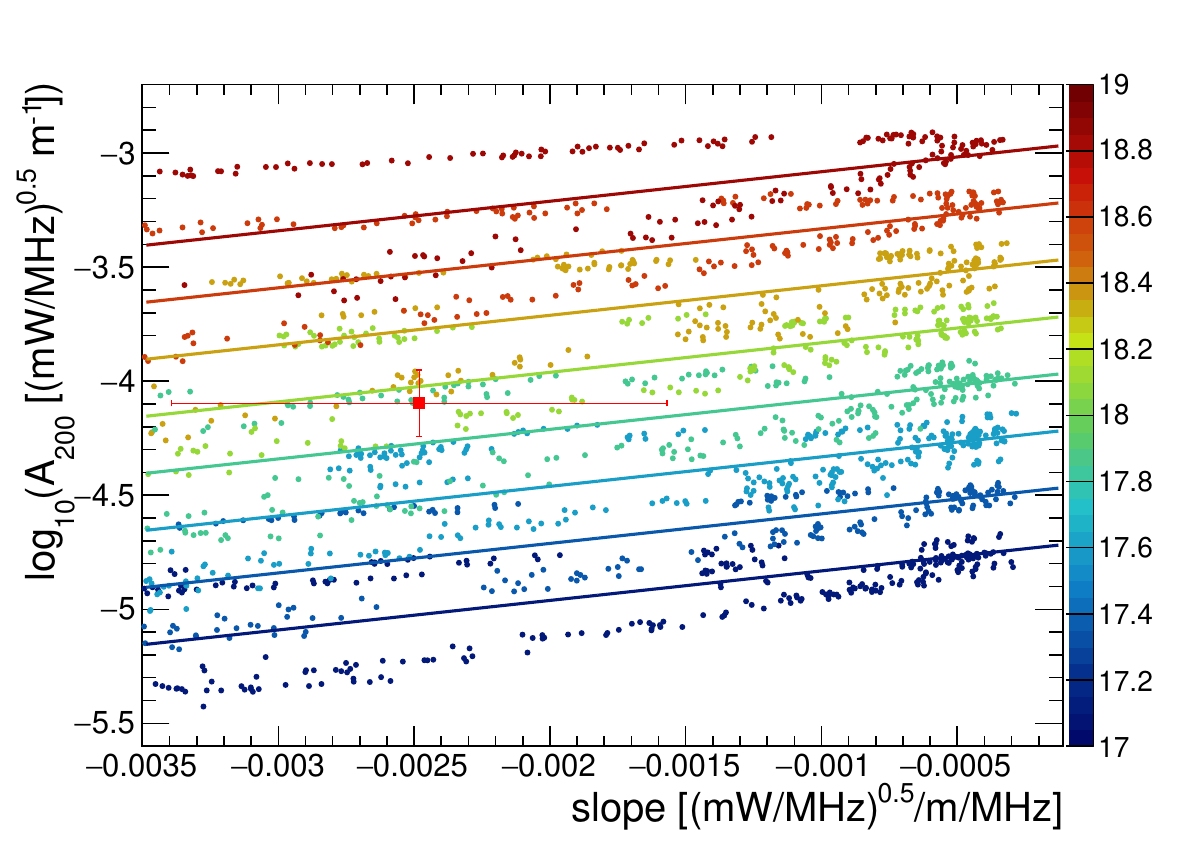} 
    	\end{subfigure}
    	\hfill
    	\begin{subfigure}{0.47\textwidth}
    		\includegraphics[width=\textwidth]{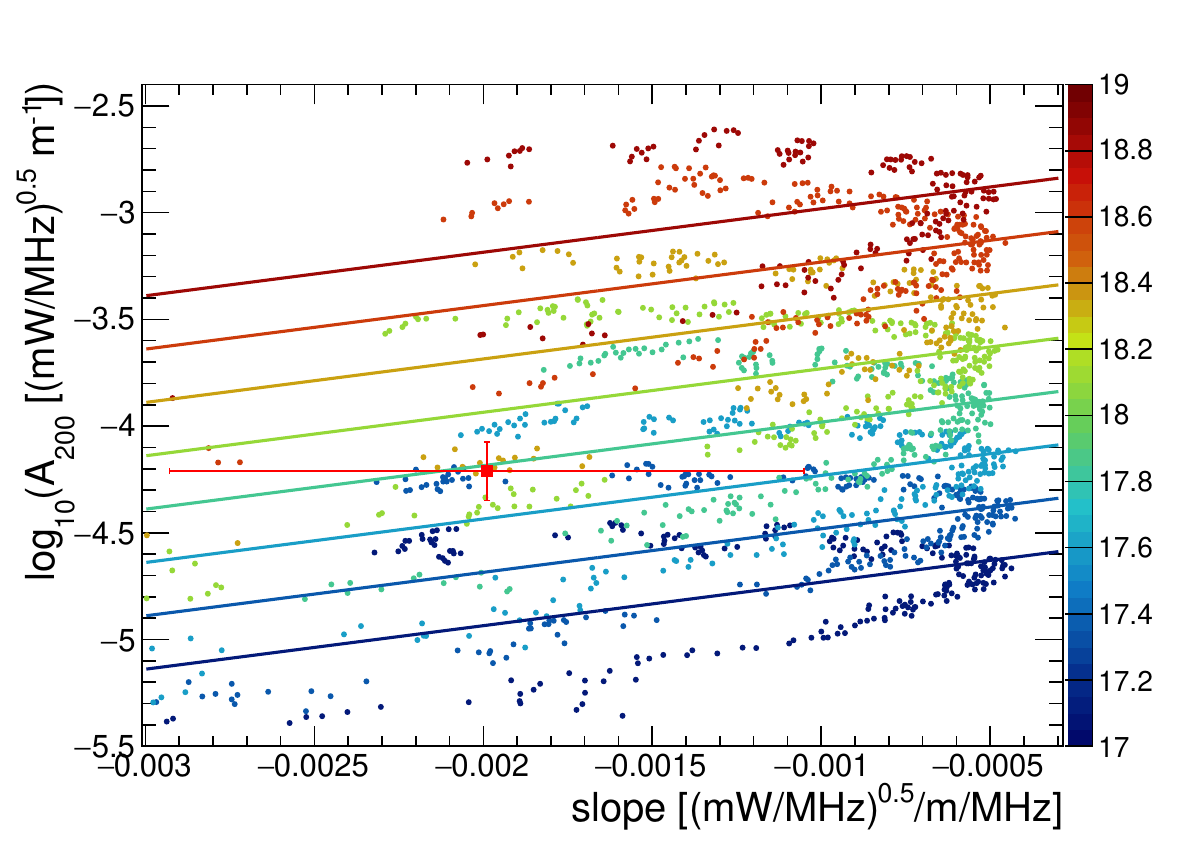} 
    	\end{subfigure}

        \caption{Each panel shows the spectral intercept versus slope distribution of simulated showers (colored markers indicating the primary energy) in the reconstructed direction of each cosmic ray event in Fig.~\ref{fig:SpecFit}, with lines from the linear fit with Eq.~\ref{eq:logA_slope} (shown in Fig.~\ref{fig:CREFit_Enorm}).  
        The measured spectral parameters of each event in Fig.~\ref{fig:SpecFit} (red marker with error bars) are superimposed for the energy measurement.
        }
        \label{fig:CREFit}
    \end{figure}

    \subsubsection{Measured Event Energy and Distribution}
   
    The estimated energies $\log \hat{E}$ of the seven CR candidates derived from Eq.~\ref{eq:CREFit} are summarized in Table \ref{tab:CR}; four examples are shown in Fig.~\ref{fig:CREFit}.
    The inverse-variance weighted mean energy of these events is $\langle \log E \rangle = 17.98 \pm 0.17 \text{(statistical)} \pm 0.10 \text{(scale)}$, or $0.95_{-0.31}^{+0.46} \si{\exa\eV}$,
    %
    where the uncertainty includes the energy spread (represented by the weighted RMS) of the events and the overall scale uncertainty.
    %
    %
    The main uncertainty is due to the geometric ambiguity in the observation angle relative to the Cherenkov ring, which increases at smaller zenith angles, and with $\sigma_{\log E} = 0.17$ $\rightarrow$ $0.72$ for $\theta = \ang{82}$  $\rightarrow$ \ang{43}.

\begin{figure}
    	\centering
    	\begin{subfigure}{0.49\textwidth}
    		\includegraphics[width=\textwidth]{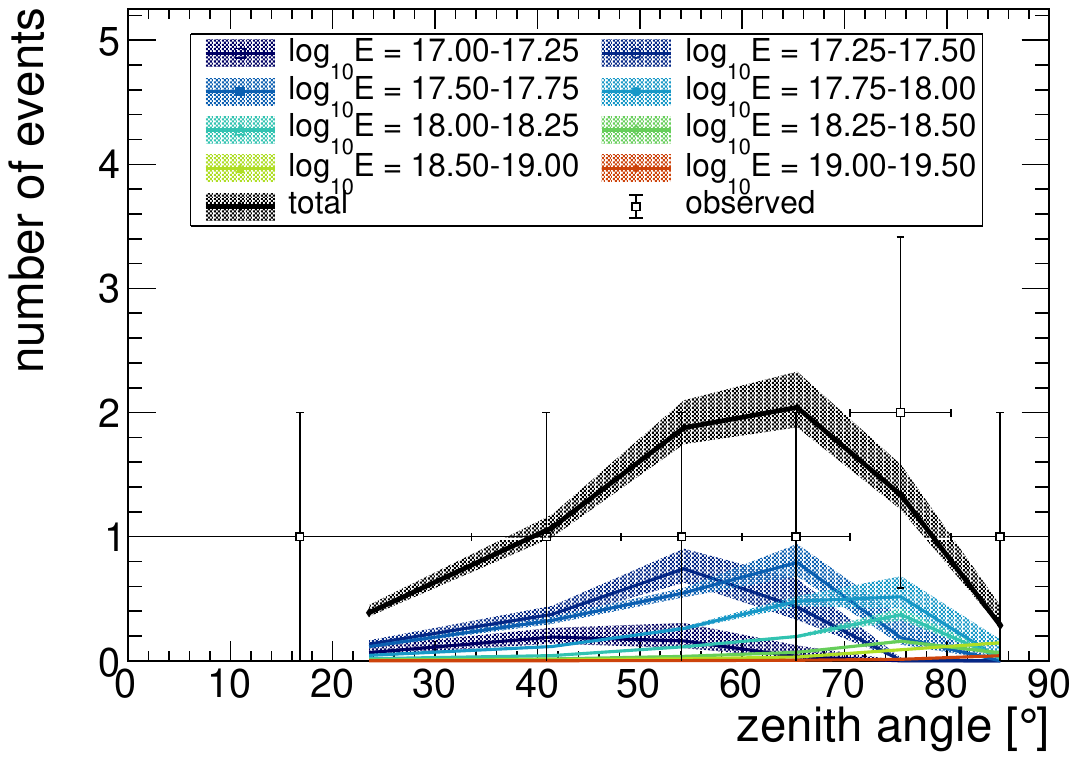}
    	\end{subfigure}
    	\begin{subfigure}{0.49\textwidth}
    		\includegraphics[width=\textwidth]{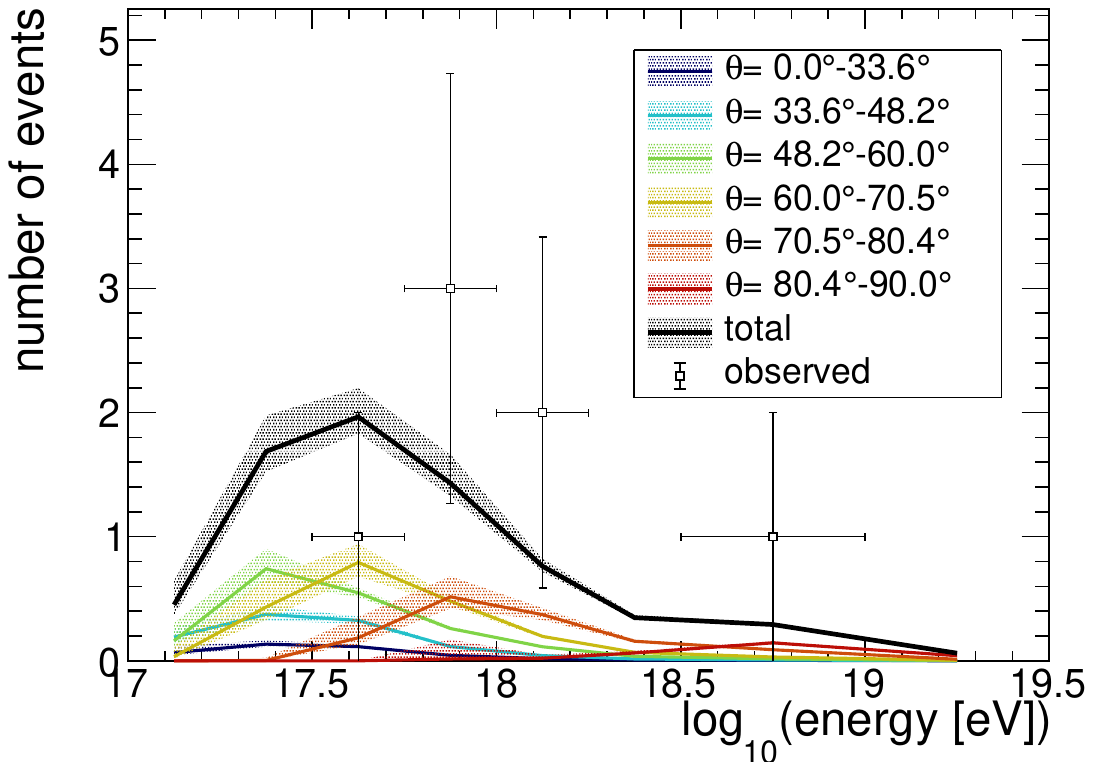}
    	\end{subfigure}
    \caption{ 
        The zenith-angle (left) and energy (right) distributions of detected cosmic ray events (markers with error bars), compared with those predicted by simulation for each energy and zenith range (colored curves), obtained by integrating the average TAROGE-M exposure (Fig.~\ref{fig:CRacceptance}) with the cosmic ray energy spectrum measured by the Pierre Auger Observatory \cite{Auger2019}. 
        %
        The expected distributions are scaled to the number of observed events to compare shapes.
    }
    \label{fig:CRdist}
    \end{figure}

    The event at \ang{25} zenith angle (run\# 25 event\# 54906) has the highest estimated energy, as it is outside of the main lobe of antennas and its momentum vector makes a smaller angle with the geomagnetic field.
    But it also has the largest uncertainty, by more than an order of magnitude, as showers very close to the detector (less than \SI{3}{\km} to shower maximum) may reach the mountain top and be clipped without full development. 
    %
    The resulting zenith-angle and energy distributions of the seven detected events are otherwise consistent with expectation, as shown in Fig.~\ref{fig:CRdist}.

    In the future, the energy measurement method will be modified, as the method adopted here requires a collection of simulated showers for each detected event, and will eventually become prohibitively computing intensive as the detector exposure increases. 
    The alternative method suggested in Ref.~\cite{Welling2019} using a parameterization will be investigated and applied to TAROGE-M to evaluate the reduction in computational demand.
    If successful, the method would eliminate the major uncertainty caused by the angular ambiguity. In simulations, an energy resolution of about \SI{20}{\percent} was reported in \cite{Welling2019} for a study modeling the configuration of an ARIANNA station. 
    %

    \begin{figure}
        \centering
    
    	\includegraphics[width=.6\textwidth]{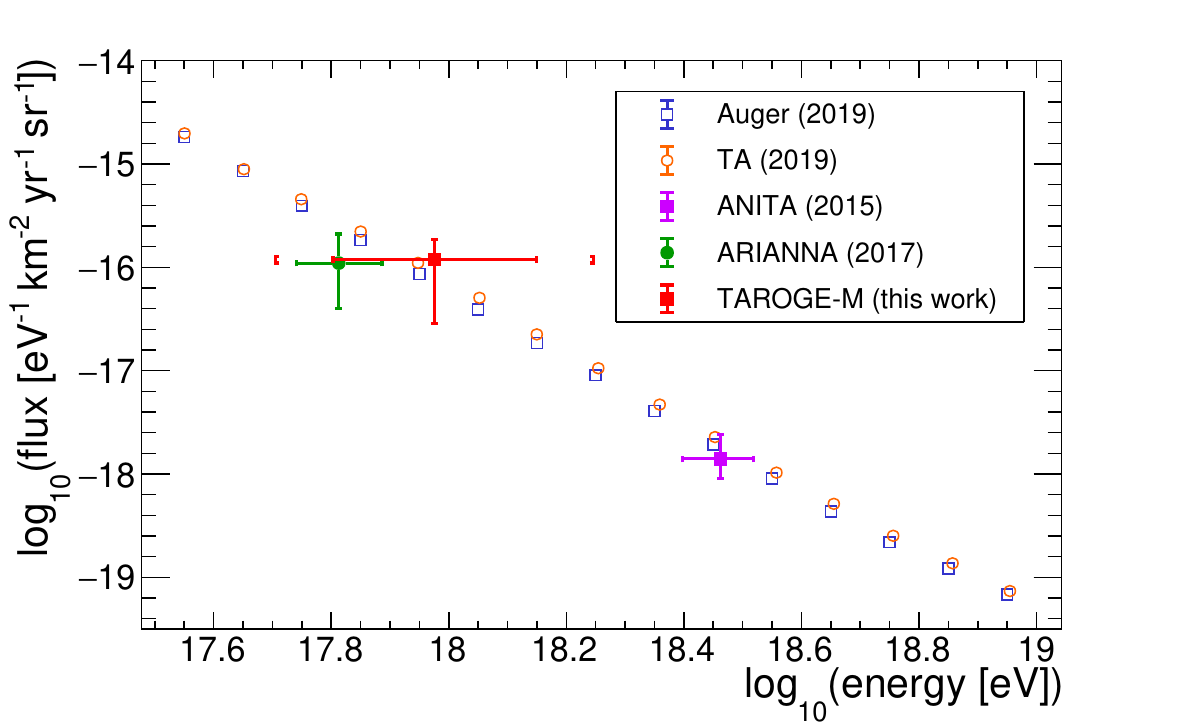}
        \caption{Cosmic ray flux measured with the TAROGE-M station using $25.3$-days of data using Eq.~\ref{eq:flux-MD} (red marker), compared to the energy spectrum measured by the Pierre Auger Observatory \protect\cite{Auger2019} and the Telescope Array \protect\cite{TA2019}, as well as those previously reported by the ANITA \protect\cite{ANITA2016a} and ARIANNA \protect\cite{ARIANNA2017} radio experiments.
        The energy error bar of TAROGE-M denotes the spread (weighted RMS) of the seven detected events, whereas the outer brackets indicate the scale uncertainty.
        }
        \label{fig:CRflux}
    \end{figure}

    \subsection{Cosmic Ray Flux Measurement}
    \label{sec:CRflux}
   
    Due to the limited number and limited range in the energy of the detected cosmic ray events, we used all the events to estimate the cosmic ray flux for a single energy bin centered at the experimentally-inferred mean energy of $\langle \log E \rangle = 17.98 \pm 0.17 \text{(stat.)} \pm 0.10 \text{(scale)}$,
   %
    The flux is estimated using Eq.~\ref{eq:rate} by dividing the number of events $N_{CR}=7$ (with Poisson error $\pm \sqrt{7})$) by the analysis efficiency $\eta=$\SI{89.5}{\percent} (see Table \ref{tab:selection}), the mean energy, the average acceptance at the energy $\langle A\Omega \rangle$, and the livetime $T_{\rm live}$, 
    %
    The acceptance from Sec.~\ref{sec:CRSim} along with the attendant systematic uncertainties are estimated at $\langle A\Omega \rangle = 2.98_{-0.97}^{+0.94} \times 10^{-1}$ \si{\square \km \steradian}.
    %
    The total livetime is $25.3$ days, excluding the high-wind samples (in Sec.~\ref{sec:highwind}) and correcting for temporal clustering (Sec.~\ref{sec:selection}).
    %
    %
    The flux at the mean energy of $\log E = 17.75-19.0$, covering all detected events is
    %
    \begin{align}
         \Phi(\langle E \rangle) = \frac{N_{CR} }{ \eta T_{\rm live}  \langle A\Omega \rangle (\langle E \rangle) \langle E \rangle   \ln 10 \Delta \log E }.
        \label{eq:flux-MI}
    \end{align}
    Using the values above and propagating the uncertainties, the estimated cosmic ray flux is $1.4_{-0.7}^{+0.7} \times 10^{-16}$ \si{ \per \eV \per \square \km \per \year \per \steradian} at $0.95_{-0.31}^{+0.46}$ \si{\exa\eV}. 
    
    Alternatively, a model-dependent flux at the mean energy can be estimated assuming that the cosmic ray flux can be described as a piece-wise power law, $\Phi(E) = \Phi(\langle E \rangle) (E/\langle E \rangle)^{\alpha}$ with the spectral index $\alpha=-3.3$ obtained from the Auger CR spectrum at \si{\exa\eV} below the ankle \cite{Auger2019}, 
     \begin{align}
         \Phi(\langle E \rangle) = \frac{N_{CR} }{ \eta T_{\rm live}  \langle E \rangle^{-\alpha} \int \langle A\Omega \rangle (E) E^{\alpha} dE },
        \label{eq:flux-MD}
    \end{align}
    integrated over the interval $\log E = 17.0-19.5$.
    This method yields a flux ( $1.2_{-0.9}^{+0.7} \times 10^{-16}$ \si{ \per \eV \per \square \km \per \year \per \steradian}) similar to the one above.
    This value is consistent with those previously reported by other experiments, as summarized in Fig.~\ref{fig:CRflux}.

    In summary, the seven candidate events have polarizations, spectral features, angular and energy distributions, and an event rate that are consistent with those of UHE cosmic rays. Hence we can conclude that the first TAROGE-M station is able to detect UHE air showers and demonstrates discovery-potential.

\section{Conclusion and Future Work} 
    %
    Radio antenna arrays on top of Antarctic mountains can not only detect air showers induced by UHE cosmic rays and tau neutrinos efficiently in near horizontal directions, but also show promising potential for reproducing the discovery of ANITA anomalous upward-going air shower events, and help in clarifying their origin.
    The initial TAROGE-M station that has been deployed in 2020 atop Mt.~Melbourne is the first realization of the approach using the geographical advantages for radio detection of air showers generated by ultra-high energy particles. 
    Despite power problems that interrupted operation in 2020, the station has been successfully calibrated to sub-degree angular resolution for event reconstruction.
    Seven UHECR events were detected by the initial TAROGE-M station in $25.3$-days of data-taking. Their polarizations, spectral features, energy, angular distributions, and estimated flux are consistent with results from simulation, as well as those from other experiments, demonstrating TAROGE-M's ability to detect UHE air showers and deliver science.

    The primary goal of TAROGE-M in the coming season is to achieve long-term operation in the austral summer by upgrading station power provision.
    %
    %
    As the deployment procedures have been proven robust, two more receiver antennas will be added to improve the angular resolution and the trigger efficiency.
    With increased knowledge about the noise background, the trigger threshold can be further relaxed, and the antenna and the filter design will be improved with less loss and dispersion, so as to lower the energy threshold for air shower detection.
    %
    Several candidate sites on Mt.~Melbourne have been found and more stations can be built with different orientations in the next 2--3 of years to provide coverage of different parts of the horizon and the sky.
    A long-range Wi-Fi link from Mt.~Melbourne to the nearby station JBS as a substitute for satellite communication is also planned for a larger data transfer bandwidth. 
    A ground-based pulser system about \SI{10}{\km} away from the mountain will be set up for long-term monitoring of the reconstruction performance below the horizon, and more drone pulser flights will be conducted for calibrating the possible interference from ground reflections in near horizontal directions.
    %
    %
    We hope that, by fully taking scientific advantage of radio detectors on Antarctic mountains, TAROGE-M can serve as a fast probe for unraveling the mystery of ANITA's anomalous events and the discovery of UHE cosmic tau neutrinos.

\section{Acknowledgements}

    We would like to acknowledge support from the Taiwan Ministry of Science and Technology (MOST) for the TAROGE-M project under grant 110-2112-M-002-037, and from the Korea Polar Research Institute (KOPRI), as well as the great logistical and field supports from on-ice Jang Bogo station crews in 2018--2019 and 2019--2020 season.
    This work was also partly supported by research grant PE22020 from KOPRI.
    %
    %
    %
    We are grateful to the U.S.~National Science Foundation-Office of Polar Programs, the U.S.~National Science Foundation-Physics Division (grant NSF-1607719) for supporting the ARIANNA array at Moore's Bay, and NSF grant NRT 1633631 and Award ID 2019597. 
    Without the invaluable contributions of the people at McMurdo Station, the ARIANNA stations would have never been built. 
    We acknowledge funding from the German research foundation (DFG) under grants NE 2031/2-1, and the Swedish Government strategic program Stand Up for Energy. 
    The authors are also thankful for the help and the facilities offered by the National Space Organization (NSPO) in Taiwan for the antenna calibration.

\bibliographystyle{JHEP.bst}
\bibliography{ref}

\end{document}